\documentclass[fleqn,usenatbib,usenames,dvipsnames]{mnras}

\usepackage[T1]{fontenc}

\usepackage{pifont}

\DeclareRobustCommand{\VAN}[3]{#2}
\let\VANthebibliography\thebibliography
\def\thebibliography{\DeclareRobustCommand{\VAN}[3]{##3}\VANthebibliography}

\usepackage{graphicx}	
\usepackage{amsmath}	
\usepackage{amssymb}	
\usepackage{subcaption}

\usepackage{tikz}

\usepackage{newtxtext,newtxmath}

\title[Binary companions triggering fragmentation]{Binary companions triggering fragmentation in self-gravitating discs}

\author[Cadman et al.]{
James Cadman$^{1,2}$,
Cassandra Hall$^{3,4}$,
Clémence Fontanive$^{5}$,
Ken Rice$^{1,2}$,
\\
$^{1}$SUPA, Institute for Astronomy, University of Edinburgh, Blackford Hill, Edinburgh, EH9 3HJ, Scotland, UK\\
$^{2}$Centre for Exoplanet Science, University of Edinburgh, Edinburgh, UK\\
$^{3}$Department of Physics and Astronomy, The University of Georgia, Athens, GA 30602, USA. \\
$^{4}$Center for Simulational Physics, The University of Georgia, Athens, GA 30602, USA.\\
$^{5}$Center for Space and Habitability, University of Bern, Bern 3012, Switzerland.
}

\date{Accepted 2021 December 30. Received 2021 December 8; in original form 2021 October 12
}

\pubyear{2021}

\begin{document}
\label{firstpage}
\pagerange{\pageref{firstpage}--\pageref{lastpage}}
\maketitle

\begin{abstract}
Observations of systems hosting close in ($<1$\,AU) giant planets and brown dwarfs ($M\gtrsim7$\,M$_{\rm Jup}$) find an excess of binary star companions, indicating that stellar multiplicity may play an important role in their formation. There is now increasing evidence that some of these objects may have formed via fragmentation in gravitationally unstable discs. We present a suite of 3D smoothed particle hydrodynamics (SPH) simulations of binary star systems with circumprimary self-gravitating discs, which include a realistic approximation to radiation transport, and extensively explore the companion's orbital parameter space for configurations which may trigger fragmentation. We identify a "sweet spot" where intermediate separation binary companions ($100$\,AU $\lesssim a\lesssim400$\,AU) can cause a marginally stable disc to fragment. The exact range of ideal binary separations is a function of the companion's eccentricity, inclination and mass. Heating is balanced by efficient cooling, and fragmentation occurs inside a spiral mode driven by the companion. Short separation, disc penetrating binary encounters ($a\lesssim100$\,AU) are prohibitive to fragmentation, as mass stripping and disc heating quench any instability. This is also true of binary companions with high orbital eccentricities ($e\gtrsim0.75$). Wide separation companions ($a\gtrsim500$\,AU) have little effect on the disc properties for the setup parameters considered here. The sweet spot found is consistent with the range of binary separations which display an excess of close in giant planets and brown dwarfs. Hence we suggest that fragmentation triggered by a binary companion may contribute to the formation of these substellar objects.

\end{abstract}




\section{Introduction}
There are both theoretical \citep{linpringle87,rice10} and observational \citep{rodriguez05,tobin12,tobin15} indications that during the earliest stages of star formation, protostellar disc masses may be a significant fraction of the mass of the central protostar.  If so, these discs would be susceptible to the growth of a gravitational instability \citep{toomre64}.

The evolution of a gravitationally unstable disc can follow two basic pathways. It can settle into a quasi-steady state \citep{paczynski78} in which spiral density waves act to drive angular momentum outwards, allowing mass to accrete onto the central protostar \citep{linpringle87,laughlin94,lodato04}. Recent observations of very young protostellar systems have shown the presence of spirals \citep{perez16}, consistent with the disc being gravitationally unstable \citep{dong2015,hall2016,meru17,hall18,cadman2020b,hall2020,paneque2021,veronesi2021}.

The other potential pathway is that the gravitational instability (GI) can lead to a disc becoming so unstable that it fragments into bound objects that could then contract to become gas giant planets, or brown dwarfs \citep{boss97,boss00,mayer02,durisen07}. This outcome, however, requires not only that the disc is gravitationally unstable, but also that it can also cool very efficiently \citep{gammie01, rice03}. Such conditions are unlikely to be satisfied in the inner parts of protostellar discs \citep{rafikov05,clarke09,rice09}, suggesting that this pathway is unlikely to play a dominant role in the in-situ formation of the known close-in gas giant planets/exoplanets \citep{boley09,johnson13}.

However, it is possible that such a process may operate in the outer parts of extended protostellar discs \citep{stamatellos09,vorobyov10}, potentially explaining the origin of some directly-imaged, wide-orbit planetary-mass and brown dwarf companions \citep{nero09,kratter10,cadman2021,humphries2021}. It has been shown that fragmentation is favoured in discs around higher mass stars \citep{cadman2020frag,haworth2020}, consistent with results from direct imaging surveys which find an excess of wide-orbit, giant planets in these systems \citep{nielsenetal19}. Population synthesis models \citep{forgan13,forgan18} suggest that the mass distribution of planets formed via GI is indeed consistent with results from direct-imaging surveys of wide-orbit giant planets and brown dwarfs \citep{Vigan17,vigan21}.

It has also been suggested that objects that form via GI on wide orbits could migrate inwards rapidly \citep{baruteau11} and potentially undergo tidal stripping \citep{nayakshin10, boley10} to produce close-in planets with a wide range of masses \citep{nayakshin15}. However, hydrodynamics simulations of such systems show that these objects either stay on wide orbits or are destroyed during the migration process \citep{hall2017}. Population synthesis models also suggest that such an outcome is relatively rare, and that most objects that form via GI will remain on wide-orbits as giant planets, or brown dwarfs \citep{forgan13,forgan18}. %
Such objects could, though, still be scattered onto highly eccentric orbits that can then tidally circularise onto close-in orbits \citep{rice15}. This will tend to form gas giant planets, or brown dwarfs, with very close-in circular orbits (with orbital properties similar to those of 'hot' Jupiters) or eccentric orbits that are still undergoing tidal circularisation.

Given that the scatterer is likely to be a companion to the host star, this motivated a search for companions to systems with close-in massive planets, or brown dwarfs \citep{fontanive19}. The results of this search did indeed indicate a binary fraction twice as high as for field stars on projected separations between 20--10,000~AU. However, only about half of these systems were consistent with high eccentricity migration through secular interactions with the outer stellar companion, the others being on orbits where the tidal circularisation timescale was far too long to explain their origin \citep{fontanive19}.

Nonetheless, even if the close-in objects were not scattered onto their current orbits, the high binary fraction for these systems suggests that the existence of a companion may still influence their formation. There are also indications that some of these objects may have formed via GI. The sample of stars studied in \cite{fontanive19}, hosting close-in companions with masses between 7--60~M$_\mathrm{Jup}$, has a mean metallicity of $\langle$[Fe/H]$\rangle$ = $-$0.12, consistent with the mean field metallicity \citep{moe19}. This is substantially lower than the mean metallicity for hosts to genuine hot Jupiters (0.2--4~M$_\mathrm{Jup}$) of $\langle$[Fe/H]$\rangle$ = 0.23 \citep{santos04,fischervalenti05}, which also do not show the same excess in multiplicity frequency \citep{ngoetal16,moekratter19}.

This lower-mass planetary population is thought to have formed via the alternative scenario for planet formation, core accretion (CA; \citealp{pollack96}). This formation mechanism shows a strong metallicity dependence in the formation of giant planets with masses above a few Jupiter masses \citep{mordasini12,jenkins17}. In contrast, the GI formation process has no metallicity dependence \citep{merubate10}, and preferentially forms massive planets or brown dwarfs \citep{kratter10,forganrice11}, with a transition at around $\sim$4--10~M$_\mathrm{Jup}$ between the two mechanisms \citep{schlaufman18}.

This suggests that only the most massive planetary and brown dwarf companions, likely forming via GI, are effected by stellar binarity.
\cite{fontanive21} recently confirmed this idea, finding that close-in exoplanets and substellar companions with masses of several Jupiter masses and above are almost exclusively observed in binary-star systems with separations of a few hundred AU or less. In contrast, sub-Jovian and wide giant planets are less frequently seen in multiple-star systems, mostly observed in binaries with wider separations, and show similar planet properties when compared to the population of planets orbiting single and binary stars \citep{fontanive21}. We therefore investigate how the presence of a companion at a few hundred AU can influence the likelihood of a disc undergoing fragmentation and forming such high-mass planetary systems.

There is little agreement in the literature as to whether binary companions or stellar flyby events can trigger fragmentation in a disc which would be marginally stable in isolation. Early work considering isothermal discs suggested that encounters during a flyby event could trigger fragmentation \citep{boffinetal98,watkinsetal98a,watkinsetal98b}. \cite{bossetal06} also found that a binary star will act to promote fragmentation, as the spiral arms driven by the companion will typically go on to form self-gravitating clumps. Other authors, however, found that tidal heating during the binary orbit generally acts to stabilise the disc against fragmentation \citep{nelson2000,mayeretal05,lodatoetal07,forganrice2009}. Whilst none of their simulations resulted in fragmentation, and the majority of their results suggest that the effect of encounters is to prohibit fragmentation, \cite{forganrice2009} find that, for some orbital parameters, their discs become more unstable over a larger range of radii, suggesting that there may be some region of parameter space which is favourable to fragmentation. It has also been shown that once a fragment forms in a GI disc, further fragmentation may be triggered as material is channelled inward causing the inner spirals to become sufficiently dense to fragment \citep{meru2015}.

In this paper we present a suite of smoothed particle hydrodynamics (SPH) simulations of binary star systems. We extensively test the parameter space of binary orbital properties for configurations which may trigger fragmentation in discs that would be marginally stable in isolation. We evolve a total of 62 discs which, to the authors' knowledge, represents the most thorough search of this parameter space to date, and in each simulation we model realistic cooling through the \cite{forganetal2009} hybrid radiative transfer method. Section~\ref{methods} details the various disc setups explored in our simulations. Section~\ref{results} presents the results obtained, which we discuss in Section~\ref{discussion}. Our conclusions are presented in Section~\ref{conclusions}.

\section{Methods -- SPH Simulations}
\label{methods}

We simulate a three-dimensional gaseous disc using SPH, a Lagrangian method where a continuous fluid is discretised as $N$ pseudo-particles \citep{benz1990,monaghan1992}. We employ the \textsc{Phantom} SPH code \citep{priceetal18}, which has been modified to include the radiative transfer method introduced in \cite{forganetal2009}; a hybrid cooling approach which combines the polytropic cooling approximation from \cite{stamatellosetal07} and flux-limited diffusion \citep[e.g.][]{mayeretal07b}. We also include the standard SPH artificial viscosity, with parameters $\alpha_{\rm SPH}=0.1$ and $\beta_{\rm SPH}=0.2$.

Each disc is initialised with $N=1\times10^6$ SPH particles, distributed such that the initial surface density profile of the disc is $\Sigma(R) = \Sigma_0 (R/R_0)^{-1.5}$ and the temperature profile is $T(R) = T_0 (R/R_0)^{-1.0}$ between $R_{0}=1$\,AU and $R_{\rm out}=100$\,AU. In each disc $T_0$ and $\Sigma_0$ are determined self-consistently, with $T_0=374$\,K for all discs set up here, and $\Sigma_0$ varying with the disc mass being considered. Any particles that fall within $R_{0}$ are accreted onto the central star, which is represented as a point mass particle. When considering binary star systems, we set up circumprimary discs only, and the companion star behaves as a gravitationally bound point mass, modelled using a sink particle.

\begin{table*}
    \centering
    \begin{tabular}{cccc}
        \hline
         $N_{\rm SPH}$ & $R_{\rm out,disc}$ & $M_{\rm *}$ & $M_{\rm disc}$ \\
         \hline
         $1\times10^6$ & 100\,AU  & 1\,M$_{\odot}$ & $[0.1, 0.2, 0.3, 0.4]$\,M$_{\odot}$ \\
    \end{tabular}
    \caption{SPH disc setup parameters for the reference run of discs with no companion star. Final states of these discs are shown in Figure \ref{fig:sphresults_referencerun}}
    \label{tab:sphsetup_referencerun}

    \centering
    \begin{tabular}{cccccccc}
        \hline
         $N_{\rm SPH}$ & $R_{\rm out,disc}$ & $M_{\rm *,primary}$ & $M_{\rm *,companion}$ & $a$ & $M_{\rm disc}$ & $e$ & $i$ \\
         \hline
         $1\times10^6$ & 100\,AU  & 1\,M$_{\odot}$ & 0.2\,M$_{\odot}$ & $[100, 250, 500, 1000]$\,AU & $[0.1, 0.2, 0.3, 0.4]$\,M$_{\odot}$ & 0 & 0$^\circ$ \\
    \end{tabular}
    \caption{SPH disc setup parameters for the initial suite of discs, where we explore the parameter space in binary semi-major axis and disc mass. Final states of these discs are shown in Figure \ref{fig:sphresults_initial}.}
    \label{tab:sphsetup_initial}

    \centering
    \begin{tabular}{cccccccc}
        \hline
         $N_{\rm SPH}$ & $R_{\rm out,disc}$ & $M_{\rm *,primary}$ & $M_{\rm *,companion}$ & $a$ & $M_{\rm disc}$ & $e$ & $i$  \\
         \hline
         $1\times10^6$ & 100\,AU  & 1\,M$_{\odot}$ & $0.2$\,M$_{\odot}$ & $[150, 200, 325, 400]$\,AU  & 0.2\,M$_{\odot}$ & 0 & 0$^\circ$ \\
    \end{tabular}
    \caption{SPH disc setup parameters where we probe the parameter space in binary semi-major axis further, considering small changes in binary semi-major axis and keeping the disc mass constant. Final states of these discs are shown in Figure \ref{fig:sphresults_separation}.}
    \label{tab:sphsetup_separation}

    \centering
    \begin{tabular}{cccccccc}
        \hline
         $N_{\rm SPH}$ & $R_{\rm out,disc}$ & $M_{\rm *,primary}$ & $M_{\rm *,companion}$ & $a$ & $M_{\rm disc}$ & $e$ & $i$ \\
         \hline
         $1\times10^6$ & 100\,AU  & 1\,M$_{\odot}$ & 0.2\,M$_{\odot}$ & $[150, 200, 250, 325, 400, 500]$\,AU & 0.2\,M$_{\odot}$ & $[0.25,0.5,0.75]$ & 0$^\circ$ \\
    \end{tabular}
    \caption{SPH disc setup parameters where we explore the parameter space in binary semi-major axis and orbital eccentricity. Final states of these discs are shown in Figure \ref{fig:sphresults_eccentricity}.}
    \label{tab:sphsetup_eccentricity}

    \centering
    \begin{tabular}{cccccccccc}
        \hline
         $N_{\rm SPH}$ & $R_{\rm out,disc}$ & $M_{\rm *,primary}$ & $M_{\rm *,companion}$ & $a$ & $M_{\rm disc}$ & $e$ & $i$ \\
         \hline
         $1\times10^6$ & 100\,AU  & 1\,M$_{\odot}$ & 0.2\,M$_{\odot}$ & $[100, 150, 200, 250]$\,AU  & 0.2\,M$_{\odot}$ & $0$ & $[30^\circ, 60^\circ, 90^\circ]$ \\
    \end{tabular}
    \caption{SPH disc setup parameters where we explore tha parameter space in binary semi-major axis and binary inclination. Final states of these discs are shown in Figure \ref{fig:sphresults_inclination}.}
    \label{tab:sphsetup_inclination}

    \centering
    \begin{tabular}{cccccccc}
        \hline
         $N_{\rm SPH}$ & $R_{\rm out,disc}$ & $M_{\rm *,primary}$ & $M_{\rm *,companion}$ & $a$ & $M_{\rm disc}$ & $e$ & $i$ \\
         \hline
         $1\times10^6$ & 100\,AU  & 1\,M$_{\odot}$ & $[0.1,0.5]$\,M$_{\odot}$ & $[150, 250, 325, 400]$\,AU  & 0.2\,M$_{\odot}$ & 0 & 0$^\circ$ \\
    \end{tabular}
    \caption{Additional SPH disc setup parameters where we explore the parameter space in binary semi-major axis and companion mass. Final states of these discs are shown in Figure \ref{fig:sphresults_msecondary}.}
    \label{tab:sphsetup_msecondary}
\end{table*}

\subsection{Suite of SPH models}

To effectively explore the parameter space in binary star separation, $a$, orbital eccentricity, $e$, orbital inclination, $i$, and companion star mass, $M_{*,\rm companion}$, we set up 4 suites of discs -- one for studying each variable individually. In each case, the disc setup parameters are selected to be close to where we find the limit for disc fragmentation to be, identified during our reference run of discs which are detailed below.

\subsubsection{Reference run of discs with no companion}\label{sec:setup_referencerun}

We initially set up a reference run of discs with no companion. This allows us to understand how our discs would evolve in isolation, whilst also being able to identify the region of parameter space where our discs are near to the limit for fragmentation. We set up 4 discs here with masses $M_{\rm disc}=0.1, 0.2, 0.3$ and $0.4$\,M$_{\odot}$, and a parent star of mass $M_{*}=1.0$\,M$_{\odot}$. The value of $\Sigma_0$ for each of these discs, and all subsequent discs of the same mass, are 322, 644, 966, 1288\,g\,cm$^{-2}$, respectively. A summary of these setup parameters can be found in Table \ref{tab:sphsetup_referencerun}.

\subsubsection{Varying binary star separation}\label{sec:setup_separation}

In the first of our suites which include a companion star, we aim to determine how binary separation affects a disc's susceptibility to fragmentation. We set up a grid of 16 systems in which we vary $a$ between 100--1000\,AU and $M_{\rm disc}$ between 0.1--0.4\,M$_{\odot}$, as for the reference runs. We then explore 4 additional cases of $a$ with finer steps in binary separation between $a=$\,150--400\,AU and a fixed disc mass $M_{\rm disc}=0.2$\,M$_{\odot}$. In all the setups we consider a primary star with mass $M_{\rm *,primary}=1.0$\,M$_{\odot}$ and a companion star with mass $M_{\rm *,companion}=0.2$\,M$_{\odot}$, hence a stellar mass ratio $q_{\rm binary}=0.2$. For this initial suite we set up binaries on circular orbits in the plane of the disc, with $e=0$ and $i=0$. These disc setups are summarised in Tables \ref{tab:sphsetup_initial} and \ref{tab:sphsetup_separation}.

Results from this initial suite of discs, where we vary the companion’s semi-major axis, were then used to inform the range of semi-major axes considered for the subsequent suites of disc simulations.

\subsubsection{Varying orbital eccentricity}\label{sec:setup_ecccentricity}

We then wish to study the effect of varying the orbital eccentricity of the binary orbit. 18 new discs are setup where we introduce eccentricities, $e=0.25$, $0.5$ and $0.75$. We vary the binary separation between $a=$\,150--500\,AU whilst keeping the disc mass constant at $M_{\rm disc}=0.2$\,M$_{\odot}$. In all cases we consider a primary star mass, $M_{\rm *,primary}=1.0$\,M$_{\odot}$, a companion star mass, $M_{\rm *,companion}=0.2$\,M$_{\odot}$, and binary orbits in the plane of the disc, with $i=0$. A summary of these disc setups is outlined in Table \ref{tab:sphsetup_eccentricity}.

\subsubsection{Varying orbital inclination}\label{sec:setup_inclination}

A third suite of discs is set up where we study the effect of varying the companion star's orbital inclination relative to the plane of the disc. We set up 12 new discs which include inclinations $i=30^{\circ}$, $60^{\circ}$ and $90^{\circ}$. We consider binary separations in the range $a=$\,100--250\,AU, whilst keeping the disc mass constant at $M_{\rm disc}=0.2$\,M$_{\odot}$. In each case we consider a primary star mass, $M_{\rm *,primary}=1.0$\,M$_{\odot}$, a companion star mass, $M_{\rm *,companion}=0.2$\,M$_{\odot}$, and circular binary orbits with $e=0$. A summary of these disc setups is outlined in Table \ref{tab:sphsetup_inclination}.

\subsubsection{Varying companion star mass}\label{sec:setup_msecondary}

Finally, we set up a suite of discs to explore the effect of varying the mass of the companion star. We set up a grid of discs which includes 8 new setups, where we introduce companion star masses $M_{\rm *,companion}=0.1$\,M$_{\odot}$ and $0.5$\,M$_{\odot}$. We vary the binary separation between $a=$\,150--400\,AU, whilst keeping the disc mass constant at $M_{\rm disc}=0.2$\,M$_{\odot}$. For all disc setups we consider a primary star mass, $M_{\rm *,primary}=1.0$\,M$_{\odot}$, and circular binary orbits in the plane of the disc, with $e=0$ and $i=0$. A summary of these disc setups is outlined in Table \ref{tab:sphsetup_msecondary}.

\section{Results}
\label{results}

Results from the final states of all the discs simulated here are summarised in Table \ref{tab:sphresultssummary}. In all cases we allow the discs to evolve for at least 5 orbital periods at the disc outer edge ($R_{\rm out}=100$\,AU), equivalent to $t=5000$\,yrs, or until fragmentation occurs. In the case of the wide orbit binary systems, for which the binary orbit is longer than 5 orbital periods at $R_{\rm out}=100$\,AU, we allow the simulations to evolve for at least a full binary orbit. We define a simulation as having fragmented when a local clump forms where the density is significantly higher than the surrounding disc gas. Typically, these clumps have densities that are a few orders of magnitude greater than the surrounding gas.

\subsection{Reference run of discs with no companion}\label{sec:resultsreferencerun}

\begin{figure*}
    \centering
    \begin{tabular}{c|ccccc}
    \centering
    & $M_{\rm disc}=0.1$\,M$_{\odot}$ & $M_{\rm disc}=0.2$\,M$_{\odot}$ & $M_{\rm disc}=0.3$\,M$_{\odot}$ & $M_{\rm disc}=0.4$\,M$_{\odot}$ & \\

    \hline\hline
    \rotatebox{90}{Reference run - no companion} \vline\vline & \includegraphics[width=0.2\textwidth]{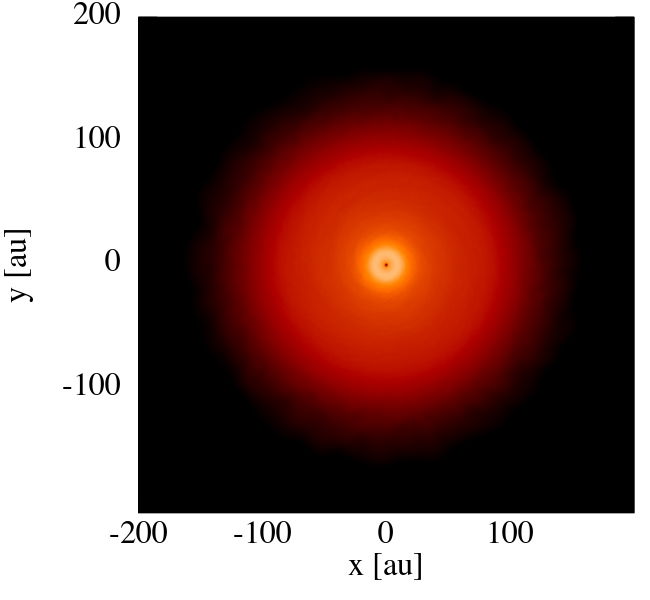} &
    \includegraphics[width=0.2\textwidth]{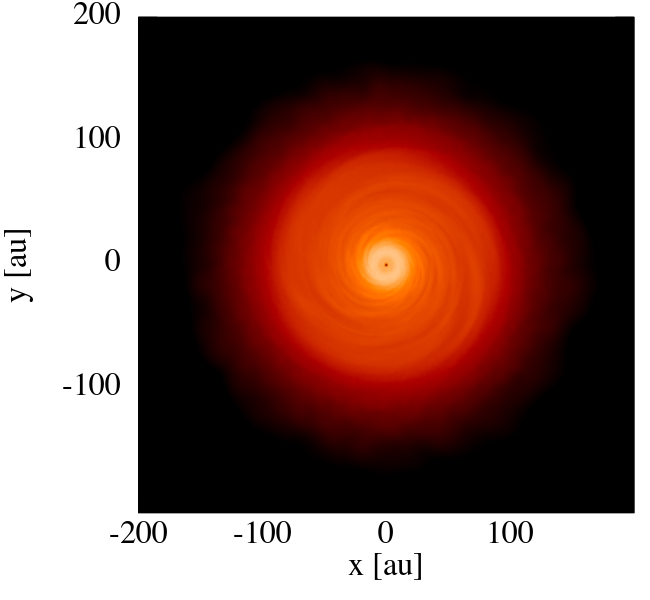} &
    \includegraphics[width=0.2\textwidth]{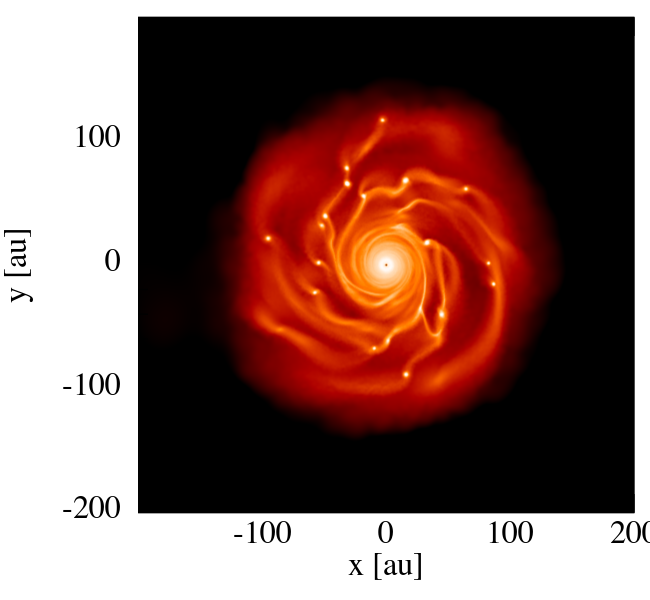} &
    \includegraphics[width=0.2\textwidth]{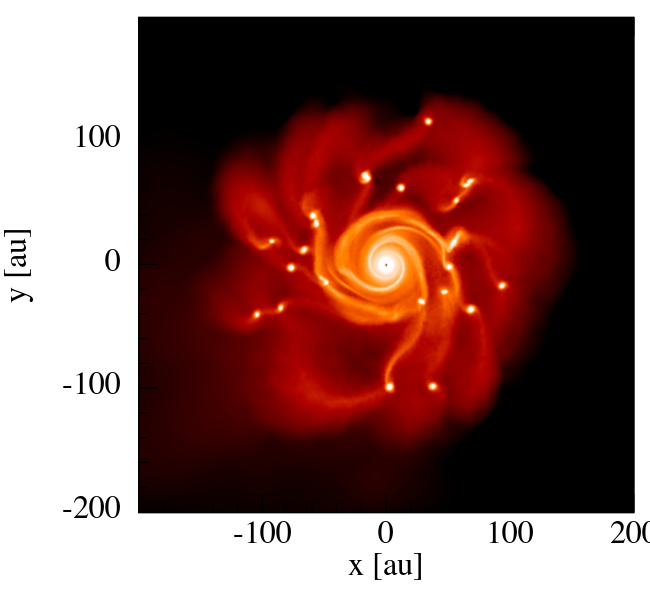} &
    \includegraphics[width=0.05\textwidth]{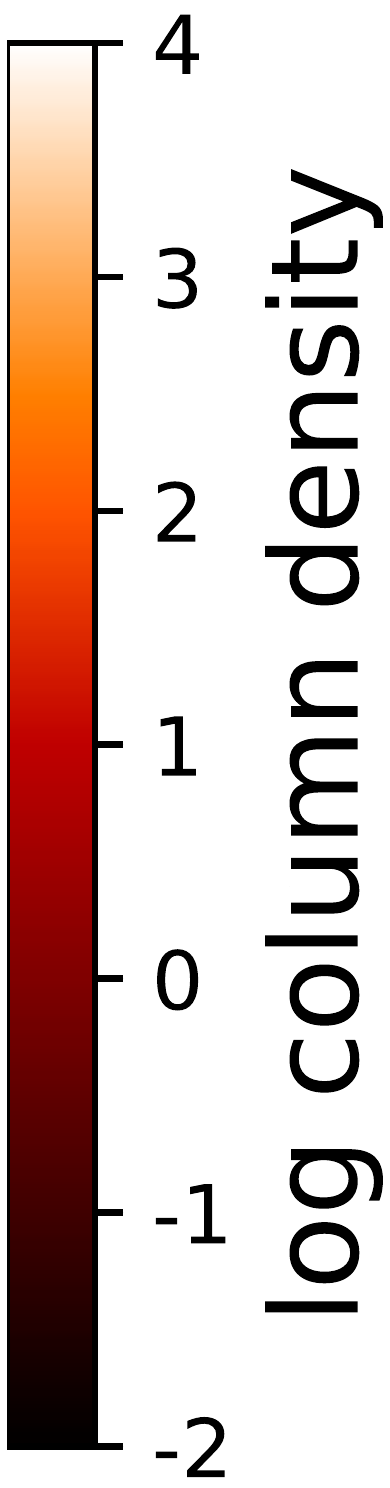} \\

    \end{tabular}

    \begin{tikzpicture}[remember picture,overlay]


      \draw[cyan,line width=5] (-0.2,0) rectangle (7.8,3.7);
    \end{tikzpicture}

    \caption{Final states of the reference run of disc setups with no companion star included. A summary of the disc setup parameters laid out in Table \ref{tab:sphsetup_referencerun}, and outlined in detail in Section \ref{sec:setup_referencerun}. Blue boxes are included to highlight the reference run discs which resulted in fragmentation.}
    \label{fig:sphresults_referencerun}
\end{figure*}
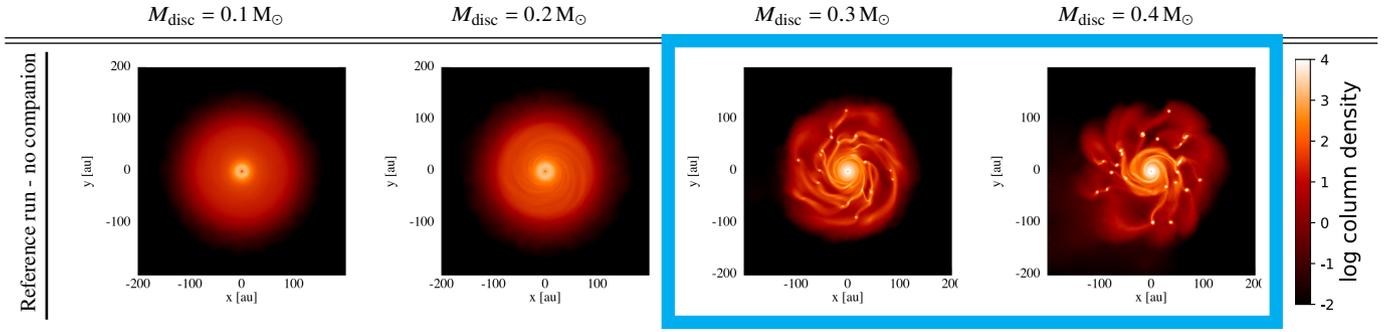

Figure \ref{fig:sphresults_referencerun} shows the final states of the reference run of discs with no companion star included. Setup parameters for these discs are outlined in Section \ref{sec:setup_referencerun} and summarised in Table \ref{tab:sphsetup_referencerun}. We find the lower mass limit for disc fragmentation to be in the range $0.2<M_{\rm disc}<0.3$\,M$_{\odot}$. The disc with $M_{\rm disc.}=0.2$\,M$_{\odot}$ is able to evolve for the full simulation time without fragmenting, whilst the disc with $M_{\rm disc}=0.3$\,M$_{\odot}$ fragments quickly, after $700$\,yrs of evolution.

In Figure \ref{fig:a250_vs_referencerun} we plot the azimuthally averaged midplane disc properties from some of the systems simulated here. We include plots of the Toomre parameter, $Q$, where \citep{toomre64},
\begin{equation}
    Q = \frac{c_{\rm s}\Omega}{\pi{\rm G}\Sigma}.
\end{equation}
A disc will become susceptible to fragmentation when $Q \lesssim 1$. We also plot the disc cooling time, $t_{\rm cool}(=u/\dot{u})$, from \cite{forganetal2009}, and the dimensionless cooling parameter, $\beta_{\rm cool}=t_{\rm cool}\Omega$, where $\Omega=\sqrt{{\rm G}M_*/r^3}$ is the Keplerian frequency at a given radius. The term $\beta_{\rm cool}$ quantifies how the disc cooling time compares to the dynamical time. Early work has shown that fragmentation may occur if the disc is able to cool on dynamical timescales, with $\beta_{\rm cool}\lesssim3$ \citep{gammie01,rice03}.

From Figure \ref{fig:a250_vs_referencerun}, we find that the system with $M_{\rm disc}=0.2$\,M$_{\odot}$ reaches a marginally stable final state with $Q_{\rm min}=1.06$ at $R=74$\,AU, and $\beta_{\rm cool,min}=9.75$ at $R=100$\,AU.

In order to ensure that the fragmentation threshold found here is independent of the discs' initial setups, we also ran each disc such that $Q \approx 1$ at $R=R_{\rm out}$, by adjusting the value of the disc aspect ratio ($H/R$) at $R=R_{\rm out}$. Again, we found the limit for fragmentation to be in the range $0.2<M_{\rm disc}<0.3$\,M$_{\odot}$.

\subsection{Varying binary separation}

\subsubsection{Initial suite of discs}

\begin{figure*}
    \centering
    \begin{tabular}{c|ccccc}
    \centering
    & $M_{\rm disc}=0.1$\,M$_{\odot}$ & $M_{\rm disc}=0.2$\,M$_{\odot}$ & $M_{\rm disc}=0.3$\,M$_{\odot}$ & $M_{\rm disc}=0.4$\,M$_{\odot}$ & \\

    \hline\hline

    \rotatebox{90}{$a=100$\,AU} \vline\vline & \includegraphics[width=0.2\linewidth]{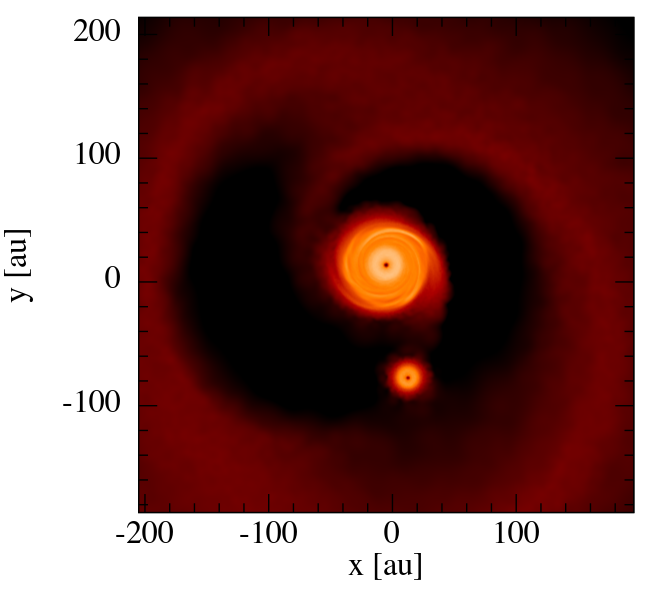} & \includegraphics[width=0.2\linewidth]{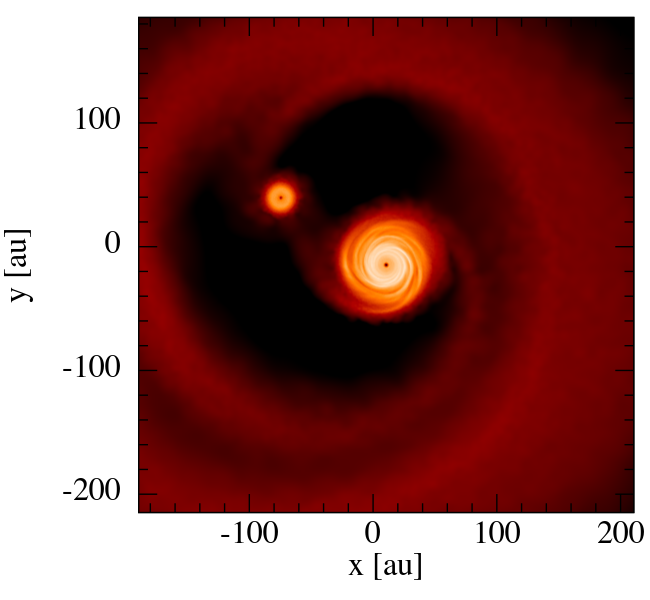} &
    \includegraphics[width=0.2\linewidth]{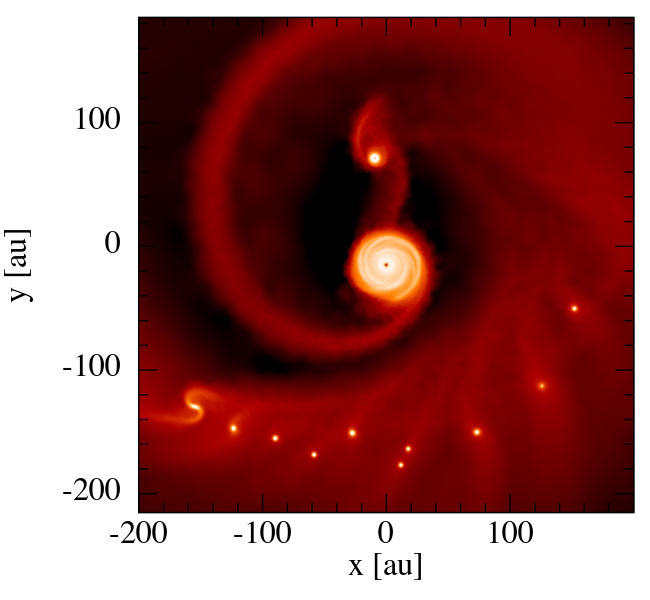} &
    \includegraphics[width=0.2\linewidth]{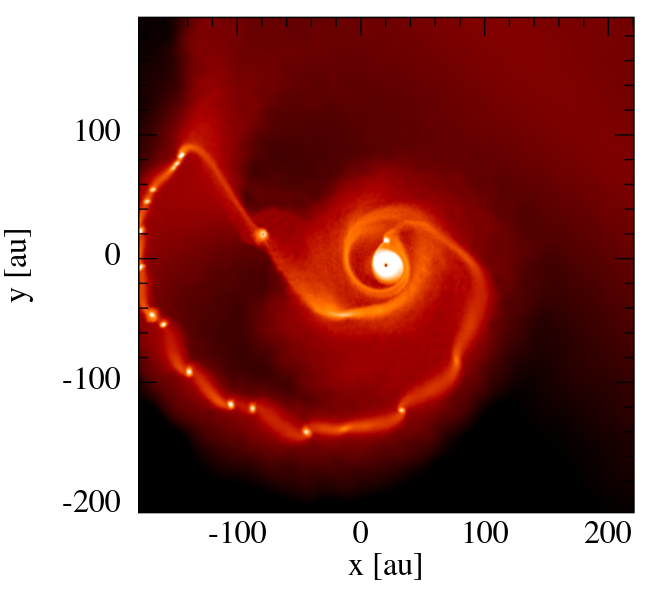} & \includegraphics[width=0.05\linewidth]{cbar.pdf} \\

    \hline

    \rotatebox{90}{$a=250$\,AU} \vline\vline & \includegraphics[width=0.2\linewidth]{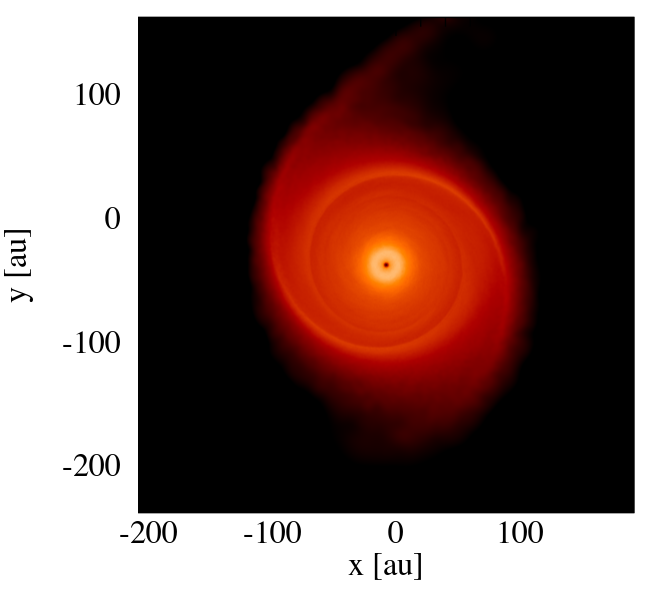} &
    \includegraphics[width=0.2\linewidth]{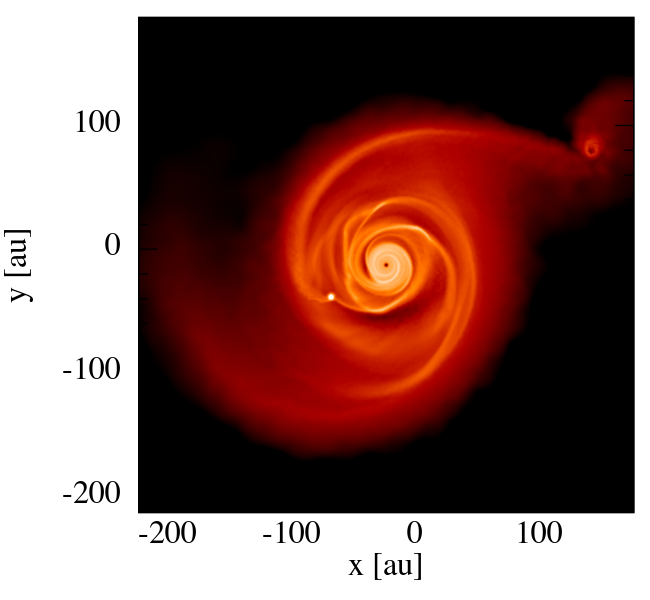} &
    \includegraphics[width=0.2\linewidth]{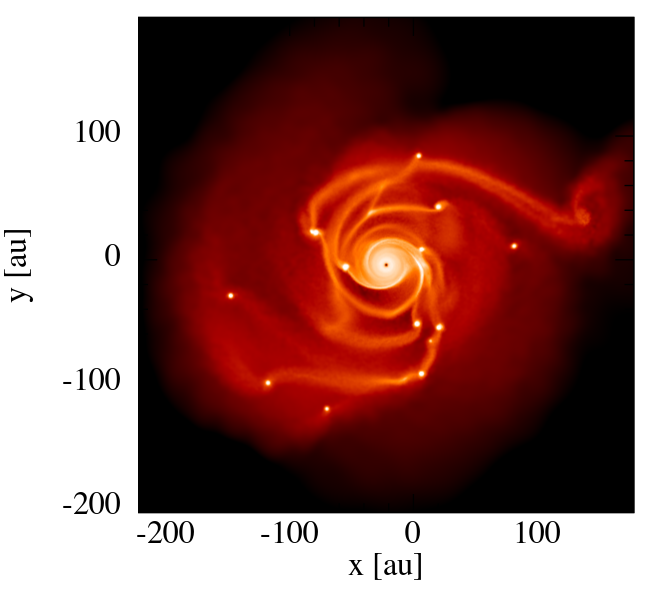} &
    \includegraphics[width=0.2\linewidth]{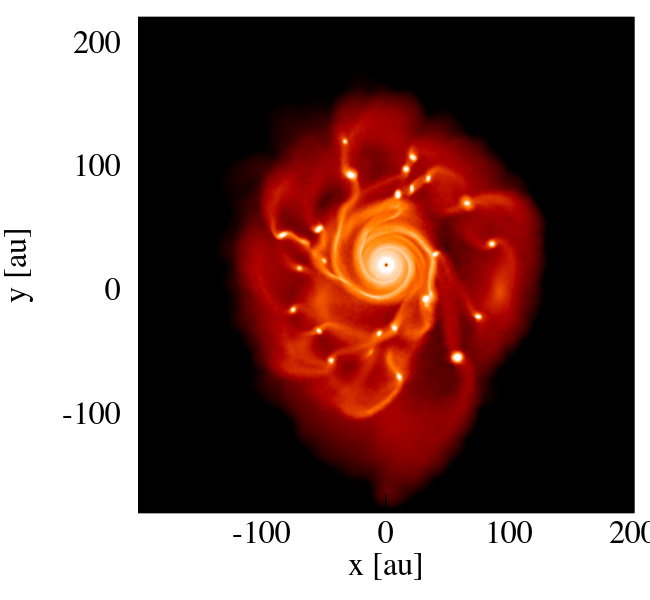} & \includegraphics[width=0.05\linewidth]{cbar.pdf} \\

    \hline

    \rotatebox{90}{$a=500$\,AU} \vline\vline & \includegraphics[width=0.2\linewidth]{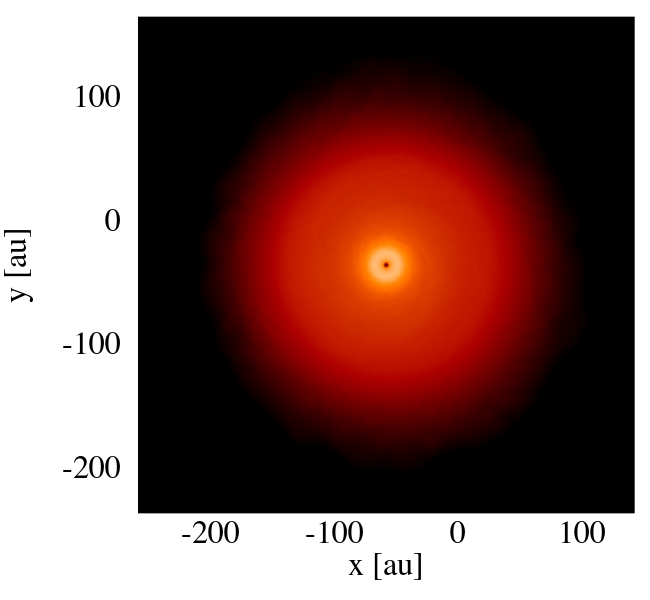} & \includegraphics[width=0.2\linewidth]{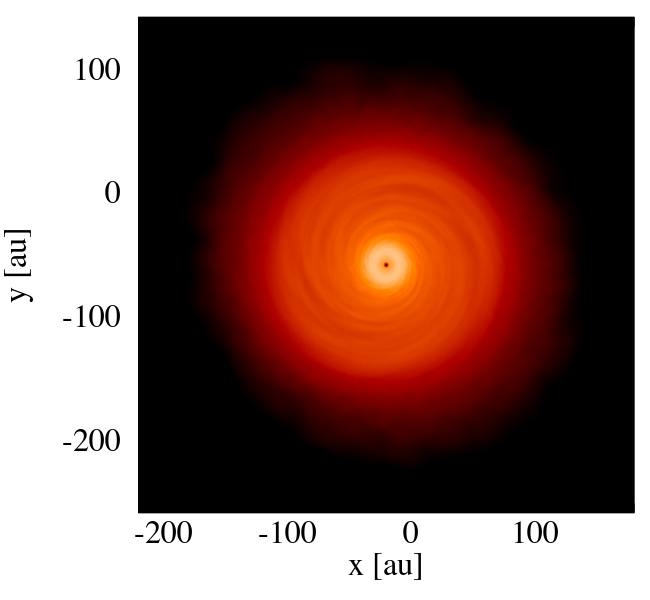} &
    \includegraphics[width=0.2\linewidth]{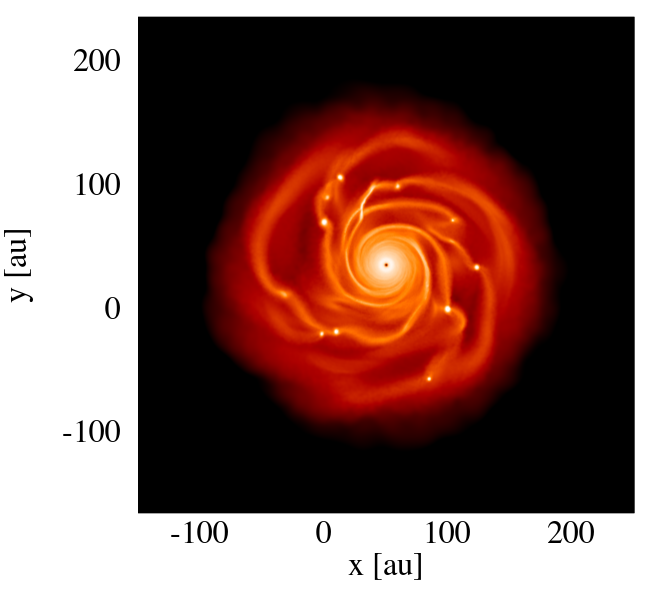} &
    \includegraphics[width=0.2\linewidth]{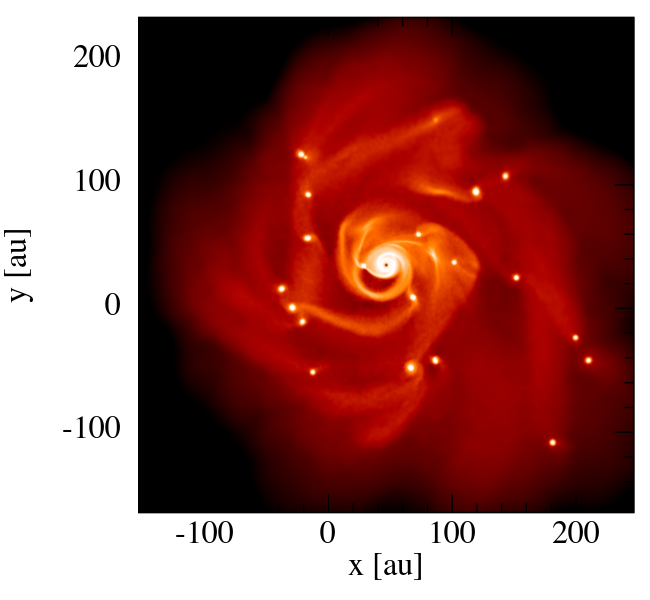} & \includegraphics[width=0.05\linewidth]{cbar.pdf}
     \\

    \hline

    \rotatebox{90}{$a=1000$\,AU} \vline\vline &
    \includegraphics[width=0.2\linewidth]{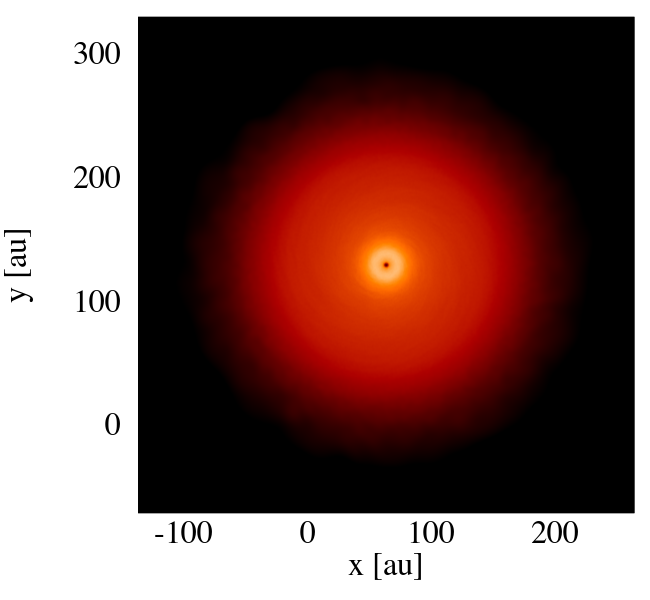} & \includegraphics[width=0.2\linewidth]{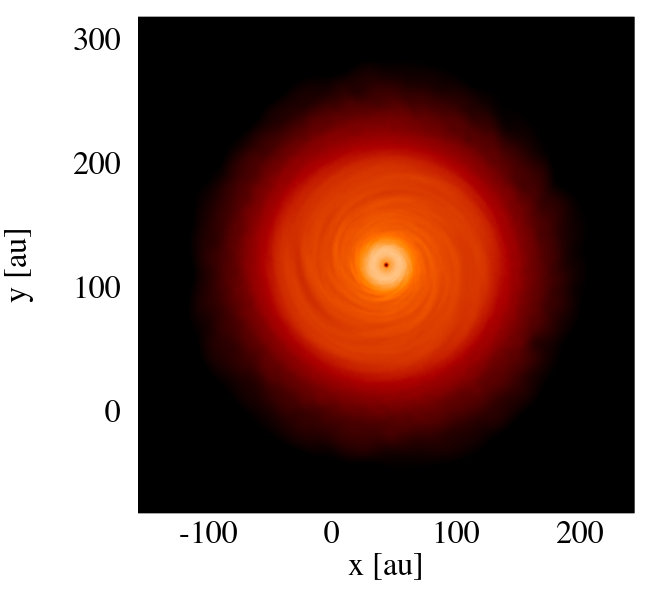} &
    \includegraphics[width=0.2\linewidth]{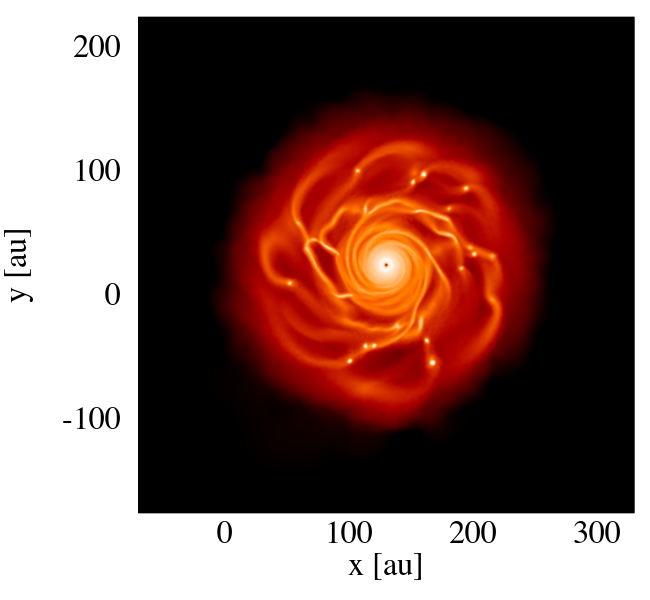} &
    \includegraphics[width=0.2\linewidth]{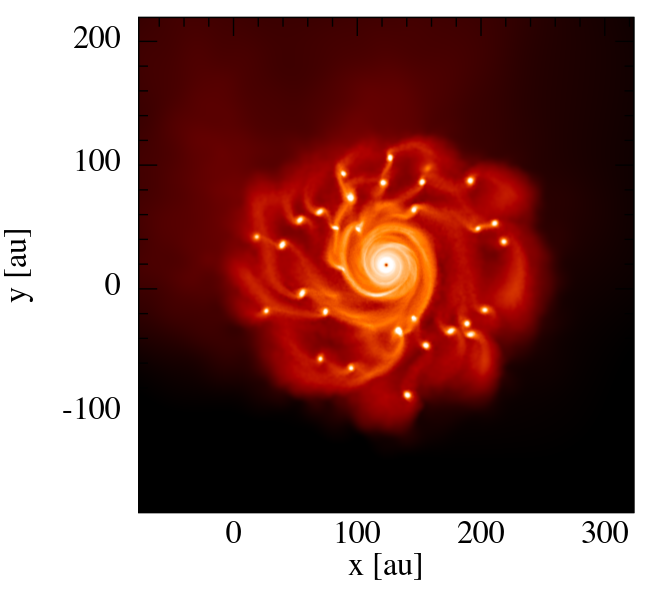} & \includegraphics[width=0.05\linewidth]{cbar.pdf} \\

    \end{tabular}

    \begin{tikzpicture}[remember picture,overlay]



      \draw[cyan,line width=5] (-0.13,0) rectangle (7.8,14.95);

      \draw[green,line width=5] (-4.1,7.4) rectangle (-0.3,11.15);
    \end{tikzpicture}

    \caption{Final states of the discs where we vary the disc mass and the semi-major axis of the companion star. Disc setup parameters are summarised in Table \ref{tab:sphsetup_initial} and outlined in detail in Section \ref{sec:setup_separation}. Blue boxes are included to highlight discs which resulted in fragmentation when their reference run analogs also fragmented. Green boxes are included to highlight disc configurations which resulted in fragmentation when their reference run analog did not fragment.}
    \label{fig:sphresults_initial}
\end{figure*}

Results from the initial suite of discs where we vary binary separation and disc mass are displayed in Figure \ref{fig:sphresults_initial}. Setup parameters for these discs are outlined in Section \ref{sec:setup_separation} and summarised in Table \ref{tab:sphsetup_initial}. As with the results in Section \ref{sec:resultsreferencerun}, we also ran each of these discs with slightly different initial $Q-$profiles to ensure our conclusions remain consistent.

When comparing the results in Figure \ref{fig:sphresults_initial} to those from the reference run in Figure \ref{fig:sphresults_referencerun}, we find one disc configuration, with $M_{\rm disc}=0.2$\,M$_{\odot}$ and $a=250$\,AU, where the simulation results in fragmentation, and its analog from the reference run, with $M_{\rm disc}=0.2$\,M$_{\odot}$ and no companion star, did not. The companion star's initial eccentricity is $e=0$, however energy exchange with disc material throughout the companion's orbit results in a periastron binary separation of $r_{\rm peri,actual}=186$\,AU. As the companion approaches and passes through periastron, an $m=2$ spiral mode propagates through the disc generating a bump in the surface density at $R\approx60$\,AU.

\begin{figure}
    \centering
    \includegraphics[width=\linewidth]{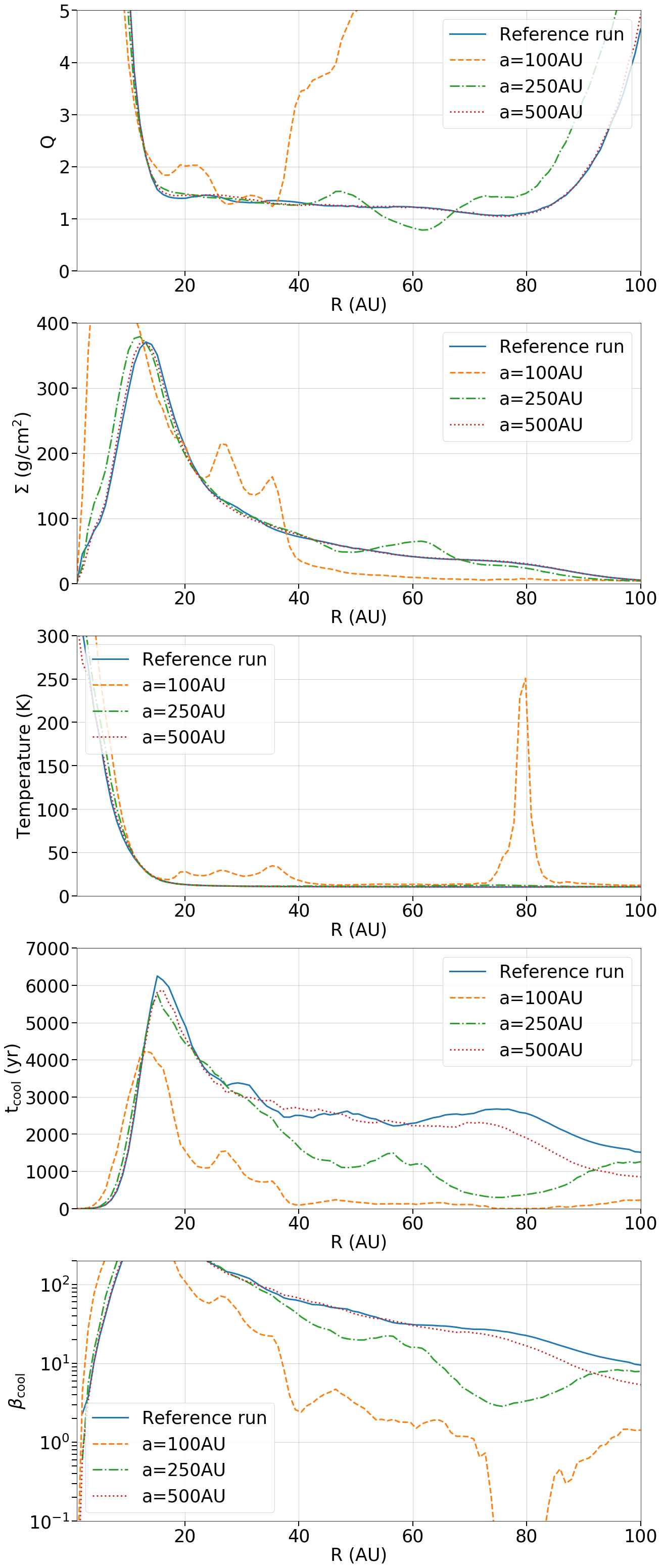}
    \caption{Azimuthally averaged midplane disc properties calculated from the $M_{\rm disc}=0.2$\,M$_{\odot}$ discs from the initial suite which includes a binary companion (setup parameters in Table \ref{tab:sphsetup_initial}, final states in Figure \ref{fig:sphresults_initial}). We plot the reference run final state, the $a=100$\,AU run at periastron, the $a=250$\,AU run immediately before fragmentation, and the $a=500$\,AU run at periastron.}
    \label{fig:a250_vs_referencerun}
\end{figure}

Comparing the disc properties in Figure \ref{fig:a250_vs_referencerun} for the system where $a=250$\,AU and $M_{\rm disc}=0.2$\,M$_{\odot}$ immediately before fragmentation occurs (at $t\approx2700$\,yrs), and the properties from the final state of the analog disc from the reference run (where $M_{\rm disc}=0.2$\,M$_{\odot}$), we observe how the surface density increases in the disc of the $a=250$\,AU system, consistent with the location of the spirals driven by the companion. Efficient cooling, evident from the drop in $t_{\rm cool}$ between $\approx$\,60--90\,AU, is able to prevent the disc temperature from increasing significantly at the spiral location. Hence $Q$ can drop to below $Q=1$ and fragmentation ensues. For the system where $a=250$\,AU, immediately before the disc fragments, we find $\beta_{\rm cool,min}=2.92$ at $R=75$\,AU.

Binaries on wide orbits ($a=500, 1000$\,AU) converge to the single star solution, where the mass limit for fragmentation is $0.2$\,M$_{\odot}<M_{\rm disc}<0.3$\,M$_{\odot}$. In Figure \ref{fig:a250_vs_referencerun} we also plot the azimuthally averaged midplane disc properties for the setup where $M_{\rm disc}=0.2$\,M$_{\odot}$ and $a=500$\,AU at the time of periastron passage. The temperature profile and $Q-$profile are almost identical to the analog disc from the reference run, with $Q_{\rm min}=1.05$ at $R=75$\,AU. Similarly, we find no significant bump in the surface density profile which would be consistent with a spiral being driven by the companion star. The $M_{\rm disc}=0.3$\,M$_{\odot}$ system with a companion at $a=500$\,AU fragments quickly, as it did around a single star.

Binaries with small semi-major axes, whose orbits result in the companion star passing through the disc ($a=100$\,AU), are found to be prohibitive to fragmentation. In Figure \ref{fig:a250_vs_referencerun} we plot the azimuthally averaged midplane properties for the disc with $a=100$\,AU and $M_{\rm disc}=0.2$\,M$_{\odot}$, at the point of periastron passage (where $r_{\rm peri,actual}=78$\,AU). As the companion moves through the disc the midplane temperature increases, whilst material is simultaneously ejected from the outer disc and channeled toward the inner disc. The final surface density profiles for the $a=100$\,AU discs, set up with $M_{\rm disc}=0.1$\,M$_{\odot}$ and $0.2$\,M$_{\odot}$, are consequently truncated at $a\approx40$\,AU, with final disc masses $M_{\rm disc}=0.06$\,M$_{\odot}$ and $0.12$\,M$_{\odot}$ respectively. Hence no fragmentation can occur. The massive discs with $M_{\rm disc}\geq0.3$\,M$_{\odot}$ are still able to fragment quickly, before completing a full binary orbital period, in a spiral arm which trails the path of the companion star.

\subsubsection{Further probing the parameter space in binary separation}

\begin{figure*}
    \centering
    \begin{tabular}{cccc}
    \centering
    $a=100$\,AU & $a=150$\,AU & $a=200$\,AU \\

    \hline\hline

    \includegraphics[width=0.2\linewidth]{a100_mdisc02.png} & \includegraphics[width=0.2\linewidth]{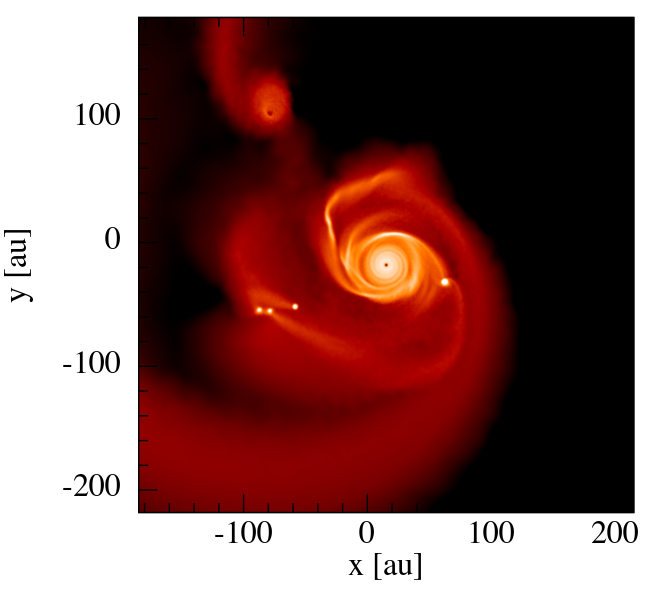} & \includegraphics[width=0.2\linewidth]{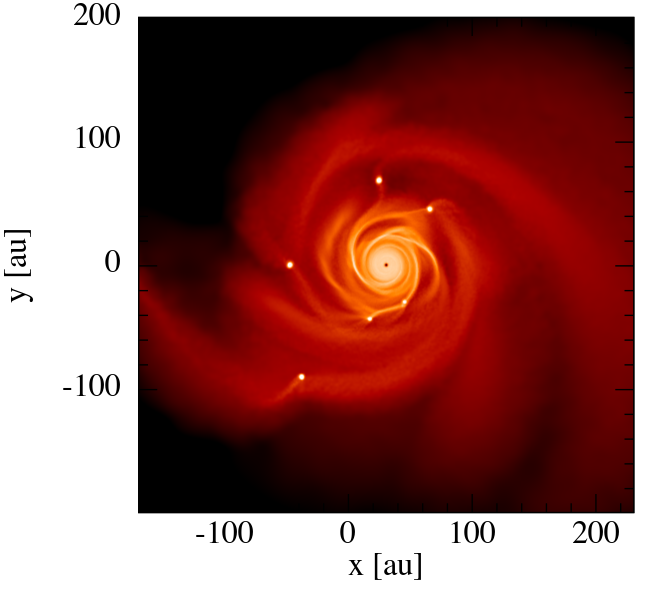} & \includegraphics[width=0.05\linewidth]{cbar.pdf} \\

    $a=250$\,AU & $a=325$\,AU & $a=400$\,AU  \\
    \hline\hline
    \includegraphics[width=0.2\linewidth]{a250_mdisc02.png} & \includegraphics[width=0.2\linewidth]{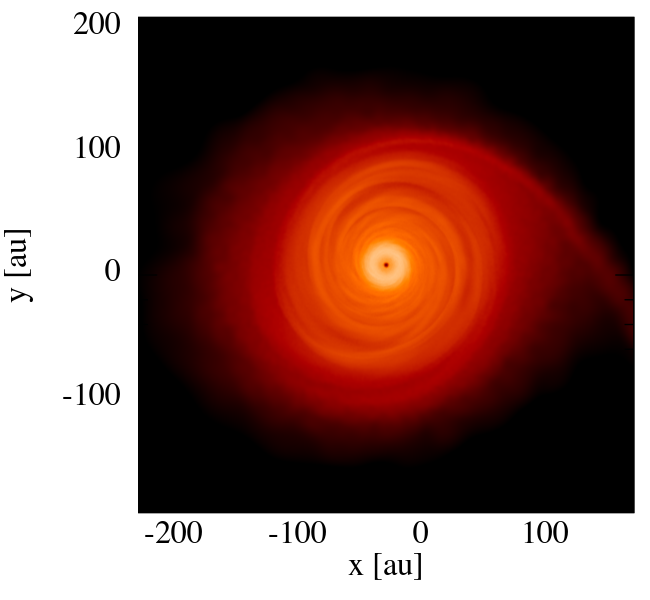} & \includegraphics[width=0.2\linewidth]{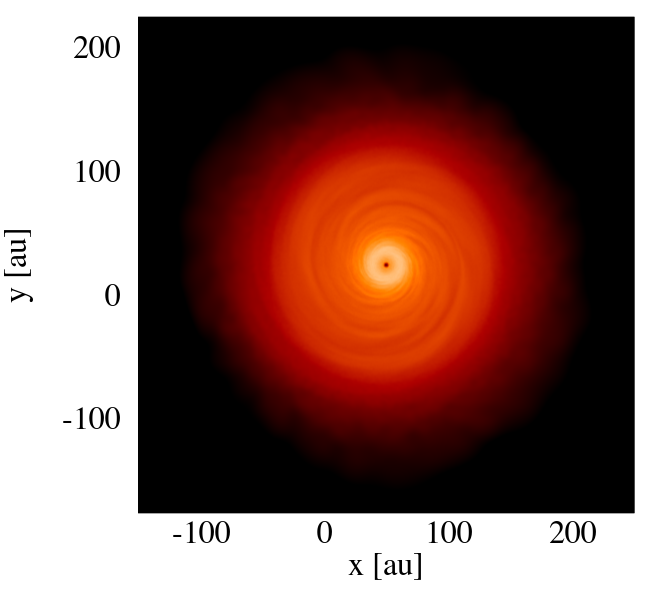} & \includegraphics[width=0.05\linewidth]{cbar.pdf} \\
    \end{tabular}

    \begin{tikzpicture}[remember picture,overlay]

      \draw[green,line width=5] (-6.65,0) rectangle (-2.7,3.7);
      \draw[green,line width=5] (-2.7,4.3) rectangle (5.35,7.85);

    \end{tikzpicture}

    \caption{Final states of the discs where we keep the disc mass constant and consider small changes in the semi-major axis of the companion star. Disc setup parameters are summarised in Table \ref{tab:sphsetup_separation} and outlined in detail in Section \ref{sec:setup_separation}. Green boxes are included to highlight disc configurations which resulted in fragmentation, when their reference run analog did not.}
    \label{fig:sphresults_separation}
\end{figure*}

As the results in Figure \ref{fig:sphresults_initial} indicate that there may be a sweet spot in binary separation at $a\approx250$\,AU which can trigger fragmentation, we ran an additional set of discs where we probe this region of parameter space with greater granularity. This consists of 4 additional discs with $M_{\rm disc}=0.2$\,M$_{\odot}$ and $a=150, 200, 325$ and $400$\,AU. Setup parameters for these are outlined in Section \ref{sec:setup_separation} and summarised in Table \ref{tab:sphsetup_separation}. Final states of these discs are shown in Figure \ref{fig:sphresults_separation}.

We find that the disc setups with binary semi-major axes between $150$\,AU\,$ \leq a \leq 250$\,AU result in fragmentation. In the configuration with $a=150$\,AU, the companion narrowly avoids passing through the outer extent of the disc, with $r_{\rm peri,actual}=116$\,AU. As the companion approaches and passes through periastron, an $m=2$ spiral mode propagates through the disc causing a significant drop in $Q$ at the inner regions of one of these spirals, between $30$\,AU\,$ \leq R \leq 50$\,AU, and 4 fragments initially form. All 4 of these fragments survive as the companion travels back towards apastron, as can be seen in the final state of the disc in Figure \ref{fig:sphresults_separation}. A similar process occurs in the system with $a=200$\,AU.

As we increase the binary separation beyond $a=250$\,AU the influence of the companion star becomes progressively weaker. Once $a \geq 325$\,AU the companion star can no longer trigger fragmentation in the disc. The spiral mode induced by the $a = 325$\,AU companion is much weaker than in the $a = 250$\,AU disc. We find that $Q$ drops slightly in the outer disc of the $a=325$\,AU system when compared to its analog from the reference run (with $M_{\rm disc}=0.2$\,M$_{\odot}$ and no companion), but not enough to push the disc over the fragmentation threshold. Once we increase the binary semi-major axis to $a=400$\,AU we find a similar result as with the $a=500$\,AU system in the previous section. The surface density profile, $Q-$profile and temperature profile at periastron passage become increasingly similar to the final state of the reference run disc with $M_{\rm disc}=0.2$\,M$_{\odot}$.

\subsection{Varying orbital eccentricity}

\begin{figure*}
    \centering
    \begin{tabular}{c|ccccc}
    \centering
    & $e=0$ & $e=0.25$ & $e=0.5$ & $e=0.75$ & \\

    \hline\hline

    \rotatebox{90}{$a=150$\,AU} \vline\vline &  \includegraphics[width=0.2\linewidth]{a150_mdisc02.png} & \includegraphics[width=0.2\linewidth]{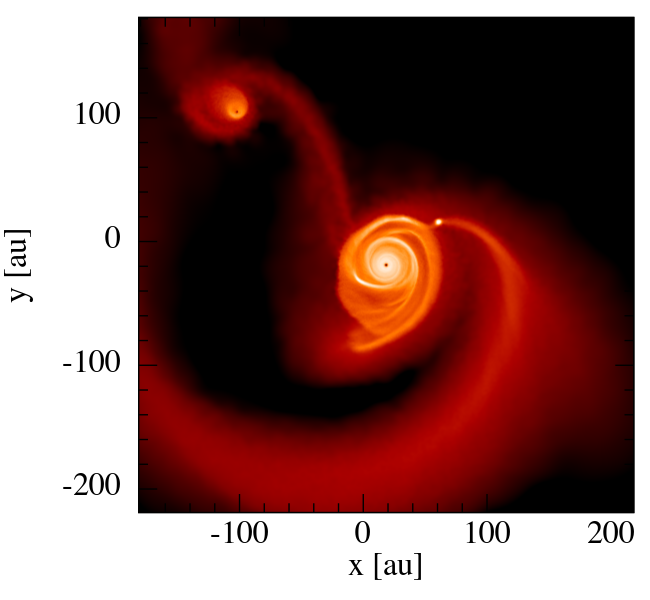} & \includegraphics[width=0.2\linewidth]{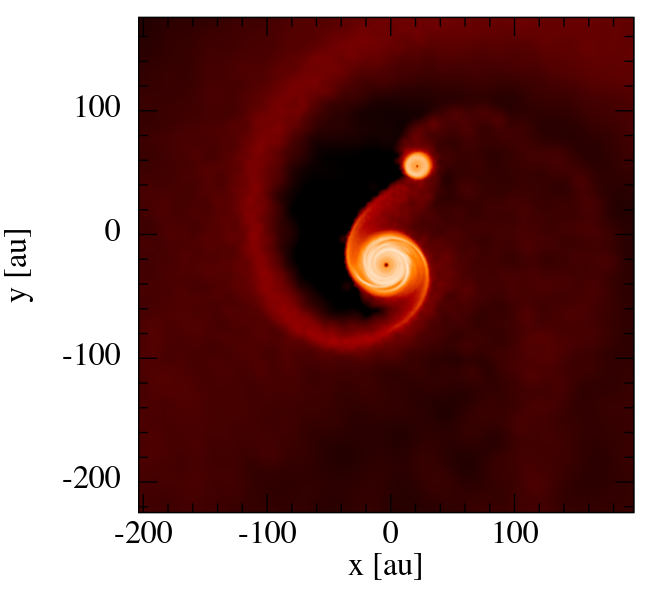} & \includegraphics[width=0.2\linewidth]{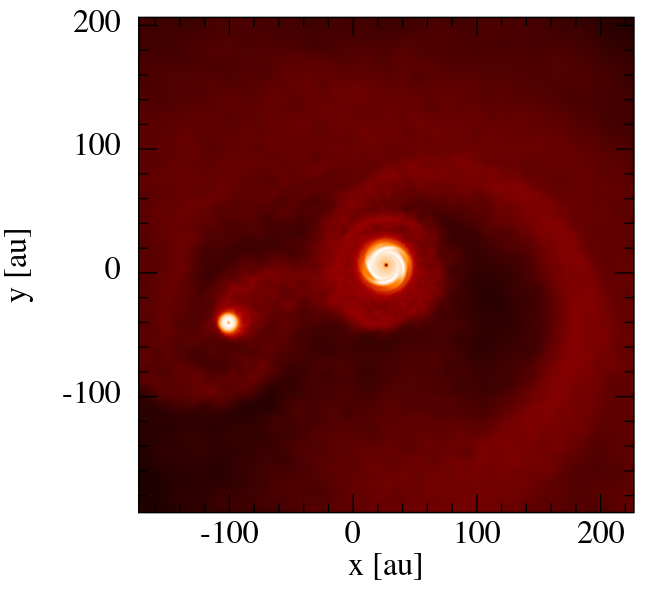} & \includegraphics[width=0.05\linewidth]{cbar.pdf} \\

    \hline

    \rotatebox{90}{$a=200$\,AU} \vline\vline & \includegraphics[width=0.2\linewidth]{a200_mdisc02.png} & \includegraphics[width=0.2\linewidth]{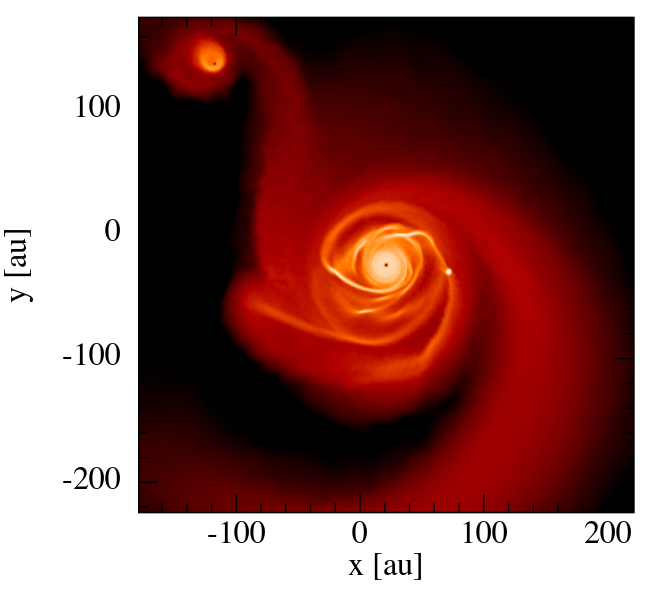} & \includegraphics[width=0.2\linewidth]{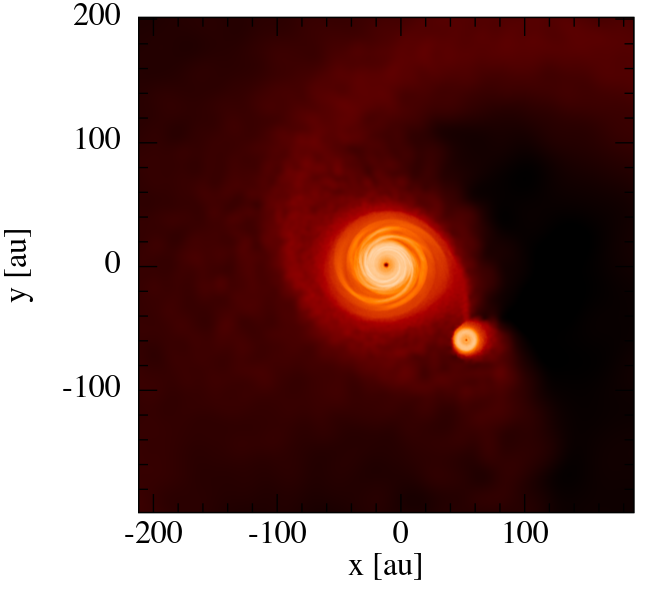} & \includegraphics[width=0.2\linewidth]{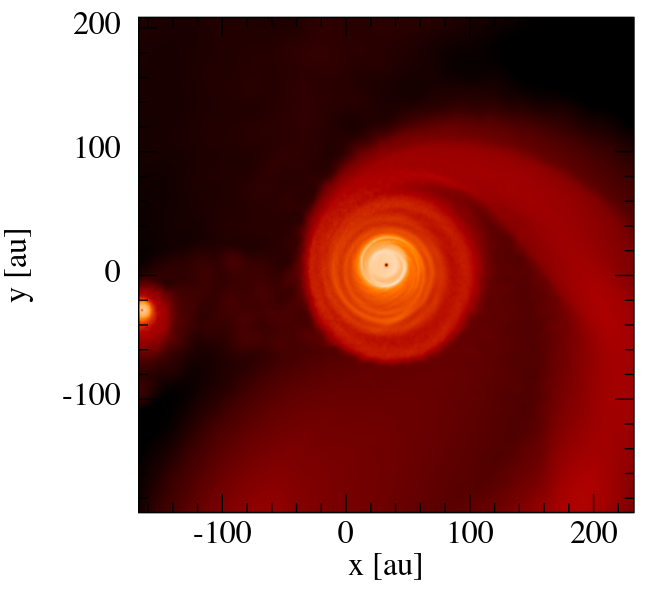} & \includegraphics[width=0.05\linewidth]{cbar.pdf} \\

    \hline

    \rotatebox{90}{$a=250$\,AU} \vline\vline & \includegraphics[width=0.2\linewidth]{a250_mdisc02.png} & \includegraphics[width=0.2\linewidth]{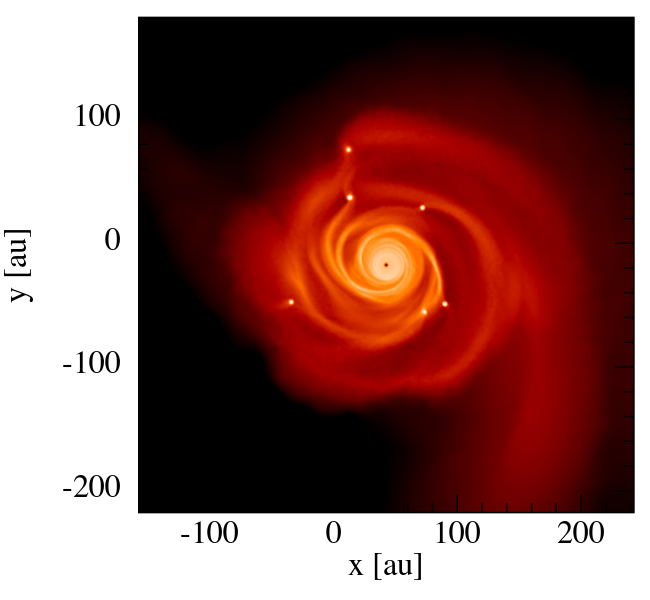} & \includegraphics[width=0.2\linewidth]{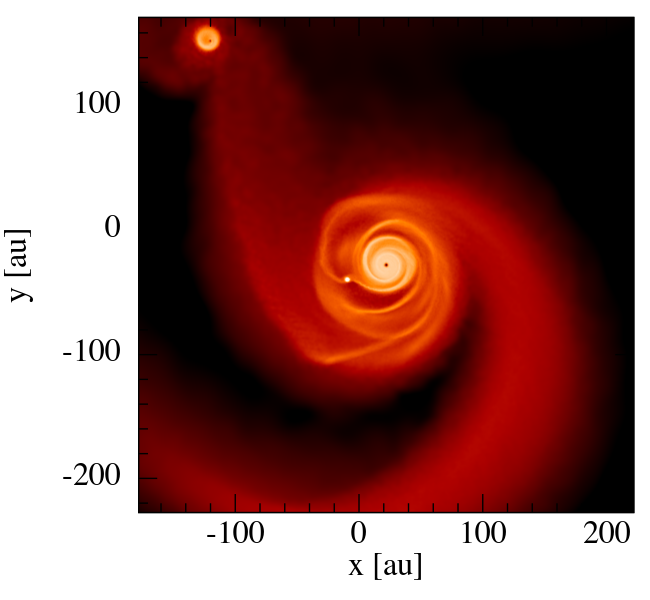} & \includegraphics[width=0.2\linewidth]{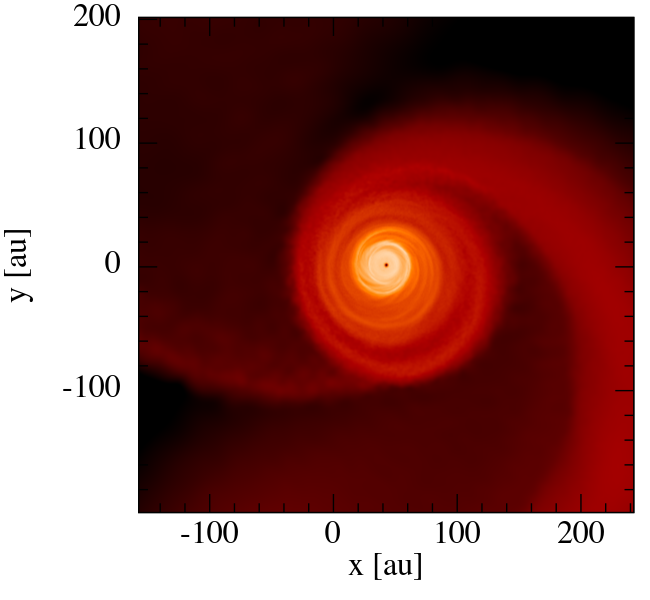} & \includegraphics[width=0.05\linewidth]{cbar.pdf} \\

    \hline

    \rotatebox{90}{$a=325$\,AU} \vline\vline &  \includegraphics[width=0.2\linewidth]{a325_mdisc02.png} & \includegraphics[width=0.2\linewidth]{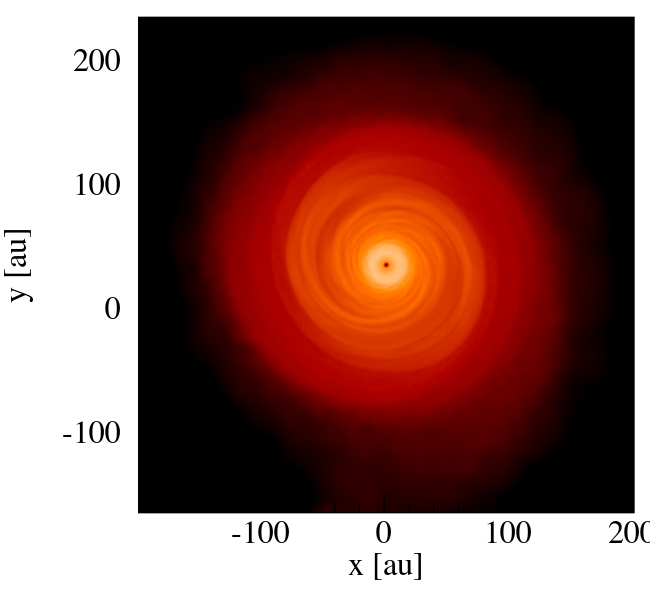} & \includegraphics[width=0.2\linewidth]{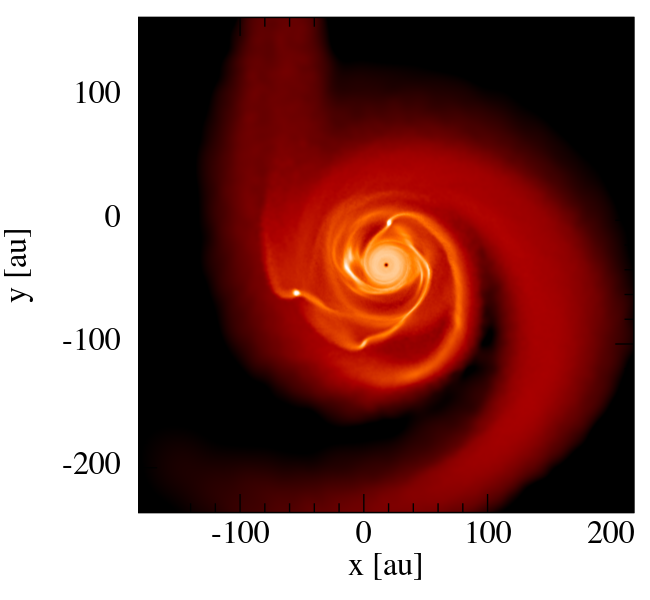}  &  \includegraphics[width=0.2\linewidth]{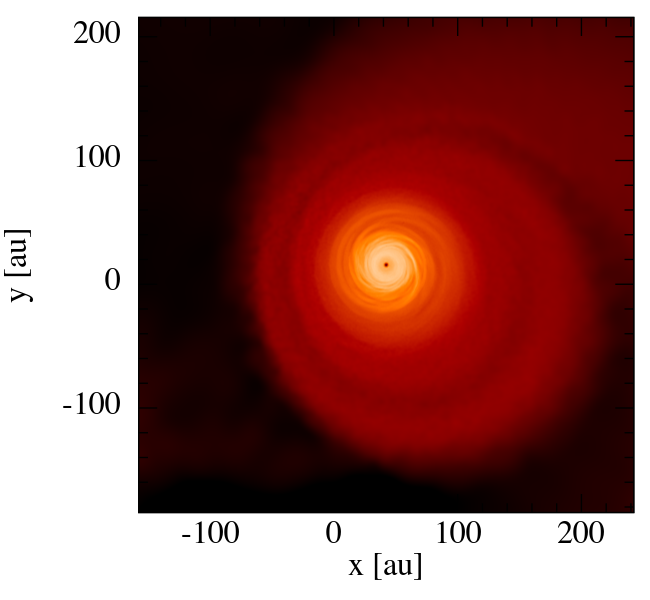} & \includegraphics[width=0.05\linewidth]{cbar.pdf} \\

    \hline

    \rotatebox{90}{$a=400$\,AU} \vline\vline & \includegraphics[width=0.2\linewidth]{a400_mdisc02.png}  &  \includegraphics[width=0.2\linewidth]{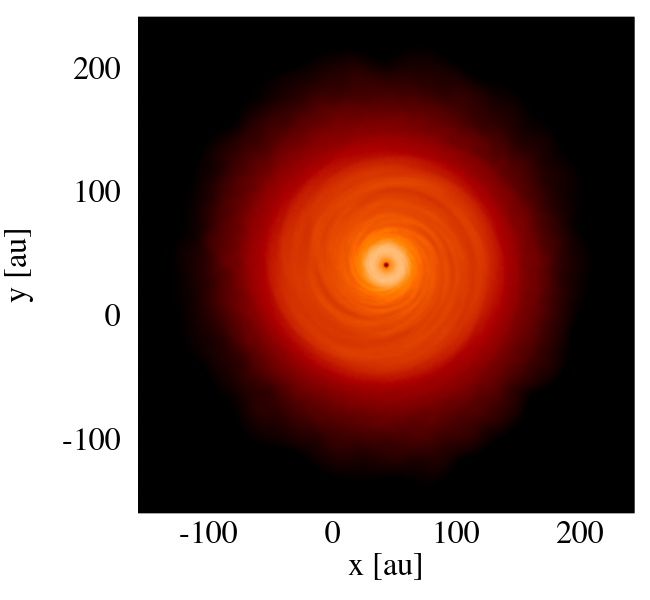} & \includegraphics[width=0.2\linewidth]{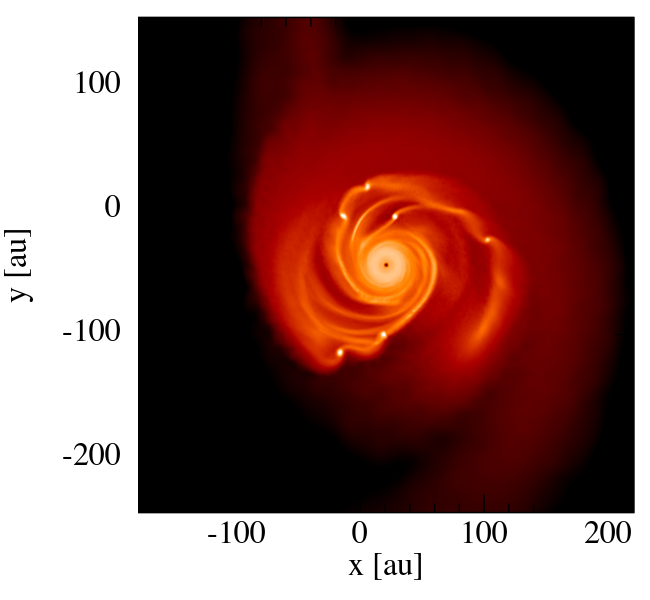} &  \includegraphics[width=0.2\linewidth]{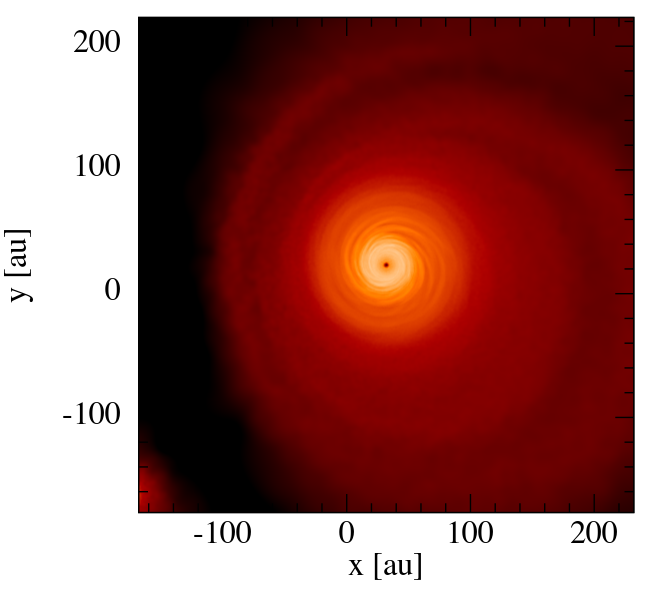} & \includegraphics[width=0.05\linewidth]{cbar.pdf} \\

    \hline

    \rotatebox{90}{$a=500$\,AU} \vline\vline & \includegraphics[width=0.2\linewidth]{a500_mdisc02.png} & \includegraphics[width=0.2\linewidth]{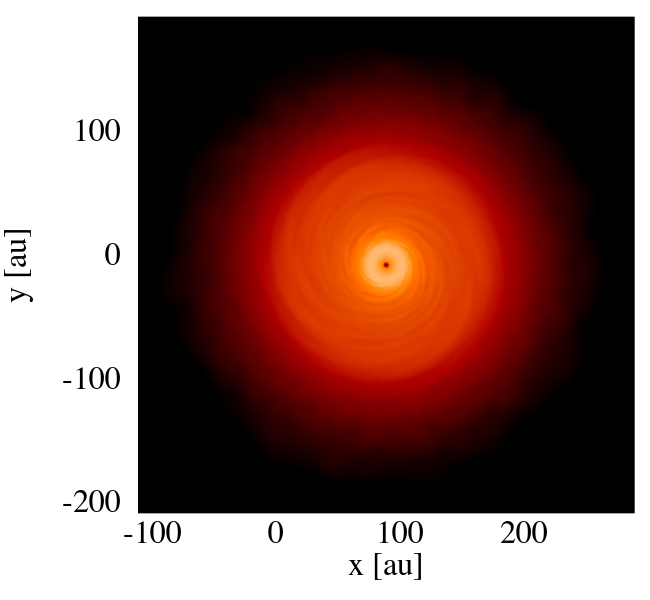} & \includegraphics[width=0.2\linewidth]{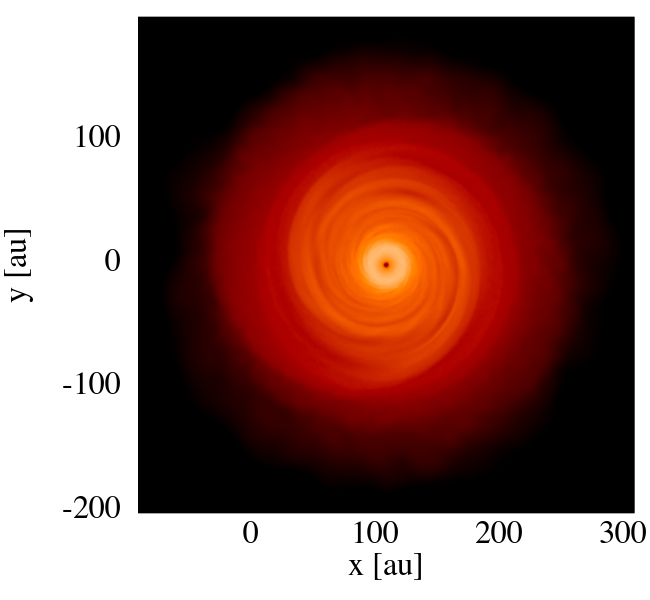} & \includegraphics[width=0.2\linewidth]{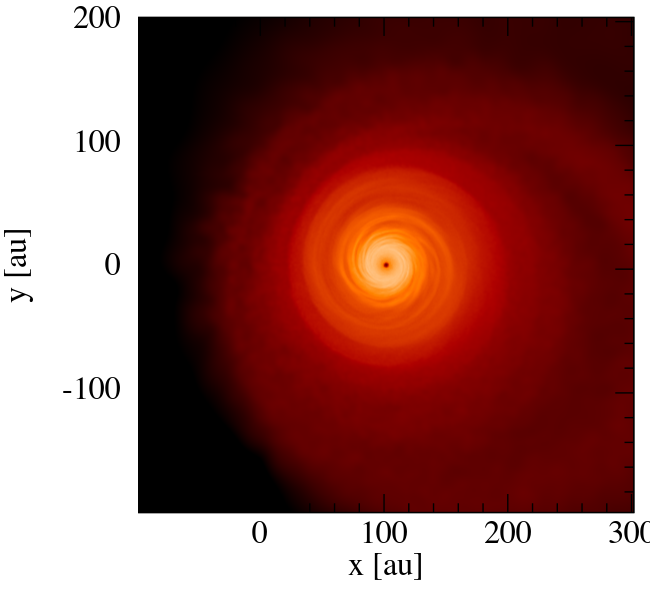} & \includegraphics[width=0.05\linewidth]{cbar.pdf} \\

    \end{tabular}

    \begin{tikzpicture}[remember picture,overlay]
      \draw[green,line width=5] (-0.2,3.7) -- (-0.2,11.2) -- (-8.1,11.2) -- (-8.1,22.45) -- (-0.2,22.45) -- (-0.2,14.9) -- (3.8,14.9) -- (3.8,3.7) -- (-0.25,3.7);
    \end{tikzpicture}

    \caption{Final states of the discs where we vary the orbital eccentricity and the semi-major axis of the companion star. Disc setup parameters are summarised in Table \ref{tab:sphsetup_eccentricity} and outlined in detail in Section \ref{sec:setup_ecccentricity}. Green boxes are included to highlight disc configurations which resulted in fragmentation, when their reference run analog did not.}
    \label{fig:sphresults_eccentricity}
\end{figure*}

So far we have only considered companion stars on circular orbits, with $e=0$. In reality it is likely that there will be some orbital eccentricity. Hence in this section we simulate 18 new discs, introducing eccentricities, $e=0.25, 0.5$ and $0.75$. We consider setup parameters found to be near to the limit for fragmentation, keeping the disc mass constant at $M_{\rm disc}=0.2$\,M$_{\odot}$, and varying $a$ and $e$ only. These setups are outlined in Section \ref{sec:setup_ecccentricity} and summarised in Table \ref{tab:sphsetup_eccentricity}. Final states of these discs are shown in Figure \ref{fig:sphresults_eccentricity}.

In the previous section we found that companion stars with semi-major axes $150$\,AU$ \leq a \leq 250$\,AU ($116$\,AU $ \leq r_{\rm peri,actual} \leq 186$\,AU) may induce fragmentation in a disc which would not fragment in isolation. When including eccentricity we find a wider range of semi-major axes are capable of inducing fragmentation, given that $r_{\rm peri,actual}$ falls roughly within the same range as found previously.

When including an eccentricity of $e=0.5$ we find that disc setups with $a=325$\,AU and $a=400$\,AU now also result in fragmentation. The periastron distances observed in these simulations are $r_{\rm peri,actual}=134$\,AU and $r_{\rm peri,actual}=163$\,AU respectively. Fragmentation occurs in a similar manner here to what was found in the previous section; an $m=2$ spiral which is generated as the companion approaches and passes through periastron becomes unstable and forms bound, self-gravitating clumps.

We find that fragmentation occurs in this suite for a slightly lower $r_{\rm peri,actual}$ than was found in the previous section. In the configuration with $a=150$\,AU and $e=0.25$, corresponding to $r_{\rm peri,actual}=92$\,AU, the companion passes through the very outer edge of the disc and a single fragment forms in the spiral $180^\circ$ from the companion's location. Disc heating at periastron is much less here than was the case when $a=100$\,AU and $e=0$ ($r_{\rm peri,actual}=78$\,AU), which was found to inhibit fragmentation. The amount of mass ejected is also significantly less here, with a final disc mass $M_{\rm disc}=0.18$\,M$_{\odot}$. What remains is a slightly more compact disc, truncated at $R\approx75$\,AU.

For any companion which passes closer than $r_{\rm peri,actual}=92$\,AU, the disc-star interaction becomes destructive. Disc material is dispersed as the companion passes through the disc, ejecting a significant amount of mass, leaving a compact, lower mass disc, thus entirely preventing fragmentation. All configurations here with $e=0.75$ suffer this fate.

An interesting case is the disc setup with $a=500$\,AU and $e=0.75$. Despite this system's actual periastron distance of $r_{\rm peri,actual}=104$\,AU falling within the sweet spot found which may induce fragmentation, no fragmentation occurs here. As the companion passes through periastron a self-gravitating clump begins to form at the inner edge of one of the spirals but immediately disperses as the companion quickly moves away from periastron. As the companion moves back toward apastron the disc stabilizes again.

\subsection{Varying orbital inclination}

\begin{figure*}
    \centering
    \begin{tabular}{c|ccccc}
    \centering &
    $i=0^\circ$ &   $i=30^\circ$ & $i=60^\circ$ &  $i=90^\circ$ & \\

    \hline\hline

     \rotatebox{90}{$a=100$\,AU} \vline\vline & \includegraphics[width=0.2\linewidth]{a100_mdisc02.png} & \includegraphics[width=0.2\linewidth]{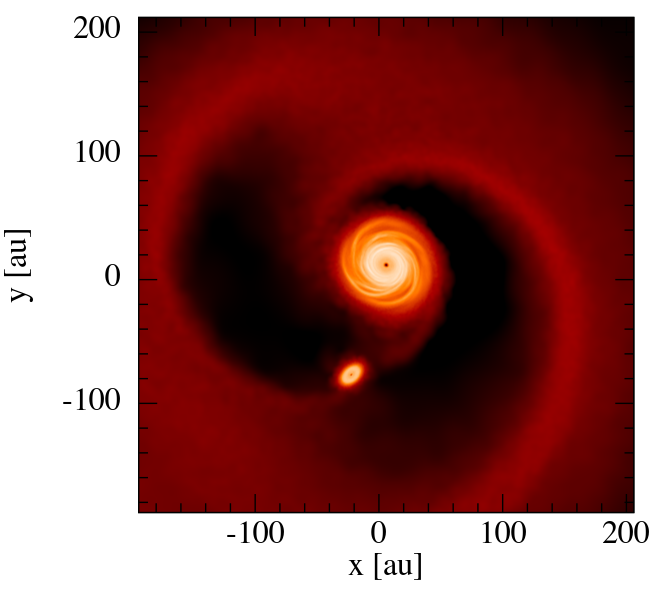} & \includegraphics[width=0.2\linewidth]{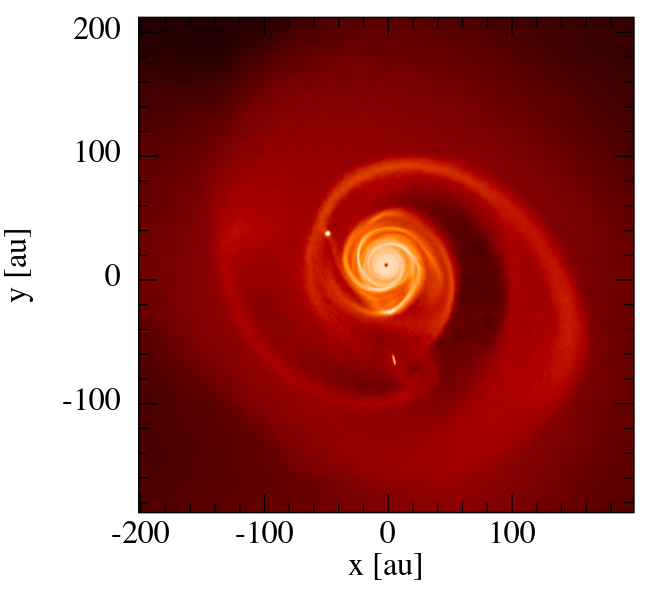} & \includegraphics[width=0.2\linewidth]{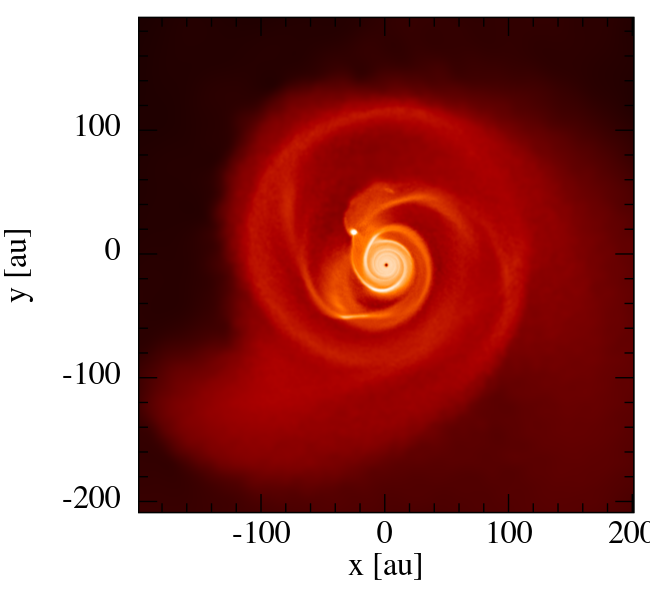} & \includegraphics[width=0.05\linewidth]{cbar.pdf} \\

    \hline

     \rotatebox{90}{$a=150$\,AU} \vline\vline & \includegraphics[width=0.2\linewidth]{a150_mdisc02.png} & \includegraphics[width=0.2\linewidth]{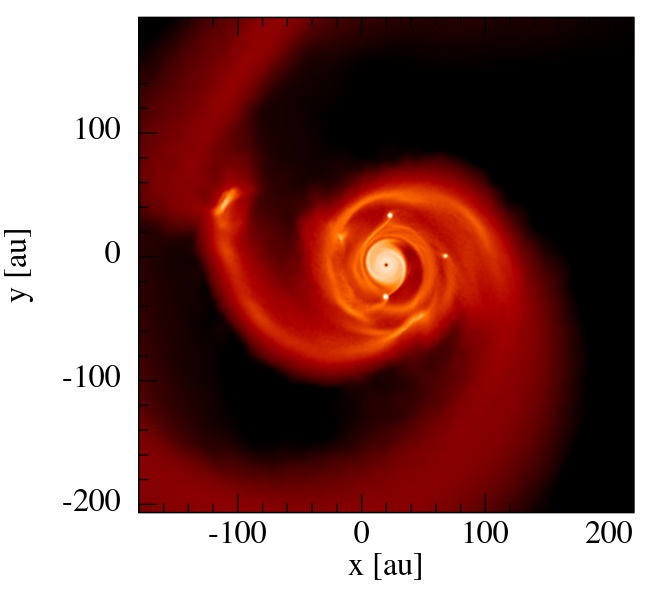} & \includegraphics[width=0.2\linewidth]{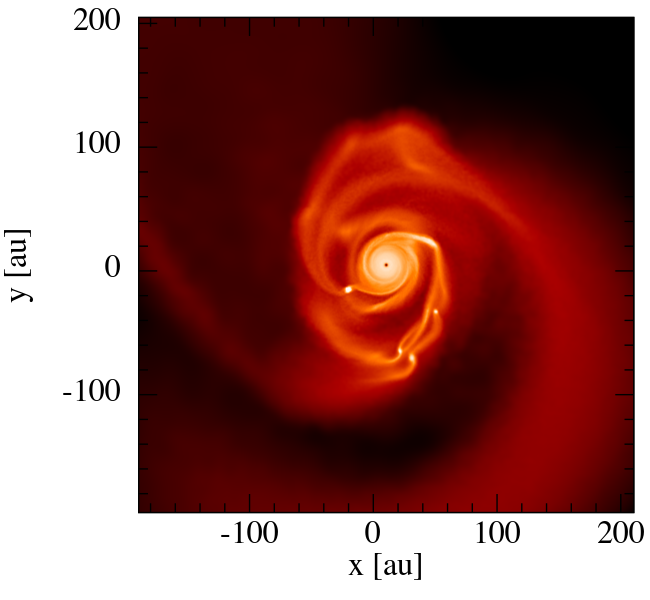} & \includegraphics[width=0.2\linewidth]{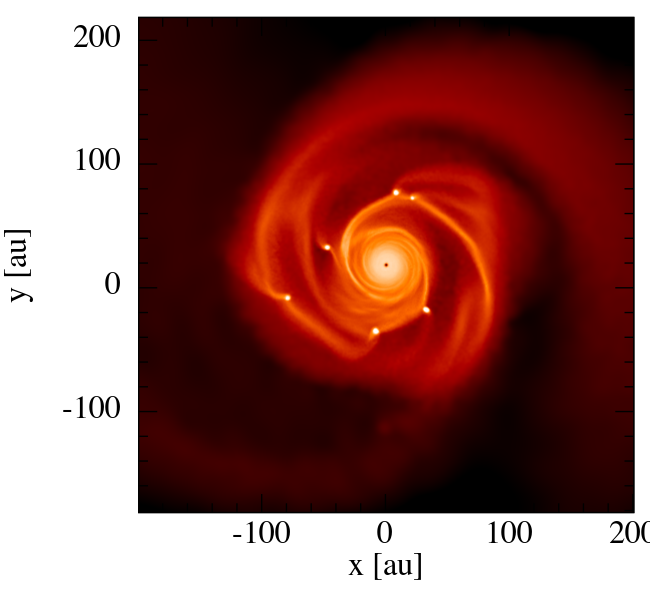} & \includegraphics[width=0.05\linewidth]{cbar.pdf} \\

    \hline

     \rotatebox{90}{$a=200$\,AU} \vline\vline & \includegraphics[width=0.2\linewidth]{a200_mdisc02.png} & \includegraphics[width=0.2\linewidth]{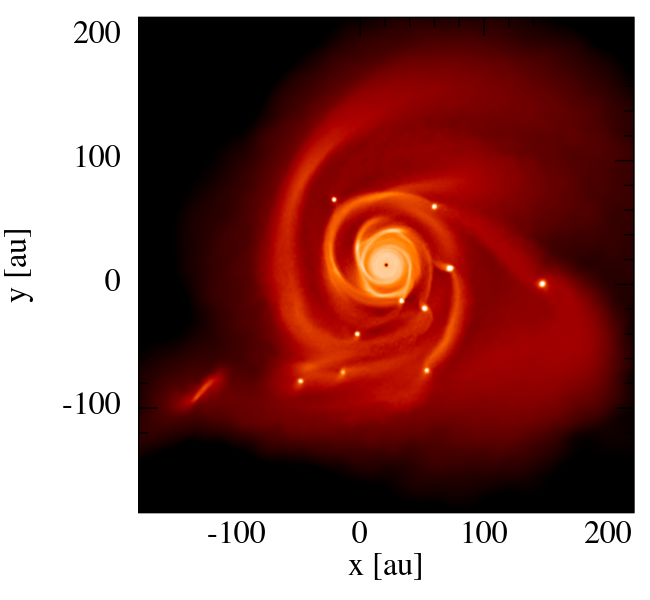} & \includegraphics[width=0.2\linewidth]{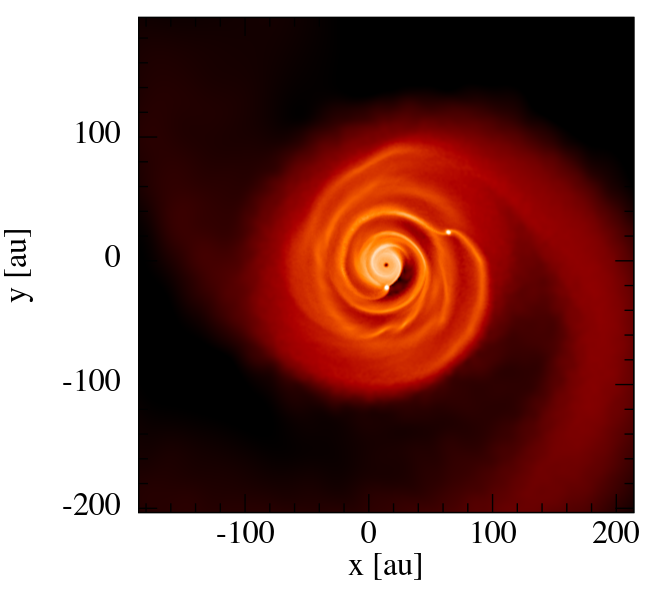} & \includegraphics[width=0.2\linewidth]{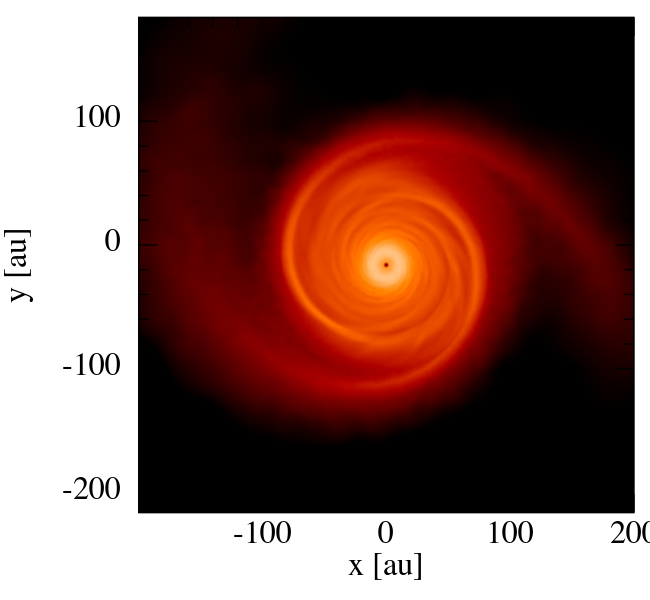} & \includegraphics[width=0.05\linewidth]{cbar.pdf} \\

    \hline

     \rotatebox{90}{$a=250$\,AU} \vline\vline & \includegraphics[width=0.2\linewidth]{a250_mdisc02.png} & \includegraphics[width=0.2\linewidth]{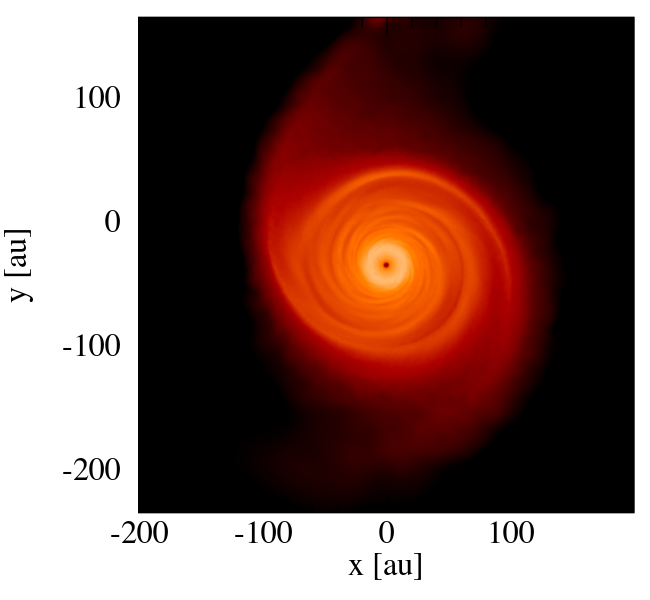} & \includegraphics[width=0.2\linewidth]{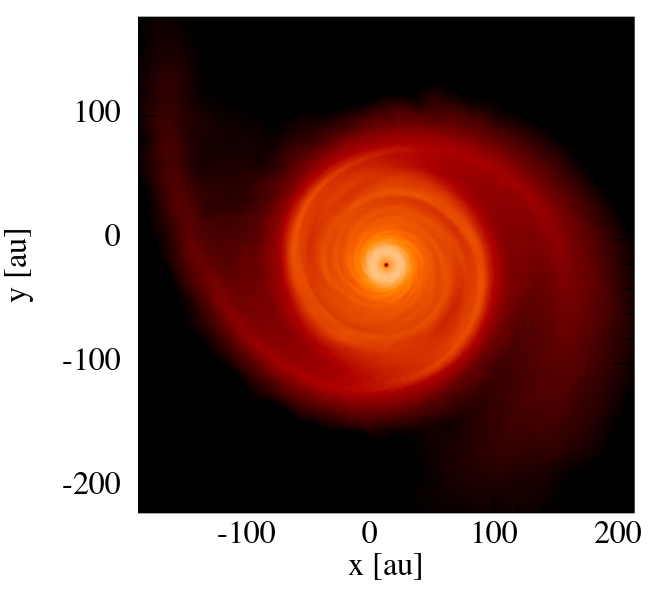} & \includegraphics[width=0.2\linewidth]{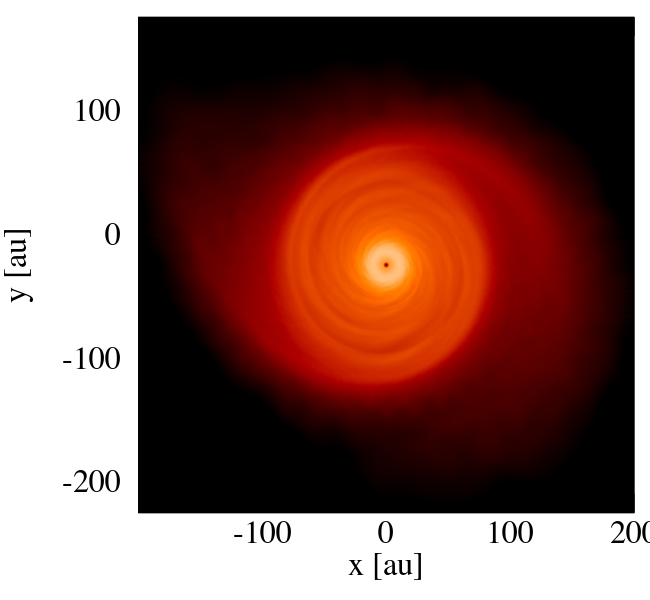} & \includegraphics[width=0.05\linewidth]{cbar.pdf} \\

    \end{tabular}

    \begin{tikzpicture}[remember picture,overlay]
      \draw[green,line width=5] (3.9,3.7) -- (-4.2,3.7) -- (-4.2,0) -- (-8.1,0) -- (-8.1,11.15) -- (-0.2,11.15) -- (-0.2,14.95) -- (7.8,14.95) -- (7.8,7.4) -- (3.9,7.4) -- (3.9,3.7);
    \end{tikzpicture}

    \caption{Final states of the discs where we vary the orbital inclination and the semi-major axis of the companion star. Disc setup parameters are summarised in Table \ref{tab:sphsetup_inclination} and outlined in detail in Section \ref{sec:setup_inclination}. Green boxes are included to highlight disc configurations which resulted in fragmentation, when their reference run analog did not.}
    \label{fig:sphresults_inclination}
\end{figure*}

In this section we consider systems with some orbital inclination relative to the plane of the disc. We introduce inclinations $i=30^\circ, 60^\circ$ and $90^\circ$, once again keeping the disc mass constant at $M_{\rm disc}=0.2$\,M$_{\odot}$ and varying the binary separation close to the limit for fragmentation. Disc setup parameters are outlined in Section \ref{sec:setup_inclination} and summarised in Table \ref{tab:sphsetup_inclination}. Final states of these discs are shown in Figure \ref{fig:sphresults_inclination}.

In the short orbit system, with $a=100$\,AU, we find that including an inclination $i \geq 60^\circ$ results in a less destructive disc-star interaction than when the companion orbits in the plane of the disc, hence fragmentation can occur. Despite their $r_{\rm peri,actual}$ being similar, the resulting surface density profile after the companion has passed through the plane of the disc is much less truncated when $i=60^\circ$ than was the case when $i=0^\circ$, extending to $R_{\rm out}\approx80$\,AU after several binary orbital periods. An $m=2$ spiral mode forms quickly, and a fragment forms at the inner edge of one the spirals, at $R=55$\,AU. The fragment's orbit is initially slightly inclined relative to the disc, with $i\approx8^\circ$, but it quickly settles into the plane of the disc after an orbital period. The final state of the $i=60^\circ$ disc is more flared compared to when $i=0^\circ$, with $H/R=0.15$ at $R=100$\,AU compared to $H/R=0.10$ at $R=100$\,AU.

In the wider orbit systems, with $a=200$\,AU and $250$\,AU, we find that including an inclination can weaken the disc-star interaction compared to when $i=0^\circ$ such that disc fragmentation no longer occurs. Considering the discs with $a=250$\,AU, as we increase the companion's inclination from $i=0^\circ$ to $i=90^\circ$ we find that the $m=2$ spiral mode generated by the companion becomes progressively less significant. When $i=30^\circ$, we observe a much smaller bump in the surface density profile when compared to the system with $i=0^\circ$. Hence the $Q-$profile remains relatively flat, and no fragmentation occurs. For the discs with $a=200$\,AU, only a single fragment forms when $i=60^\circ$, and increasing the inclination to $i=90^\circ$ prevents fragmentation from happening altogether.

\subsection{Varying companion mass}

The gravitational influence of the companion star on the disc will vary with the star's mass. Hence we setup 8 new discs where we explore the parameter space in $M_{\rm *,companion}$ and $a$ for configurations which result in fragmentation. In particular, we focus on how varying the companion star's mass affects the binary separations which are capable of inducing fragmentation.

\begin{figure*}
    \centering
    \begin{tabular}{c|cccc}
    \centering &
    $M_{\rm *,companion}=0.1$\,M$_{\odot}$ &  $M_{\rm *,companion}=0.2$\,M$_{\odot}$  &  $M_{\rm *,companion}=0.5$\,M$_{\odot}$ & \\

    \hline\hline

     \rotatebox{90}{$a=150$\,AU} \vline\vline & \includegraphics[width=0.2\linewidth]{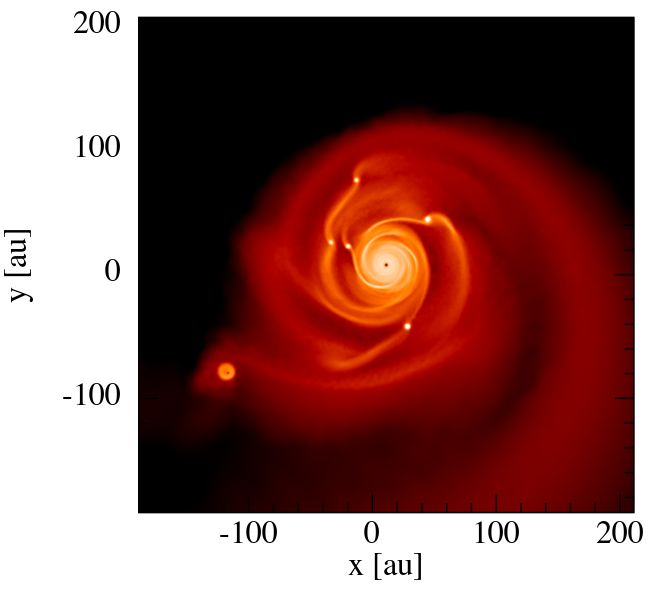} & \includegraphics[width=0.2\linewidth]{a150_mdisc02.png} & \includegraphics[width=0.2\linewidth]{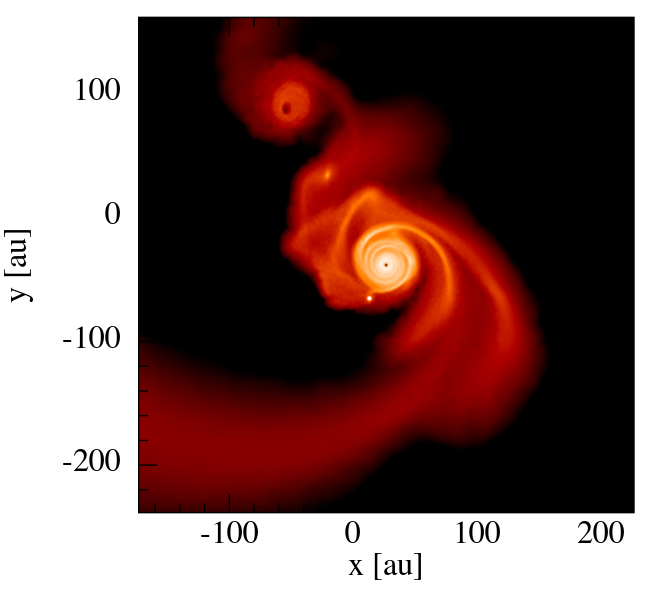} & \includegraphics[width=0.05\linewidth]{cbar.pdf} \\

    \hline

     \rotatebox{90}{$a=250$\,AU} \vline\vline & \includegraphics[width=0.2\linewidth]{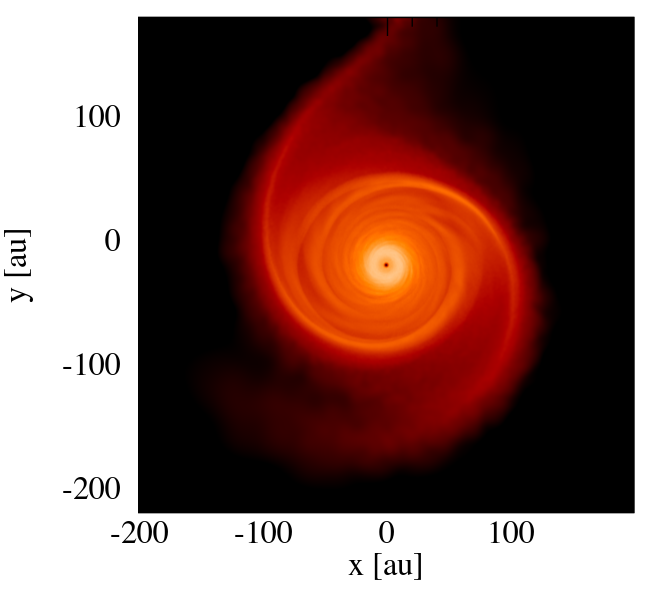} & \includegraphics[width=0.2\linewidth]{a250_mdisc02.png} & \includegraphics[width=0.2\linewidth]{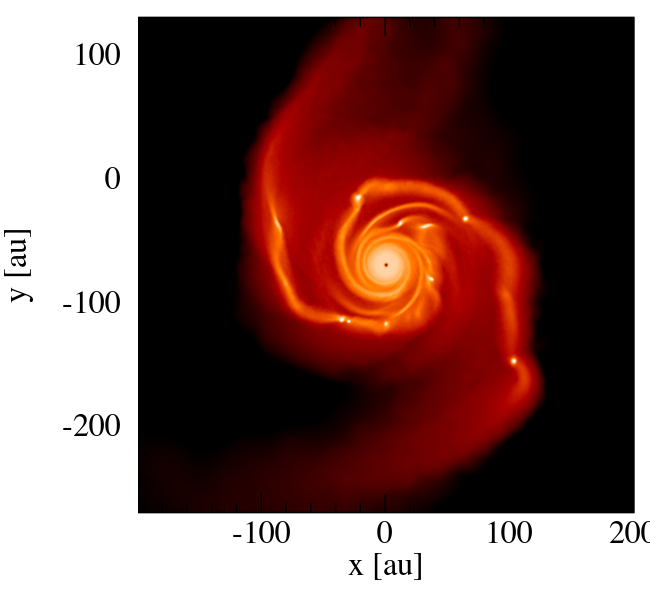} & \includegraphics[width=0.05\linewidth]{cbar.pdf} \\

    \hline

     \rotatebox{90}{$a=325$\,AU} \vline\vline & \includegraphics[width=0.2\linewidth]{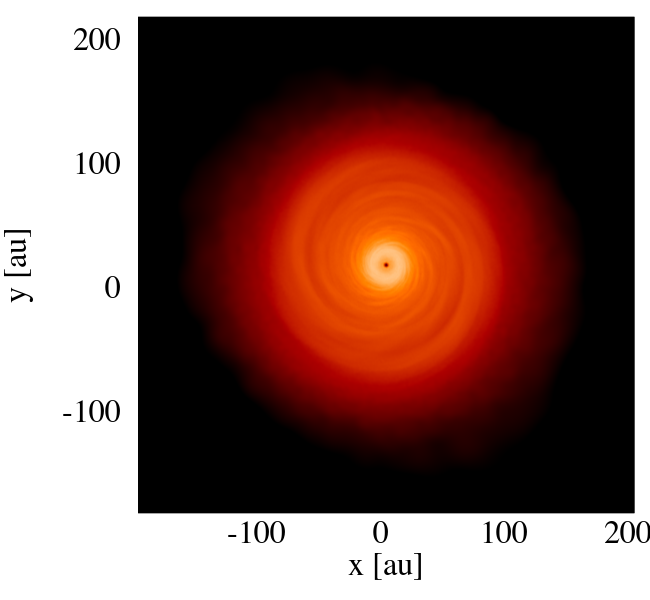} & \includegraphics[width=0.2\linewidth]{a325_mdisc02.png} & \includegraphics[width=0.2\linewidth]{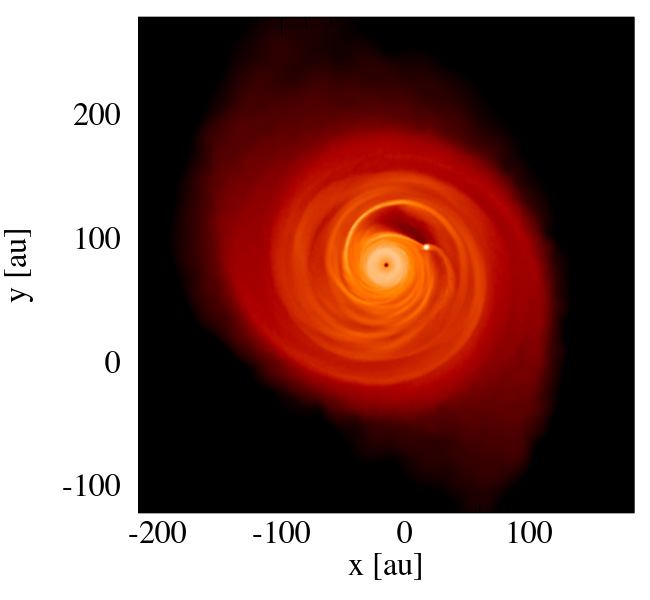} & \includegraphics[width=0.05\linewidth]{cbar.pdf} \\

    \hline

    \rotatebox{90}{$a=400$\,AU} \vline\vline & \includegraphics[width=0.2\linewidth]{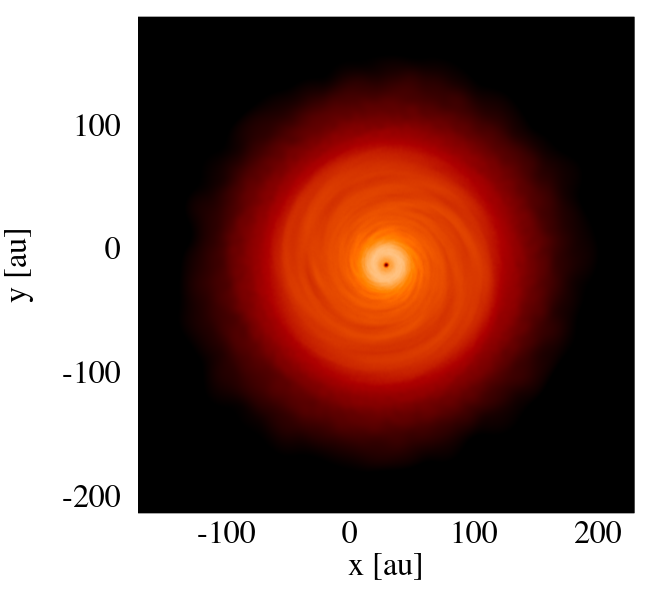} & \includegraphics[width=0.2\linewidth]{a400_mdisc02.png} & \includegraphics[width=0.2\linewidth]{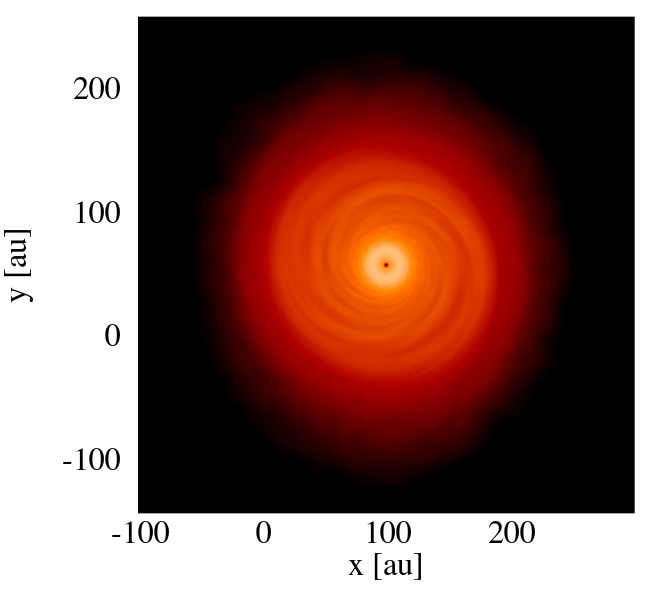} & \includegraphics[width=0.05\linewidth]{cbar.pdf} \\

    \end{tabular}

    \begin{tikzpicture}[remember picture,overlay]
      \draw[green,line width=5] (-6.2,14.95) -- (5.8,14.95) -- (5.8,3.7) -- (1.8,3.7) -- (1.8,7.4) -- (-2.2,7.4) -- (-2.2,11.15) -- (-6.2,11.15) -- (-6.2,14.95);
    \end{tikzpicture}

    \caption{Final states of the discs where we vary the mass and the semi-major axis of the companion star. Disc setup parameters are summarised in Table \ref{tab:sphsetup_msecondary} and outlined in detail in Section \ref{sec:setup_msecondary}. Green boxes are included to highlight disc configurations which resulted in fragmentation, when their reference run analog did not.}
    \label{fig:sphresults_msecondary}
\end{figure*}

Discs are setup with parameters close to the fragmentation limit found until now, keeping the disc mass constant at $M_{\rm disc}=0.2$\,M$_{\odot}$ and varying $M_{\rm *,companion}$ and $a$ only. We introduce new companion masses $M_{\rm *,companion}=0.1$\,M$_{\odot}$ and $M_{\rm *,companion}=0.5$\,M$_{\odot}$. These setups are outlined in Section \ref{sec:setup_msecondary} and summarised in Table \ref{tab:sphsetup_msecondary}. The final states of these discs are shown in Figure \ref{fig:sphresults_msecondary}.

Considering the discs with $a=150$\,AU, we find that both of the new setups, with $M_{\rm *,companion}=0.1$\,M$_{\odot}$ and $M_{\rm *,companion}=0.5$\,M$_{\odot}$, result in fragmentation, with a trend for the discs to fragment faster with increasing $M_{\rm *,companion}$. We find $t_{\rm frag}=900, 750$ and $550$\,yrs when $M_{\rm *,companion}=0.1, 0.2$ and $0.5$\,M$_{\odot}$ respectively.

When $a=250$\,AU we find that decreasing $M_{\rm *,companion}$ from $0.2$\,M$_{\odot}$ to $0.1$\,M$_{\odot}$ no longer results in fragmentation. A weaker $m=2$ spiral mode is driven by the $0.1$\,M$_{\odot}$ companion, which results in an increase in $\Sigma$ and a decrease in $Q$ at the spiral location, but the change is not significant enough to induce fragmentation. Instead, the spiral mode persists for two full binary orbits until the simulation ends. For the discs which do fragment, we again find a trend for discs to fragment earlier with increasing companion mass, finding $t_{\rm frag}=3050$ and $1550$\,yrs for $M_{\rm *,companion}=0.2$ and $0.5$\,M$_{\odot}$ respectively.

Considering the discs with $a=325$\,AU we find that increasing $M_{\rm *,companion}$ from $0.2$\,M$_{\odot}$ to $0.5$\,M$_{\odot}$ causes the disc to fragment, as the more massive star generates a stronger spiral mode. A single fragment forms after $4000$\,yrs in the disc with $a=325$\,AU and $M_{\rm *,companion}=0.5$\,M$_{\odot}$. Of all the discs simulated here, this is the widest periastron separation ($r_{\rm peri,actual}=256$\,AU) for which we find fragmentation can be triggered.

None of the configurations where $a=400$\,AU result in fragmentation, for any of the new companion star masses considered here.

\section{Discussion}
\label{discussion}

\subsection{Summary of results}

\begin{figure}
    \centering
    \includegraphics[width=\linewidth]{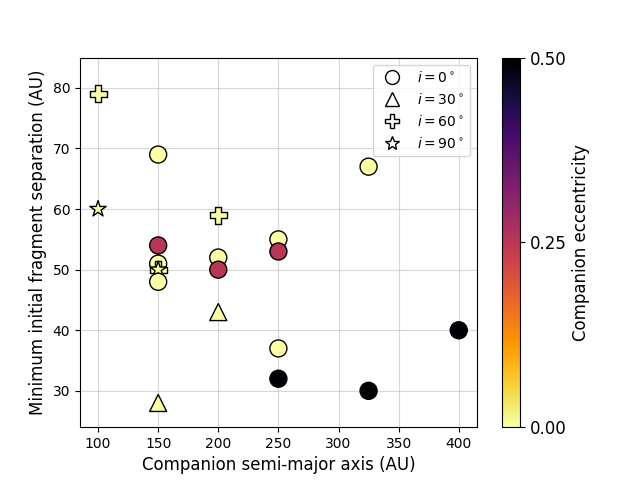}
    \caption{Summary of model parameters found to trigger fragmentation throughout this work, including the minimum radius at which fragments formed in each disc. A total of 20 systems presented here resulted in fragmentation. Companion orbital eccentricities are distinguished by different colors. Companion orbital inclinations are distinguished by different plot markers.}
    \label{fig:initial_fragment_separations}
\end{figure}

We have identified a "sweet spot" in the orbital parameter space of binary stars which may trigger fragmentation in a disc which does not fragment in isolation. We find that the disc-star interaction for intermediate separation binaries can be beneficial for fragmentation, with the exact range of ideal semi-major axes being a function of the orbital eccentricity, inclination and companion star mass. A plot summarising the companion's orbital parameters which are found to trigger fragmentation is included in Figure \ref{fig:initial_fragment_separations}, where we also highlight the minimum radius at which fragments formed in each disc.

In general, the companion will drive an $m=2$ spiral through the disc, and fragmentation occurs at the inner region of one, or both, of the spirals as a result of the enhanced surface density pushing the disc over the fragmentation threshold. Heating of the disc induced by the companion is balanced by efficient cooling (see Figure \ref{fig:a250_vs_referencerun}) and the instability is able to grow until fragmentation occurs.

We find that this is true for intermediate separation binaries with $150$\,AU $\leq a \leq 250$\,AU ($116$\,AU $\leq r_{\rm peri,actual} \leq 186$\,AU), when considering circular binary orbits in the plane of the disc.

For wide orbit binaries the spiral induced by the companion becomes progressively weaker with increasing binary separation. When $a \gtrsim 400$\,AU and $M_{\rm disc}=0.2$\,M$_{\odot}$, the disc's final surface density profile and $Q-$profile are almost identical to the counterpart disc from the reference run with no companion.

Very short separation binary encounters, where the companion passes through the outer edge of the disc, become prohibitive to fragmentation. As the companion star passes through the disc, material is ejected and the remaining surface density profile is modified to be much steeper in the inner disc, and truncated at a distance slightly smaller than the distance of periastron passage. Hence a much more compact and lower mass disc remains, and no fragmentation can occur.

When including an eccentricity in the binary orbit, we find a similar range in $r_{\rm peri,actual}$ capable of triggering fragmentation. From the suite considering non-circular orbits with moderate eccentricities ($e=0.25, 0.5$) we find that semi-major axes $150$\,AU $\leq a \leq 400$\,AU ($92$\,AU $\leq r_{\rm peri,actual} \leq 163$\,AU) can induce fragmentation. When considering highly eccentric orbits, with $e=0.75$, none of our simulations fragment. This is generally because the high eccentricity causes the companion to pass through the disc at periastron passage.

When including an orbital inclination for the companion, we find its influence to become progressively lesser as we move its orbit away from the plane of the disc. When $i=60^\circ$ we find the sweet spot in binary semi-major axis to be between $100$\,AU $\leq a \leq 200$\,AU ($74$\,AU $\leq r_{\rm peri, actual}\leq 152$\,AU), which is reduced to being between $100$\,AU $\leq a \leq 150$\,AU ($75$\,AU $\leq r_{\rm peri, actual}\leq 116$\,AU) when considering companions with $i=90^\circ$. High inclination binary companions ($i=60^\circ, 90^\circ$) which pass through the disc outer edge are less destructive than when the binary orbit is in the plane of the disc, hence fragmentation can occur for slightly shorter separations when $i=60^\circ$ and $i=90^\circ$ compared to when $i=0^\circ$.

The sweet spot found in binary separation is broadened as we increase the companion star's mass from $0.2$\,M$_{\odot}$ to $0.5$\,M$_{\odot}$, as the higher mass companion drives a stronger spiral mode through the disc. We find companions with semi-major axes as wide as $a=325$\,AU ($r_{\rm peri,actual}=256$\,AU) can trigger fragmentation when $M_{\rm *,companion}=0.5$\,M$_{\odot}$. Equally, when considering less massive companions, the sweet spot in binary separation is narrowed. Only one of our simulations, with $a=150$\,AU ($r_{\rm peri,actual}=110$\,AU), results in fragmentation when $M_{\rm *,companion}=0.1$\,M$_{\odot}$. In the disc configurations which fragment for more than one value of $M_{\rm *,companion}$ (when $a=150$\,AU and $250$\,AU), we find that the discs fragment faster with increasing companion mass.

\subsection{Comparison to previous theoretical work}

Previous work considering the possibility of fragmentation induced by the presence of a binary star companion consists of three key papers in \cite{nelson2000}, \cite{mayeretal05} and \cite{bossetal06}, with their results discussed in the review paper \cite{mayeretal07}. \cite{nelson2000} and \cite{mayeretal05} found that the presence of a companion suppresses any instability due to significant tidal heating in spiral shock waves, thus stabilising the disc. \cite{bossetal06}, however, concluded that binary companions may promote fragmentation, finding that spiral waves generated from the tidal interaction between the disc and the companion would typically go on to form dense, self-gravitating clumps. In \cite{mayeretal07} the authors largely attribute the differences in their results to the use of an artificial viscosity in \cite{nelson2000} and \cite{mayeretal05}, which isn't included in \cite{bossetal06}, and would contribute significantly toward heating of the disc in the presence of a shock wave, given a sufficiently large artificial viscosity was included.

Various authors have also investigated the role of stellar flyby events in promoting or suppressing fragmentation in discs which would be marginally stable in isolation. The early work of \cite{boffinetal98}, \cite{watkinsetal98a} and \cite{watkinsetal98b} found that, when considering isothermal discs, previously non-fragmenting discs would fragment during star-disc and disc-disc interactions. However later models which included more realistic cooling found that heating of the disc during the stellar encounter was sufficient to stabilise it against fragmentation \citep{lodatoetal07,forganrice2009}.

Until now, most work has considered either simple cooling prescriptions, where the cooling time is proportional to the local orbital time \citep[$\beta$-cooling,][]{gammie01,rice03}, isothermal discs, or have not included an artificial viscosity which will capture heating from shocks. Including an algorithm to approximate radiation transport in our models \citep{forganetal2009} allows us to model realistic disc cooling, hence we can realistically capture whether the disc is able to radiate away the additional energy generated through tidal heating during the binary encounter.

The aforementioned works generally also considered much more compact discs than we have modelled here \citep[in the case of][]{nelson2000,mayeretal05,bossetal06,lodatoetal07}, or discs with $R_{\rm out}=1000$\,AU \citep[in the case of][]{boffinetal98,watkinsetal98a,watkinsetal98b}. However, owing to the simpler methods used to model disc cooling, their models are scale-free and can be scaled to different physical units for comparison with the results here. Hence when comparing to the works of \cite{boffinetal98,watkinsetal98a,watkinsetal98b,nelson2000,mayeretal05,bossetal06,lodatoetal07} we can use their ratios $r_{\rm peri}/r_{\rm out,disc}$  for direct comparison to our results. This is not the case for the results from \cite{forganrice2009}, who use the \cite{forganetal2009} hybrid radiative transfer method, and considered $R_{\rm out}=40$\,AU discs. For the coplanar binary encounters considered here, we find that companions with $0.92 \leq r_{\rm peri}/r_{\rm out,disc} \leq 1.86$ trigger fragmentation. Only \cite{lodatoetal07} considered binary encounters within this range, finding that no fragmentation occurred in their simulations.

Of the previous works which include a similar cooling approximation as we use here, we find that our results broadly indicate the same thing. \cite{forganrice2009}in  find that small separation, disc-penetrating encounters heat the disc material, whilst angular momentum transport and mass stripping result in a more stable disc configuration after the encounter. Large separation encounters have very little effect, becoming less significant as the periastron distance increases. However intermediate separation encounters may modify the surface density profile of the disc, without causing significant heating, such that the disc is more unstable over a larger range of radii after the encounter. None of the discs in \cite{forganrice2009} fragment, but the authors suggest that there could be some region of parameter space in periastron distance which may act to promote fragmentation.

\cite{meru2015}, who used the flux-limited diffusion approximation \citep[e.g.][]{mayeretal07b}, also found that further fragmentation may be triggered in a disc which has fragmented already. The fragment which has initially formed causes material to be channelled inwards, increasing the density of the inner spirals, causing fragmentation.

Here, we have presented a suite of simulations which model realistic cooling using the \cite{forganrice2009} radiative transfer approach. We find that efficient cooling is able to prevent the disc temperature from increasing significantly during the binary's periastron passage, and fragmentation can occur in the spiral regions of enhanced surface density which are driven by the companion.

\subsection{Comparison to observations}

Binaries are often neglected from observational and theoretical exoplanetary science, as they complicate the modelling of planet formation, as well as the detection and characterisation of planetary systems. Most of the work (theoretical and observational) conducted so far on planets in binaries has focused on close-in binaries (separations of tens of AU), generally agreeing that tight binaries ($<$\,50--100\,AU) hinder planet formation \citep{bergfors13,kraus12,kraus16,kaib13}.
However, the first planets discovered in binary systems showed distinct orbital and physical properties from the rest of the planetary population, hinting at the possibility that binary companions could dramatically reorient the orbital configuration of planetary systems \citep{zuckermazeh02}. Observations of binary star systems suggest that stellar multiplicity at wider separations may play a key role in the formation of high-mass gas giant planets and brown dwarfs. Various surveys have found an excess of outer companions to stars with massive hot Jupiters or short-period stellar and substellar companions when compared to field stars, suggesting that binary star systems on separations of a few hundred AU may be favourable sites for the formation of these inner companions.

Beginning with their survey of solar-type spectroscopic binaries (SB), \cite{tokovinin06} found an excess of wide tertiary stellar companions for SBs with periods from 1--30\,days, rising to a frequency of $96\%$ for SBs with periods $<3$\,days.
In their series of "Friends of hot Jupiters" papers, \cite{ngoetal16} searched for stellar companions to 77 systems hosting hot Jupiters. They found that $47\pm7\%$ of stars hosting hot Jupiters have a binary companion with separations between 50--20,000\,AU (a value 3 times higher than found for field stars), although \cite{moekratter19} concluded that this excess was not significant after accounting for remaining statistical biases. Nonetheless, \cite{ngoetal16} still observed a significant deficit of tight binary companions, with separations 50--100\,AU, compared to wider systems, consistent with the idea that shorter-period binaries may be detrimental to planet formation \citep{wang14,kraus16}.

Using direct imaging data, \cite{fontanive19} searched for wide-orbit binary companions to 38 stars known to host very massive hot Jupiters or brown dwarfs (7--60\,M$_\mathrm{Jup}$) on short periods ($<$1\,AU), finding a binary fraction close to 80\% for these systems on separations of 20--10,000\,AU, twice as high as for field stars, with a significance confirmed in \cite{moekratter19}. Again, they observed a lack of binaries with separations of tens of AU, and instead found an excess of intermediate separation binaries, with a peak in binary separation at $\sim$250\,AU. The binary frequency for massive giant exoplanets and brown dwarfs ($M>7$\,M$_{\rm Jup}$) was found to be higher than for lower mass hot Jupiters (0.2--4\,M$_\mathrm{Jup}$), suggesting that the stellar companion's influence may facilitate the formation of high-mass giant planets and brown dwarfs. The systems probed in \cite{fontanive19} also have a lower mean metallicity, consistent with that of the field \citep{moe19}, compared to hosts to genuine hot Jupiters like those studied in \cite{ngoetal16}. Given the strong correlation seen between metallicity and the ability to form gas giant planets via core accretion \citep{mordasini12,jenkins17}, the high-mass inner substellar companions targeted in \cite{fontanive19} are therefore likely to have formed via GI rather than CA as for the lower-mass hot Jupiters.

Recently, \cite{fontanive21} reached similar conclusions, finding that giant planets have a substantially larger raw stellar multiplicity fraction than sub-Jovian planets, and that this trend further increases up to a $\sim$30\% raw binary fraction for massive planet and brown dwarfs ($M>7$\,M$_{\rm Jup}$) on very short orbital separations ($<0.5$\,AU), with the most massive and shortest-period substellar companions almost exclusively observed in multiple-star environments. These systems thus appear to follow the architectures of stellar spectroscopic binaries, systematically observed as part of hierarchical triple systems \citep{tokovinin06}. Notably, \cite{fontanive21} showed that these extreme inner companions, with few analogues in (seemingly) single-star systems, were predominantly found to be in binaries with separations of few hundred AU, and mostly on separations $<250$\,AU (despite a strong incompleteness at these separations) for substellar companions with masses above 7\,M$_{\rm Jup}$, consistent with results from \cite{fontanive19}. In comparison, they found a peak around 600\,AU (subject to the same incompleteness biases) for binaries hosting lower-mass planets or warm and cool gas giants on wider orbital separations, and these systems showed similar planet properties to the population of exoplanets orbiting single stars.

These results suggest that very wide binaries have no meaningful impact on the architectures of planetary systems, and confirm the idea that very tight binary systems have a negative impact on planet formation. In particular, it appears that binaries with tens to a few hundred AU separations prevent planet formation for sub-Jovian and giant planets with masses up to a few M$_\mathrm{Jup}$, while wider binaries can harbour such planets but without affecting their orbital properties. This indicates that the exoplanet population issued from core accretion only exists in binary configurations that are not disruptive to planet formation and do not influence the resulting planet properties. On the other hand, the higher-mass population of giant planets and brown dwarfs on short-period orbits, likely formed via GI, are predominantly seen in intermediate separation binaries of few hundred AU separations \citep{fontanive21}, which must thus play a role in their existence.

Here, we show with simulations of self-gravitating discs that such intermediate separation binary companions may assist in the formation of giant planets through means of the gravitational instability. We find that when introducing a stellar companion at a few hundred AU into a disc configuration which would previously not fragment, fragmentation may be induced as the companion drives strong spirals which push the disc over the limit for instability. We find this to be true for binaries with semi-major axes between $\sim$100--400\,AU for the explored parameter space, with some dependency on binary orbital eccentricity, inclination and companion mass. This is consistent with the binary projected separations observed by \cite{fontanive19} and \cite{fontanive21}, a peak in the observed distribution at around $250$\,AU. We also note that in our simulations the orbital properties of these intermediate separation binaries remain mostly unchanged after a full orbital period, as the companions do not pass directly through the disc hence the drag that they experience from the disc material is minimal.

Shorter-period binaries and highly-eccentric systems inhibit fragmentation as the disc-star interaction becomes destructive. As the companion passes through periastron it will pass through the disc, leaving a compact, lower mass disc remaining. Hence we would expect a lower frequency of GI-born planets within tight binary systems on separations of tens of AU, which is also consistent with the shortfall of such systems in observations \citep{wang14,kraus16,ngoetal16,fontanive19}.

\subsection{Outlook and implications for short-period, massive planets}

Our work provides a viable formation pathway for the high-mass giant planets and brown dwarfs observed around components of multiple star systems \citep{fontanive19,fontanive21}. However, these substellar companions are actually observed on very short orbital periods ($<$1\,AU), much tighter than the typical formation locations from disc fragmentation.

In some cases, these objects could have been scattered by the binary stellar companion onto highly eccentric orbits and then been tidally circularised onto their current orbits \citep{rice15}. However, as discussed earlier, this is only possible for a subset of the systems presented in \citet{fontanive19}.

Another possibility is that these objects may have naturally migrated to their current locations. \cite{baruteau11} showed that fragments forming in young, massive discs will undergo rapid, type I migration before having chance to open a gap, and may be able to reach the inner disc within a few orbital periods. However, it is uncertain as to what fraction of fragments will survive this migration, and what their eventual masses will be after tidal downsizing \citep{nayakshin10,boley10}.

We also find indications that fragmentation triggered by the binary companion may be occurring closer in than is usually found for discs in isolation. In Figure \ref{fig:a250_vs_referencerun}, considering the disc with an $a=250$\,AU companion, the $Q-$profile reaches a minimum of $Q=0.79$ at $R=63$\,AU, hence the first fragment initially forms at $R=56$\,AU. In Figure \ref{fig:initial_fragment_separations}, we plot the minimum separation at which fragments form in each of our discs, including all systems with $M_{\rm disc}=0.2$\,M$_{\odot}$ which resulted in fragmentation. Of the 20 discs included in the plot, we find 9 systems produce fragments within $R=50$\,AU, and 2 form fragments within $R=30$\,AU.

It may then be that a combination of scattering, fragments forming close in, and rapid inward migration can produce the giant planets and brown dwarfs observed on very short orbital periods. Whilst \cite{baruteau11} considered the subsequent migration of single fragments forming in a self-gravitating discs, it is not known how a binary companion or the formation of multiple fragments may affect this. We leave this question as subject of future work.


\section{Conclusions}
\label{conclusions}

Observations of systems with close-in massive planetary and brown dwarf companions suggest that almost all host a binary stellar companion on a wider orbit \citep{fontanive19}. Also, the properties of the close-in objects are consistent with them having formed via fragmentation in a gravitationally unstable disc \citep{fontanive21}.  However, disc fragmentation is only likely to operate in the outer parts of such discs, requiring that these objects somehow move from where they formed onto the close-in orbits they now occupy.

In some cases, the close-in object could have been scattered by the binary stellar companion and then undergone tidal circularisation onto its current close-in orbit \citep{rice15,fontanive19}.  However, in many cases the tidal circularisation timescale is far too long for this to be a viable pathway for these systems. That such systems still typically host binary stellar companions suggests that these stellar companions still play a role in their formation.

To investigate this, we have conducted a series of 3D SPH simulations of self-gravitating discs with a binary stellar companion, exploring the companion's orbital parameter space for configurations which may trigger fragmentation in a marginally gravitationally unstable disc. We find a "sweet spot" in which intermediate separation binaries can induce fragmentation, with the exact set of ideal orbital parameters being a function of the companion's semi-major axis, eccentricity, inclination and mass.

Radiation transport is modelled using the \cite{forganetal2009} hybrid approach. For the discs modelled here, with outer radii $R_{\rm out}=100$\,AU, we find that efficient cooling during intermediate separation ($100$\,AU $\lesssim a \lesssim 400$\,AU) binary encounters allows disc fragmentation to occur in a spiral region of enhanced surface density driven by the companion star. Short separation disc-penetrating ($a \lesssim 100$\,AU) encounters are generally destructive, as mass stripping and disc heating entirely wipe out any instability. This is also true of highly eccentric binary orbits, which result in the companion passing through the disc. However, highly inclined ($i \gtrsim 60^\circ$) disc-penetrating encounters can be less destructive, allowing shorter separation encounters to trigger fragmentation than when the binary orbit is in the plane of the disc. Wide orbit binary encounters ($a \gtrsim 500$\,AU) have little effect on the disc properties, with the companion's influence becoming progressively lesser with increasing binary separation.

The range of binary separations found to promote fragmentation is consistent with the projected separations of the systems which display an excess of close-in giant planets and brown dwarfs \citep{wang14,kraus16,ngoetal16,fontanive19,fontanive21}. As our results show that intermediate separation binary systems could be favourable sites for the formation of massive substellar objects, we suggest that triggered fragmentation may contribute to the excess of massive planets and brown dwarfs observed around these systems. The question now remains how these fragments, initially formed on wide orbits, might have migrated to the very short separations ($< 1$\,AU) where they are now currently observed, and will be the subject of future work.

\begin{table*}
    \centering
    \begin{tabular}{ccccccccccc}
        \hline
         & $a$ & $M_{\rm disc}$ & $M_{\rm *,primary}$ & $M_{\rm *,companion}$ & $e$ & $i$ & $r_{\rm peri, calc}$ & $r_{\rm peri, actual}$ & Fragmented? & $t_{\rm frag}$ \\
         \hline\hline
         \underline{Reference runs:} & - & $0.1$\,M$_{\odot}$ & $1$\,M$_{\odot}$ & - & - & - & - & - & \ding{55} & - \\
         & - & $0.2$\,M$_{\odot}$ & $1$\,M$_{\odot}$ & - & - & - & - & - & \ding{55} & - \\
         & - & $0.3$\,M$_{\odot}$ & $1$\,M$_{\odot}$ & - & - & - & - & - & \ding{51} & 700\,yrs \\
         & - & $0.4$\,M$_{\odot}$ & $1$\,M$_{\odot}$ & - & - & - & - & - & \ding{51} & 500\,yrs \\
         \hline
         \underline{Varying binary separation:} & $100$\,AU & $0.1$\,M$_{\odot}$ & $1$\,M$_{\odot}$ & $0.2$\,M$_{\odot}$ & 0 & 0$^\circ$ & $100$\,AU & $87$\,AU & \ding{55} & - \\
         & $100$\,AU & $0.2$\,M$_{\odot}$ & $1$\,M$_{\odot}$ & $0.2$\,M$_{\odot}$ & 0 & 0$^\circ$ & $100$\,AU & $78$\,AU & \ding{55} & - \\
         & $100$\,AU & $0.3$\,M$_{\odot}$ & $1$\,M$_{\odot}$ & $0.2$\,M$_{\odot}$ & 0 & 0$^\circ$ & $100$\,AU & $70$\,AU & \ding{51} & 600\,yrs \\
         & $100$\,AU & $0.4$\,M$_{\odot}$ & $1$\,M$_{\odot}$ & $0.2$\,M$_{\odot}$ & 0 & 0$^\circ$ & $100$\,AU & $64$\,AU & \ding{51} & 500\,yrs \\
         & $250$\,AU & $0.1$\,M$_{\odot}$ & $1$\,M$_{\odot}$ & $0.2$\,M$_{\odot}$ & 0 & 0$^\circ$ & $250$\,AU & $213$\,AU & \ding{55} & - \\
         & $250$\,AU & $0.2$\,M$_{\odot}$ & $1$\,M$_{\odot}$ & $0.2$\,M$_{\odot}$ & 0 & 0$^\circ$ & $250$\,AU & $186$\,AU &  {\color{Green} \ding{51}\ding{51}} & 3050\,yrs \\
         & $250$\,AU & $0.3$\,M$_{\odot}$ & $1$\,M$_{\odot}$ & $0.2$\,M$_{\odot}$ & 0 & 0$^\circ$ & $250$\,AU & NR & \ding{51} & 720\,yrs \\
         & $250$\,AU & $0.4$\,M$_{\odot}$ & $1$\,M$_{\odot}$ & $0.2$\,M$_{\odot}$ & 0 & 0$^\circ$ & $250$\,AU & NR & \ding{51} & 540\,yrs \\
         & $500$\,AU & $0.1$\,M$_{\odot}$ & $1$\,M$_{\odot}$ & $0.2$\,M$_{\odot}$ & 0 & 0$^\circ$ & $500$\,AU & $430$\,AU & \ding{55} & - \\
         & $500$\,AU & $0.2$\,M$_{\odot}$ & $1$\,M$_{\odot}$ & $0.2$\,M$_{\odot}$ & 0 & 0$^\circ$ & $500$\,AU & $373$\,AU & \ding{55} & - \\
         & $500$\,AU & $0.3$\,M$_{\odot}$ & $1$\,M$_{\odot}$ & $0.2$\,M$_{\odot}$ & 0 & 0$^\circ$ & $500$\,AU & NR & \ding{51} & 680\,yrs \\
         & $500$\,AU & $0.4$\,M$_{\odot}$ & $1$\,M$_{\odot}$ & $0.2$\,M$_{\odot}$ & 0 & 0$^\circ$ & $500$\,AU & NR & \ding{51} & 445\,yrs \\
         & $1000$\,AU & $0.1$\,M$_{\odot}$ & $1$\,M$_{\odot}$ & $0.2$\,M$_{\odot}$ & 0 & 0$^\circ$ & $1000$\,AU & 862\,AU & \ding{55} & - \\
         & $1000$\,AU & $0.2$\,M$_{\odot}$ & $1$\,M$_{\odot}$ & $0.2$\,M$_{\odot}$ & 0 & 0$^\circ$ & $1000$\,AU & 758\,AU & \ding{55} & - \\
         & $1000$\,AU & $0.3$\,M$_{\odot}$ & $1$\,M$_{\odot}$ & $0.2$\,M$_{\odot}$ & 0 & 0$^\circ$ & $1000$\,AU & NR & \ding{51} & 650\,yrs \\
         & $1000$\,AU & $0.4$\,M$_{\odot}$ & $1$\,M$_{\odot}$ & $0.2$\,M$_{\odot}$ & 0 & 0$^\circ$ & $1000$\,AU & NR & \ding{51} & 500\,yrs \\
         \hline
         \underline{Additional separation runs:} & $150$\,AU & $0.2$\,M$_{\odot}$ & $1$\,M$_{\odot}$ & $0.2$\,M$_{\odot}$ & 0 & 0$^\circ$ & $150$\,AU & $116$\,AU &  {\color{Green} \ding{51}\ding{51}} & 750\,yrs \\
         & $200$\,AU & $0.2$\,M$_{\odot}$ & $1$\,M$_{\odot}$ & $0.2$\,M$_{\odot}$ & 0 & 0$^\circ$ & $200$\,AU & $150$\,AU &  {\color{Green} \ding{51}\ding{51}} & 1500\,yrs \\
         & $325$\,AU & $0.2$\,M$_{\odot}$ & $1$\,M$_{\odot}$ & $0.2$\,M$_{\odot}$ & 0 & 0$^\circ$ & $325$\,AU & $240$\,AU & \ding{55} & - \\
         & $400$\,AU & $0.2$\,M$_{\odot}$ & $1$\,M$_{\odot}$ & $0.2$\,M$_{\odot}$ & 0 & 0$^\circ$ & $400$\,AU & $299$\,AU & \ding{55} & - \\
         \hline
         \underline{Varying eccentricity:} & $150$\,AU & $0.2$\,M$_{\odot}$ & $1$\,M$_{\odot}$ & $0.2$\,M$_{\odot}$ & 0.25 & 0$^\circ$ & $113$\,AU & $92$\,AU &  {\color{Green} \ding{51}\ding{51}} & 900\,yrs \\
         & $150$\,AU & $0.2$\,M$_{\odot}$ & $1$\,M$_{\odot}$ & $0.2$\,M$_{\odot}$ & 0.5 & 0$^\circ$ & $75$\,AU & $63$\,AU & \ding{55} & - \\
         & $150$\,AU & $0.2$\,M$_{\odot}$ & $1$\,M$_{\odot}$ & $0.2$\,M$_{\odot}$ & 0.75 & 0$^\circ$ & $38$\,AU & $39$\,AU & \ding{55} & - \\
         & $200$\,AU & $0.2$\,M$_{\odot}$ & $1$\,M$_{\odot}$ & $0.2$\,M$_{\odot}$ & 0.25 & 0$^\circ$ & $150$\,AU & $119$\,AU & {\color{Green} \ding{51}\ding{51}} & 1300\,yrs \\
         & $200$\,AU & $0.2$\,M$_{\odot}$ & $1$\,M$_{\odot}$ & $0.2$\,M$_{\odot}$ & 0.5 & 0$^\circ$ & $100$\,AU & $81$\,AU & \ding{55} & - \\
         & $200$\,AU & $0.2$\,M$_{\odot}$ & $1$\,M$_{\odot}$ & $0.2$\,M$_{\odot}$ & 0.75 & 0$^\circ$ & $50$\,AU & $46$\,AU & \ding{55} & - \\
         & $250$\,AU & $0.2$\,M$_{\odot}$ & $1$\,M$_{\odot}$ & $0.2$\,M$_{\odot}$ & 0.25 & 0$^\circ$ & $188$\,AU & $147$\,AU & {\color{Green} \ding{51}\ding{51}} & 2000\,yrs \\
         & $250$\,AU & $0.2$\,M$_{\odot}$ & $1$\,M$_{\odot}$ & $0.2$\,M$_{\odot}$ & 0.5 & 0$^\circ$ & $125$\,AU & $102$\,AU & {\color{Green} \ding{51}\ding{51}} & 1770\,yrs \\
         & $250$\,AU & $0.2$\,M$_{\odot}$ & $1$\,M$_{\odot}$ & $0.2$\,M$_{\odot}$ & 0.75 & 0$^\circ$ & $63$\,AU & $54$\,AU & \ding{55} & - \\
         & $325$\,AU & $0.2$\,M$_{\odot}$ & $1$\,M$_{\odot}$ & $0.2$\,M$_{\odot}$ & 0.25 & 0$^\circ$ & $244$\,AU & $200$\,AU & \ding{55} & - \\
         & $325$\,AU & $0.2$\,M$_{\odot}$ & $1$\,M$_{\odot}$ & $0.2$\,M$_{\odot}$ & 0.5 & 0$^\circ$ & $163$\,AU & $134$\,AU & {\color{Green} \ding{51}\ding{51}} & 2750\,yrs \\
         & $325$\,AU & $0.2$\,M$_{\odot}$ & $1$\,M$_{\odot}$ & $0.2$\,M$_{\odot}$ & 0.75 & 0$^\circ$ & $81$\,AU & $69$\,AU & \ding{55} & - \\
         & $400$\,AU & $0.2$\,M$_{\odot}$ & $1$\,M$_{\odot}$ & $0.2$\,M$_{\odot}$ & 0.25 & 0$^\circ$ & $300$\,AU & $236$\,AU & \ding{55} & - \\
         & $400$\,AU & $0.2$\,M$_{\odot}$ & $1$\,M$_{\odot}$ & $0.2$\,M$_{\odot}$ & 0.5 & 0$^\circ$ & $200$\,AU & $163$\,AU & {\color{Green} \ding{51}\ding{51}} & 3800\,yrs \\
         & $400$\,AU & $0.2$\,M$_{\odot}$ & $1$\,M$_{\odot}$ & $0.2$\,M$_{\odot}$ & 0.75 & 0$^\circ$ & $100$\,AU & $84$\,AU & \ding{55} & - \\
         & $500$\,AU & $0.2$\,M$_{\odot}$ & $1$\,M$_{\odot}$ & $0.2$\,M$_{\odot}$ & 0.25 & 0$^\circ$ & $375$\,AU & $295$\,AU & \ding{55} & - \\
         & $500$\,AU & $0.2$\,M$_{\odot}$ & $1$\,M$_{\odot}$ & $0.2$\,M$_{\odot}$ & 0.5 & 0$^\circ$ & $250$\,AU & $204$\,AU & \ding{55} & - \\
         & $500$\,AU & $0.2$\,M$_{\odot}$ & $1$\,M$_{\odot}$ & $0.2$\,M$_{\odot}$ & 0.75 & 0$^\circ$ & $125$\,AU & $104$\,AU & \ding{55} & - \\
         \hline
         \underline{Varying inclination:} & $100$\,AU & $0.2$\,M$_{\odot}$ & $1$\,M$_{\odot}$ & $0.2$\,M$_{\odot}$ & 0 & 30$^\circ$ & $100$\,AU & $77$\,AU & \ding{55} & - \\
         & $100$\,AU & $0.2$\,M$_{\odot}$ & $1$\,M$_{\odot}$ & $0.2$\,M$_{\odot}$ & 0 & 60$^\circ$ & $100$\,AU & $74$\,AU & {\color{Green} \ding{51}\ding{51}} & 835\,yrs \\
         & $100$\,AU & $0.2$\,M$_{\odot}$ & $1$\,M$_{\odot}$ & $0.2$\,M$_{\odot}$ & 0 & 90$^\circ$ & $100$\,AU & $75$\,AU & {\color{Green} \ding{51}\ding{51}} & 835\,yrs \\
         & $150$\,AU & $0.2$\,M$_{\odot}$ & $1$\,M$_{\odot}$ & $0.2$\,M$_{\odot}$ & 0 & 30$^\circ$ & $150$\,AU & $114$\,AU & {\color{Green} \ding{51}\ding{51}} & 720\,yrs \\
         & $150$\,AU & $0.2$\,M$_{\odot}$ & $1$\,M$_{\odot}$ & $0.2$\,M$_{\odot}$ & 0 & 60$^\circ$ & $150$\,AU & $115$\,AU & {\color{Green} \ding{51}\ding{51}} & 1045\,yrs \\
         & $150$\,AU & $0.2$\,M$_{\odot}$ & $1$\,M$_{\odot}$ & $0.2$\,M$_{\odot}$ & 0 & 90$^\circ$ & $150$\,AU & $116$\,AU & {\color{Green} \ding{51}\ding{51}} & 1340\,yrs \\
         & $200$\,AU & $0.2$\,M$_{\odot}$ & $1$\,M$_{\odot}$ & $0.2$\,M$_{\odot}$ & 0 & 30$^\circ$ & $200$\,AU & $151$\,AU & {\color{Green} \ding{51}\ding{51}} & 1600\,yrs \\
         & $200$\,AU & $0.2$\,M$_{\odot}$ & $1$\,M$_{\odot}$ & $0.2$\,M$_{\odot}$ & 0 & 60$^\circ$ & $200$\,AU & $152$\,AU & {\color{Green} \ding{51}\ding{51}} & 2840\,yrs \\
         & $200$\,AU & $0.2$\,M$_{\odot}$ & $1$\,M$_{\odot}$ & $0.2$\,M$_{\odot}$ & 0 & 90$^\circ$ & $200$\,AU & $153$\,AU & \ding{55} & - \\
         & $250$\,AU & $0.2$\,M$_{\odot}$ & $1$\,M$_{\odot}$ & $0.2$\,M$_{\odot}$ & 0 & 30$^\circ$ & $250$\,AU & $187$\,AU & \ding{55} & - \\
         & $250$\,AU & $0.2$\,M$_{\odot}$ & $1$\,M$_{\odot}$ & $0.2$\,M$_{\odot}$ & 0 & 60$^\circ$ & $250$\,AU & $189$\,AU & \ding{55} & - \\
         & $250$\,AU & $0.2$\,M$_{\odot}$ & $1$\,M$_{\odot}$ & $0.2$\,M$_{\odot}$ & 0 & 90$^\circ$ & $250$\,AU & $190$\,AU & \ding{55} & - \\
         \hline
         \underline{Varying companion mass:} & $150$\,AU & $0.2$\,M$_{\odot}$ & $1$\,M$_{\odot}$ & $0.1$\,M$_{\odot}$ & 0 & 0$^\circ$ & $150$\,AU & $110$\,AU & {\color{Green} \ding{51}\ding{51}} & 900\,yrs \\
         & $150$\,AU & $0.2$\,M$_{\odot}$ & $1$\,M$_{\odot}$ & $0.5$\,M$_{\odot}$ & 0 & 0$^\circ$ & $150$\,AU & $112$\,AU & {\color{Green} \ding{51}\ding{51}} & 550\,yrs \\
         & $250$\,AU & $0.2$\,M$_{\odot}$ & $1$\,M$_{\odot}$ & $0.1$\,M$_{\odot}$ & 0 & 0$^\circ$ & $250$\,AU & $181$\,AU & \ding{55} & - \\
         & $250$\,AU & $0.2$\,M$_{\odot}$ & $1$\,M$_{\odot}$ & $0.5$\,M$_{\odot}$ & 0 & 0$^\circ$ & $250$\,AU & $197$\,AU & {\color{Green} \ding{51}\ding{51}} & 1550\,yrs \\
         & $325$\,AU & $0.2$\,M$_{\odot}$ & $1$\,M$_{\odot}$ & $0.1$\,M$_{\odot}$ & 0 & 0$^\circ$ & $325$\,AU & $237$\,AU & \ding{55} & - \\
         & $325$\,AU & $0.2$\,M$_{\odot}$ & $1$\,M$_{\odot}$ & $0.5$\,M$_{\odot}$ & 0 & 0$^\circ$ & $325$\,AU & $256$\,AU & {\color{Green} \ding{51}\ding{51}} & 4000\,yrs \\
         & $400$\,AU & $0.2$\,M$_{\odot}$ & $1$\,M$_{\odot}$ & $0.1$\,M$_{\odot}$ & 0 & 0$^\circ$ & $400$\,AU & $292$\,AU & \ding{55} & - \\
         & $400$\,AU & $0.2$\,M$_{\odot}$ & $1$\,M$_{\odot}$ & $0.5$\,M$_{\odot}$ & 0 & 0$^\circ$ & $400$\,AU & $315$\,AU & \ding{55} & - \\
    \end{tabular}
    \caption{Summary of all discs simulated here, and whether they did or did not fragment. Black ticks: disc fragmented and so did its reference run analog. Double green ticks: disc fragmented when the reference run analog did not. Black crosses: did not fragment. $r_{\rm peri,actual}=$NR ("not reached") denotes systems where the simulation ended before reaching $r_{\rm peri,actual}$, as the disc had already fragmented.}
    \label{tab:sphresultssummary}
\end{table*}

\section*{Acknowledgements}

The authors would like to thank the anonymous referee for their useful comments in helping to improve the work presented here. The simulations presented here were carried out using high performance computing facilities funded by the Scottish Universities Physics Alliance (SUPA). 2D surface density plots were generated using \textsc{SPLASH} \citep{price07}.
CF acknowledges support from the Center for Space and Habitability (CSH). This work has been carried out within the framework of the NCCR PlanetS supported by the Swiss National Science Foundation. KR is grateful for support from the UK STFC via grant ST/V000594/1.

\section*{Data Availability}

The model data generated in this study will be shared on request to the corresponding author.



\bibliographystyle{mnras}
\bibliography{main} 

\begin{thebibliography}{}
\makeatletter
\relax
\def\mn@urlcharsother{\let\do\@makeother \do\$\do\&\do\#\do\^\do\_\do\%\do\~}
\def\mn@doi{\begingroup\mn@urlcharsother \@ifnextchar [ {\mn@doi@}
  {\mn@doi@[]}}
\def\mn@doi@[#1]#2{\def\@tempa{#1}\ifx\@tempa\@empty \href
  {http://dx.doi.org/#2} {doi:#2}\else \href {http://dx.doi.org/#2} {#1}\fi
  \endgroup}
\def\mn@eprint#1#2{\mn@eprint@#1:#2::\@nil}
\def\mn@eprint@arXiv#1{\href {http://arxiv.org/abs/#1} {{\tt arXiv:#1}}}
\def\mn@eprint@dblp#1{\href {http://dblp.uni-trier.de/rec/bibtex/#1.xml}
  {dblp:#1}}
\def\mn@eprint@#1:#2:#3:#4\@nil{\def\@tempa {#1}\def\@tempb {#2}\def\@tempc
  {#3}\ifx \@tempc \@empty \let \@tempc \@tempb \let \@tempb \@tempa \fi \ifx
  \@tempb \@empty \def\@tempb {arXiv}\fi \@ifundefined
  {mn@eprint@\@tempb}{\@tempb:\@tempc}{\expandafter \expandafter \csname
  mn@eprint@\@tempb\endcsname \expandafter{\@tempc}}}

\bibitem[\protect\citeauthoryear{{Baruteau}, {Meru}  \&
  {Paardekooper}}{{Baruteau} et~al.}{2011}]{baruteau11}
{Baruteau} C.,  {Meru} F.,   {Paardekooper} S.-J.,  2011, \mn@doi [\mnras]
  {10.1111/j.1365-2966.2011.19172.x}, \href
  {https://ui.adsabs.harvard.edu/abs/2011MNRAS.416.1971B} {416, 1971}

\bibitem[\protect\citeauthoryear{{Benz}}{{Benz}}{1990}]{benz1990}
{Benz} W.,  1990, in {Buchler} J.~R.,  ed., Numerical Modelling of Nonlinear
  Stellar Pulsations Problems and Prospects. p.~269

\bibitem[\protect\citeauthoryear{{Bergfors} et~al.,}{{Bergfors}
  et~al.}{2013}]{bergfors13}
{Bergfors} C.,  et~al., 2013, \mn@doi [\mnras] {10.1093/mnras/sts019}, \href
  {https://ui.adsabs.harvard.edu/abs/2013MNRAS.428..182B} {428, 182}

\bibitem[\protect\citeauthoryear{{Boffin}, {Watkins}, {Bhattal}, {Francis}  \&
  {Whitworth}}{{Boffin} et~al.}{1998}]{boffinetal98}
{Boffin} H.~M.~J.,  {Watkins} S.~J.,  {Bhattal} A.~S.,  {Francis} N.,
  {Whitworth} A.~P.,  1998, \mn@doi [\mnras]
  {10.1046/j.1365-8711.1998.01986.x}, \href
  {https://ui.adsabs.harvard.edu/abs/1998MNRAS.300.1189B} {300, 1189}

\bibitem[\protect\citeauthoryear{{Boley}}{{Boley}}{2009}]{boley09}
{Boley} A.~C.,  2009, \mn@doi [\apjl] {10.1088/0004-637X/695/1/L53}, \href
  {https://ui.adsabs.harvard.edu/abs/2009ApJ...695L..53B} {695, L53}

\bibitem[\protect\citeauthoryear{{Boley}, {Hayfield}, {Mayer}  \&
  {Durisen}}{{Boley} et~al.}{2010}]{boley10}
{Boley} A.~C.,  {Hayfield} T.,  {Mayer} L.,   {Durisen} R.~H.,  2010, \mn@doi
  [\icarus] {10.1016/j.icarus.2010.01.015}, \href
  {https://ui.adsabs.harvard.edu/abs/2010Icar..207..509B} {207, 509}

\bibitem[\protect\citeauthoryear{{Boss}}{{Boss}}{1997}]{boss97}
{Boss} A.~P.,  1997, \mn@doi [Science] {10.1126/science.276.5320.1836}, \href
  {https://ui.adsabs.harvard.edu/abs/1997Sci...276.1836B} {276, 1836}

\bibitem[\protect\citeauthoryear{{Boss}}{{Boss}}{2000}]{boss00}
{Boss} A.~P.,  2000, \mn@doi [\apjl] {10.1086/312737}, \href
  {https://ui.adsabs.harvard.edu/abs/2000ApJ...536L.101B} {536, L101}

\bibitem[\protect\citeauthoryear{{Boss}}{{Boss}}{2006}]{bossetal06}
{Boss} A.~P.,  2006, \mn@doi [\apj] {10.1086/500530}, \href
  {https://ui.adsabs.harvard.edu/abs/2006ApJ...641.1148B} {641, 1148}

\bibitem[\protect\citeauthoryear{{Cadman}, {Rice}, {Hall}, {Haworth}  \&
  {Biller}}{{Cadman} et~al.}{2020a}]{cadman2020frag}
{Cadman} J.,  {Rice} K.,  {Hall} C.,  {Haworth} T.~J.,   {Biller} B.,  2020a,
  \mn@doi [\mnras] {10.1093/mnras/staa187}, \href
  {https://ui.adsabs.harvard.edu/abs/2020MNRAS.492.5041C} {492, 5041}

\bibitem[\protect\citeauthoryear{{Cadman}, {Hall}, {Rice}, {Harries}  \&
  {Klaassen}}{{Cadman} et~al.}{2020b}]{cadman2020b}
{Cadman} J.,  {Hall} C.,  {Rice} K.,  {Harries} T.~J.,   {Klaassen} P.~D.,
  2020b, \mn@doi [\mnras] {10.1093/mnras/staa2596}, \href
  {https://ui.adsabs.harvard.edu/abs/2020MNRAS.498.4256C} {498, 4256}

\bibitem[\protect\citeauthoryear{{Cadman}, {Rice}  \& {Hall}}{{Cadman}
  et~al.}{2021}]{cadman2021}
{Cadman} J.,  {Rice} K.,   {Hall} C.,  2021, \mn@doi [\mnras]
  {10.1093/mnras/stab905}, \href
  {https://ui.adsabs.harvard.edu/abs/2021MNRAS.504.2877C} {504, 2877}

\bibitem[\protect\citeauthoryear{{Clarke}}{{Clarke}}{2009}]{clarke09}
{Clarke} C.~J.,  2009, \mn@doi [\mnras] {10.1111/j.1365-2966.2009.14774.x},
  \href {https://ui.adsabs.harvard.edu/abs/2009MNRAS.396.1066C} {396, 1066}

\bibitem[\protect\citeauthoryear{{Dong}, {Hall}, {Rice}  \& {Chiang}}{{Dong}
  et~al.}{2015}]{dong2015}
{Dong} R.,  {Hall} C.,  {Rice} K.,   {Chiang} E.,  2015, \mn@doi [\apjl]
  {10.1088/2041-8205/812/2/L32}, \href
  {https://ui.adsabs.harvard.edu/abs/2015ApJ...812L..32D} {812, L32}

\bibitem[\protect\citeauthoryear{{Durisen}, {Boss}, {Mayer}, {Nelson}, {Quinn}
  \& {Rice}}{{Durisen} et~al.}{2007}]{durisen07}
{Durisen} R.~H.,  {Boss} A.~P.,  {Mayer} L.,  {Nelson} A.~F.,  {Quinn} T.,
  {Rice} W.~K.~M.,  2007, in {Reipurth} B.,  {Jewitt} D.,   {Keil} K.,  eds,
  Protostars and Planets V. p.~607 (\mn@eprint {arXiv} {astro-ph/0603179})

\bibitem[\protect\citeauthoryear{{Fischer} \& {Valenti}}{{Fischer} \&
  {Valenti}}{2005}]{fischervalenti05}
{Fischer} D.~A.,  {Valenti} J.,  2005, \mn@doi [\apj] {10.1086/428383}, \href
  {https://ui.adsabs.harvard.edu/abs/2005ApJ...622.1102F} {622, 1102}

\bibitem[\protect\citeauthoryear{{Fontanive} \& {Bardalez
  Gagliuffi}}{{Fontanive} \& {Bardalez Gagliuffi}}{2021}]{fontanive21}
{Fontanive} C.,  {Bardalez Gagliuffi} D.,  2021, \mn@doi [Frontiers in
  Astronomy and Space Sciences] {10.3389/fspas.2021.625250}, \href
  {https://ui.adsabs.harvard.edu/abs/2021FrASS...8...16F} {8, 16}

\bibitem[\protect\citeauthoryear{{Fontanive}, {Rice}, {Bonavita}, {Lopez},
  {Mu{\v{z}}i{\'c}}, {}  \& {Biller}}{{Fontanive} et~al.}{2019}]{fontanive19}
{Fontanive} C.,  {Rice} K.,  {Bonavita} M.,  {Lopez} E.,  {Mu{\v{z}}i{\'c}} {}
  K.,   {Biller} B.,  2019, \mn@doi [\mnras] {10.1093/mnras/stz671}, \href
  {https://ui.adsabs.harvard.edu/abs/2019MNRAS.485.4967F} {485, 4967}

\bibitem[\protect\citeauthoryear{{Forgan} \& {Rice}}{{Forgan} \&
  {Rice}}{2009}]{forganrice2009}
{Forgan} D.,  {Rice} K.,  2009, \mn@doi [\mnras]
  {10.1111/j.1365-2966.2009.15596.x}, \href
  {https://ui.adsabs.harvard.edu/abs/2009MNRAS.400.2022F} {400, 2022}

\bibitem[\protect\citeauthoryear{{Forgan} \& {Rice}}{{Forgan} \&
  {Rice}}{2011}]{forganrice11}
{Forgan} D.,  {Rice} K.,  2011, \mn@doi [\mnras]
  {10.1111/j.1365-2966.2011.19380.x}, \href
  {https://ui.adsabs.harvard.edu/abs/2011MNRAS.417.1928F} {417, 1928}

\bibitem[\protect\citeauthoryear{{Forgan} \& {Rice}}{{Forgan} \&
  {Rice}}{2013}]{forgan13}
{Forgan} D.,  {Rice} K.,  2013, \mn@doi [\mnras] {10.1093/mnras/stt672}, \href
  {https://ui.adsabs.harvard.edu/abs/2013MNRAS.432.3168F} {432, 3168}

\bibitem[\protect\citeauthoryear{{Forgan}, {Rice}, {Stamatellos}  \&
  {Whitworth}}{{Forgan} et~al.}{2009}]{forganetal2009}
{Forgan} D.,  {Rice} K.,  {Stamatellos} D.,   {Whitworth} A.,  2009, \mn@doi
  [\mnras] {10.1111/j.1365-2966.2008.14373.x}, \href
  {https://ui.adsabs.harvard.edu/abs/2009MNRAS.394..882F} {394, 882}

\bibitem[\protect\citeauthoryear{{Forgan}, {Hall}, {Meru}  \& {Rice}}{{Forgan}
  et~al.}{2018}]{forgan18}
{Forgan} D.~H.,  {Hall} C.,  {Meru} F.,   {Rice} W.~K.~M.,  2018, \mn@doi
  [\mnras] {10.1093/mnras/stx2870}, \href
  {https://ui.adsabs.harvard.edu/abs/2018MNRAS.474.5036F} {474, 5036}

\bibitem[\protect\citeauthoryear{{Gammie}}{{Gammie}}{2001}]{gammie01}
{Gammie} C.~F.,  2001, \mn@doi [\apj] {10.1086/320631}, \href
  {https://ui.adsabs.harvard.edu/abs/2001ApJ...553..174G} {553, 174}

\bibitem[\protect\citeauthoryear{{Hall}, {Forgan}, {Rice}, {Harries},
  {Klaassen}  \& {Biller}}{{Hall} et~al.}{2016}]{hall2016}
{Hall} C.,  {Forgan} D.,  {Rice} K.,  {Harries} T.~J.,  {Klaassen} P.~D.,
  {Biller} B.,  2016, \mn@doi [\mnras] {10.1093/mnras/stw296}, \href
  {https://ui.adsabs.harvard.edu/abs/2016MNRAS.458..306H} {458, 306}

\bibitem[\protect\citeauthoryear{{Hall}, {Forgan}  \& {Rice}}{{Hall}
  et~al.}{2017}]{hall2017}
{Hall} C.,  {Forgan} D.,   {Rice} K.,  2017, \mn@doi [\mnras]
  {10.1093/mnras/stx1244}, \href
  {https://ui.adsabs.harvard.edu/abs/2017MNRAS.470.2517H} {470, 2517}

\bibitem[\protect\citeauthoryear{{Hall}, {Rice}, {Dipierro}, {Forgan},
  {Harries}  \& {Alexander}}{{Hall} et~al.}{2018}]{hall18}
{Hall} C.,  {Rice} K.,  {Dipierro} G.,  {Forgan} D.,  {Harries} T.,
  {Alexander} R.,  2018, \mn@doi [\mnras] {10.1093/mnras/sty550}, \href
  {https://ui.adsabs.harvard.edu/abs/2018MNRAS.477.1004H} {477, 1004}

\bibitem[\protect\citeauthoryear{{Hall} et~al.,}{{Hall}
  et~al.}{2020}]{hall2020}
{Hall} C.,  et~al., 2020, \mn@doi [\apj] {10.3847/1538-4357/abac17}, \href
  {https://ui.adsabs.harvard.edu/abs/2020ApJ...904..148H} {904, 148}

\bibitem[\protect\citeauthoryear{{Haworth}, {Cadman}, {Meru}, {Hall},
  {Albertini}, {Forgan}, {Rice}  \& {Owen}}{{Haworth}
  et~al.}{2020}]{haworth2020}
{Haworth} T.~J.,  {Cadman} J.,  {Meru} F.,  {Hall} C.,  {Albertini} E.,
  {Forgan} D.,  {Rice} K.,   {Owen} J.~E.,  2020, \mn@doi [\mnras]
  {10.1093/mnras/staa883}, \href
  {https://ui.adsabs.harvard.edu/abs/2020MNRAS.494.4130H} {494, 4130}

\bibitem[\protect\citeauthoryear{{Humphries}, {Hall}, {Haworth}  \&
  {Nayakshin}}{{Humphries} et~al.}{2021}]{humphries2021}
{Humphries} J.,  {Hall} C.,  {Haworth} T.~J.,   {Nayakshin} S.,  2021, \mn@doi
  [\mnras] {10.1093/mnras/staa2411}, \href
  {https://ui.adsabs.harvard.edu/abs/2021MNRAS.502..953H} {502, 953}

\bibitem[\protect\citeauthoryear{{Jenkins} et~al.,}{{Jenkins}
  et~al.}{2017}]{jenkins17}
{Jenkins} J.~S.,  et~al., 2017, \mn@doi [\mnras] {10.1093/mnras/stw2811}, \href
  {http://adsabs.harvard.edu/abs/2017MNRAS.466..443J} {466, 443}

\bibitem[\protect\citeauthoryear{{Johnson} \& {Li}}{{Johnson} \&
  {Li}}{2013}]{johnson13}
{Johnson} J.~L.,  {Li} H.,  2013, \mn@doi [\mnras] {10.1093/mnras/stt229},
  \href {https://ui.adsabs.harvard.edu/abs/2013MNRAS.431..972J} {431, 972}

\bibitem[\protect\citeauthoryear{{Kaib}, {Raymond}  \& {Duncan}}{{Kaib}
  et~al.}{2013}]{kaib13}
{Kaib} N.~A.,  {Raymond} S.~N.,   {Duncan} M.,  2013, \mn@doi [\nat]
  {10.1038/nature11780}, \href
  {https://ui.adsabs.harvard.edu/abs/2013Natur.493..381K} {493, 381}

\bibitem[\protect\citeauthoryear{{Kratter}, {Murray-Clay}  \&
  {Youdin}}{{Kratter} et~al.}{2010}]{kratter10}
{Kratter} K.~M.,  {Murray-Clay} R.~A.,   {Youdin} A.~N.,  2010, \mn@doi [\apj]
  {10.1088/0004-637X/710/2/1375}, \href
  {https://ui.adsabs.harvard.edu/abs/2010ApJ...710.1375K} {710, 1375}

\bibitem[\protect\citeauthoryear{{Kraus}, {Ireland}, {Hillenbrand}  \&
  {Martinache}}{{Kraus} et~al.}{2012}]{kraus12}
{Kraus} A.~L.,  {Ireland} M.~J.,  {Hillenbrand} L.~A.,   {Martinache} F.,
  2012, \mn@doi [\apj] {10.1088/0004-637X/745/1/19}, \href
  {https://ui.adsabs.harvard.edu/abs/2012ApJ...745...19K} {745, 19}

\bibitem[\protect\citeauthoryear{{Kraus}, {Ireland}, {Huber}, {Mann}  \&
  {Dupuy}}{{Kraus} et~al.}{2016}]{kraus16}
{Kraus} A.~L.,  {Ireland} M.~J.,  {Huber} D.,  {Mann} A.~W.,   {Dupuy} T.~J.,
  2016, \mn@doi [\aj] {10.3847/0004-6256/152/1/8}, \href
  {https://ui.adsabs.harvard.edu/abs/2016AJ....152....8K} {152, 8}

\bibitem[\protect\citeauthoryear{{Laughlin} \& {Bodenheimer}}{{Laughlin} \&
  {Bodenheimer}}{1994}]{laughlin94}
{Laughlin} G.,  {Bodenheimer} P.,  1994, \mn@doi [\apj] {10.1086/174909}, \href
  {https://ui.adsabs.harvard.edu/abs/1994ApJ...436..335L} {436, 335}

\bibitem[\protect\citeauthoryear{{Lin} \& {Pringle}}{{Lin} \&
  {Pringle}}{1987}]{linpringle87}
{Lin} D.~N.~C.,  {Pringle} J.~E.,  1987, \mn@doi [\mnras]
  {10.1093/mnras/225.3.607}, \href
  {https://ui.adsabs.harvard.edu/abs/1987MNRAS.225..607L} {225, 607}

\bibitem[\protect\citeauthoryear{{Lodato} \& {Rice}}{{Lodato} \&
  {Rice}}{2004}]{lodato04}
{Lodato} G.,  {Rice} W.~K.~M.,  2004, \mn@doi [\mnras]
  {10.1111/j.1365-2966.2004.07811.x}, \href
  {https://ui.adsabs.harvard.edu/abs/2004MNRAS.351..630L} {351, 630}

\bibitem[\protect\citeauthoryear{{Lodato}, {Meru}, {Clarke}  \&
  {Rice}}{{Lodato} et~al.}{2007}]{lodatoetal07}
{Lodato} G.,  {Meru} F.,  {Clarke} C.~J.,   {Rice} W.~K.~M.,  2007, \mn@doi
  [\mnras] {10.1111/j.1365-2966.2006.11211.x}, \href
  {https://ui.adsabs.harvard.edu/abs/2007MNRAS.374..590L} {374, 590}

\bibitem[\protect\citeauthoryear{{Mayer}, {Quinn}, {Wadsley}  \&
  {Stadel}}{{Mayer} et~al.}{2002}]{mayer02}
{Mayer} L.,  {Quinn} T.,  {Wadsley} J.,   {Stadel} J.,  2002, \mn@doi [Science]
  {10.1126/science.1077635}, \href
  {https://ui.adsabs.harvard.edu/abs/2002Sci...298.1756M} {298, 1756}

\bibitem[\protect\citeauthoryear{{Mayer}, {Wadsley}, {Quinn}  \&
  {Stadel}}{{Mayer} et~al.}{2005}]{mayeretal05}
{Mayer} L.,  {Wadsley} J.,  {Quinn} T.,   {Stadel} J.,  2005, \mn@doi [\mnras]
  {10.1111/j.1365-2966.2005.09468.x}, \href
  {https://ui.adsabs.harvard.edu/abs/2005MNRAS.363..641M} {363, 641}

\bibitem[\protect\citeauthoryear{{Mayer}, {Boss}  \& {Nelson}}{{Mayer}
  et~al.}{2007a}]{mayeretal07}
{Mayer} L.,  {Boss} A.,   {Nelson} A.~F.,  2007a, arXiv e-prints, \href
  {https://ui.adsabs.harvard.edu/abs/2007arXiv0705.3182M} {p. arXiv:0705.3182}

\bibitem[\protect\citeauthoryear{{Mayer}, {Lufkin}, {Quinn}  \&
  {Wadsley}}{{Mayer} et~al.}{2007b}]{mayeretal07b}
{Mayer} L.,  {Lufkin} G.,  {Quinn} T.,   {Wadsley} J.,  2007b, \mn@doi [\apjl]
  {10.1086/518433}, \href
  {https://ui.adsabs.harvard.edu/abs/2007ApJ...661L..77M} {661, L77}

\bibitem[\protect\citeauthoryear{{Meru}}{{Meru}}{2015}]{meru2015}
{Meru} F.,  2015, \mn@doi [\mnras] {10.1093/mnras/stv2128}, \href
  {https://ui.adsabs.harvard.edu/abs/2015MNRAS.454.2529M} {454, 2529}

\bibitem[\protect\citeauthoryear{{Meru} \& {Bate}}{{Meru} \&
  {Bate}}{2010}]{merubate10}
{Meru} F.,  {Bate} M.~R.,  2010, \mn@doi [\mnras]
  {10.1111/j.1365-2966.2010.16867.x}, \href
  {https://ui.adsabs.harvard.edu/abs/2010MNRAS.406.2279M} {406, 2279}

\bibitem[\protect\citeauthoryear{{Meru}, {Juh{\'a}sz}, {Ilee}, {Clarke},
  {Rosotti}  \& {Booth}}{{Meru} et~al.}{2017}]{meru17}
{Meru} F.,  {Juh{\'a}sz} A.,  {Ilee} J.~D.,  {Clarke} C.~J.,  {Rosotti} G.~P.,
   {Booth} R.~A.,  2017, \mn@doi [\apjl] {10.3847/2041-8213/aa6837}, \href
  {https://ui.adsabs.harvard.edu/abs/2017ApJ...839L..24M} {839, L24}

\bibitem[\protect\citeauthoryear{{Moe} \& {Kratter}}{{Moe} \&
  {Kratter}}{2019}]{moekratter19}
{Moe} M.,  {Kratter} K.~M.,  2019, arXiv e-prints, \href
  {https://ui.adsabs.harvard.edu/abs/2019arXiv191201699M} {p. arXiv:1912.01699}

\bibitem[\protect\citeauthoryear{{Moe}, {Kratter}  \& {Badenes}}{{Moe}
  et~al.}{2019}]{moe19}
{Moe} M.,  {Kratter} K.~M.,   {Badenes} C.,  2019, \mn@doi [\apj]
  {10.3847/1538-4357/ab0d88}, \href
  {https://ui.adsabs.harvard.edu/abs/2019ApJ...875...61M} {875, 61}

\bibitem[\protect\citeauthoryear{{Monaghan}}{{Monaghan}}{1992}]{monaghan1992}
{Monaghan} J.~J.,  1992, \mn@doi [\araa] {10.1146/annurev.aa.30.090192.002551},
  \href {https://ui.adsabs.harvard.edu/abs/1992ARA&A..30..543M} {30, 543}

\bibitem[\protect\citeauthoryear{{Mordasini}, {Alibert}, {Benz}, {Klahr}  \&
  {Henning}}{{Mordasini} et~al.}{2012}]{mordasini12}
{Mordasini} C.,  {Alibert} Y.,  {Benz} W.,  {Klahr} H.,   {Henning} T.,  2012,
  \mn@doi [\aap] {10.1051/0004-6361/201117350}, \href
  {https://ui.adsabs.harvard.edu/abs/2012A&A...541A..97M} {541, A97}

\bibitem[\protect\citeauthoryear{{Nayakshin}}{{Nayakshin}}{2010}]{nayakshin10}
{Nayakshin} S.,  2010, \mn@doi [\mnras] {10.1111/j.1745-3933.2010.00923.x},
  \href {https://ui.adsabs.harvard.edu/abs/2010MNRAS.408L..36N} {408, L36}

\bibitem[\protect\citeauthoryear{{Nayakshin} \& {Fletcher}}{{Nayakshin} \&
  {Fletcher}}{2015}]{nayakshin15}
{Nayakshin} S.,  {Fletcher} M.,  2015, \mn@doi [\mnras]
  {10.1093/mnras/stv1354}, \href
  {https://ui.adsabs.harvard.edu/abs/2015MNRAS.452.1654N} {452, 1654}

\bibitem[\protect\citeauthoryear{{Nelson}}{{Nelson}}{2000}]{nelson2000}
{Nelson} A.~F.,  2000, \mn@doi [\apjl] {10.1086/312752}, \href
  {https://ui.adsabs.harvard.edu/abs/2000ApJ...537L..65N} {537, L65}

\bibitem[\protect\citeauthoryear{{Nero} \& {Bjorkman}}{{Nero} \&
  {Bjorkman}}{2009}]{nero09}
{Nero} D.,  {Bjorkman} J.~E.,  2009, \mn@doi [\apjl]
  {10.1088/0004-637X/702/2/L163}, \href
  {https://ui.adsabs.harvard.edu/abs/2009ApJ...702L.163N} {702, L163}

\bibitem[\protect\citeauthoryear{{Ngo} et~al.,}{{Ngo} et~al.}{2016}]{ngoetal16}
{Ngo} H.,  et~al., 2016, \mn@doi [\apj] {10.3847/0004-637X/827/1/8}, \href
  {https://ui.adsabs.harvard.edu/abs/2016ApJ...827....8N} {827, 8}

\bibitem[\protect\citeauthoryear{{Nielsen} et~al.,}{{Nielsen}
  et~al.}{2019}]{nielsenetal19}
{Nielsen} E.~L.,  et~al., 2019, \mn@doi [\aj] {10.3847/1538-3881/ab16e9}, \href
  {https://ui.adsabs.harvard.edu/abs/2019AJ....158...13N} {158, 13}

\bibitem[\protect\citeauthoryear{{Paczynski}}{{Paczynski}}{1978}]{paczynski78}
{Paczynski} B.,  1978, \actaa, \href
  {https://ui.adsabs.harvard.edu/abs/1978AcA....28...91P} {28, 91}

\bibitem[\protect\citeauthoryear{{Paneque-Carre{\~n}o}
  et~al.,}{{Paneque-Carre{\~n}o} et~al.}{2021}]{paneque2021}
{Paneque-Carre{\~n}o} T.,  et~al., 2021, \mn@doi [\apj]
  {10.3847/1538-4357/abf243}, \href
  {https://ui.adsabs.harvard.edu/abs/2021ApJ...914...88P} {914, 88}

\bibitem[\protect\citeauthoryear{{P{\'e}rez} et~al.,}{{P{\'e}rez}
  et~al.}{2016}]{perez16}
{P{\'e}rez} L.~M.,  et~al., 2016, \mn@doi [Science] {10.1126/science.aaf8296},
  \href {https://ui.adsabs.harvard.edu/abs/2016Sci...353.1519P} {353, 1519}

\bibitem[\protect\citeauthoryear{{Pollack}, {Hubickyj}, {Bodenheimer},
  {Lissauer}, {Podolak}  \& {Greenzweig}}{{Pollack} et~al.}{1996}]{pollack96}
{Pollack} J.~B.,  {Hubickyj} O.,  {Bodenheimer} P.,  {Lissauer} J.~J.,
  {Podolak} M.,   {Greenzweig} Y.,  1996, \mn@doi [\icarus]
  {10.1006/icar.1996.0190}, \href
  {http://adsabs.harvard.edu/abs/1996Icar..124...62P} {124, 62}

\bibitem[\protect\citeauthoryear{{Price}}{{Price}}{2007}]{price07}
{Price} D.~J.,  2007, \mn@doi [\pasa] {10.1071/AS07022}, \href
  {https://ui.adsabs.harvard.edu/abs/2007PASA...24..159P} {24, 159}

\bibitem[\protect\citeauthoryear{{Price} et~al.,}{{Price}
  et~al.}{2018}]{priceetal18}
{Price} D.~J.,  et~al., 2018, \mn@doi [\pasa] {10.1017/pasa.2018.25}, \href
  {https://ui.adsabs.harvard.edu/abs/2018PASA...35...31P} {35, e031}

\bibitem[\protect\citeauthoryear{{Rafikov}}{{Rafikov}}{2005}]{rafikov05}
{Rafikov} R.~R.,  2005, \mn@doi [\apjl] {10.1086/428899}, \href
  {https://ui.adsabs.harvard.edu/abs/2005ApJ...621L..69R} {621, L69}

\bibitem[\protect\citeauthoryear{{Rice} \& {Armitage}}{{Rice} \&
  {Armitage}}{2009}]{rice09}
{Rice} W.~K.~M.,  {Armitage} P.~J.,  2009, \mn@doi [\mnras]
  {10.1111/j.1365-2966.2009.14879.x}, \href
  {https://ui.adsabs.harvard.edu/abs/2009MNRAS.396.2228R} {396, 2228}

\bibitem[\protect\citeauthoryear{{Rice}, {Armitage}, {Bate}  \&
  {Bonnell}}{{Rice} et~al.}{2003}]{rice03}
{Rice} W.~K.~M.,  {Armitage} P.~J.,  {Bate} M.~R.,   {Bonnell} I.~A.,  2003,
  \mn@doi [\mnras] {10.1046/j.1365-8711.2003.06253.x}, \href
  {https://ui.adsabs.harvard.edu/abs/2003MNRAS.339.1025R} {339, 1025}

\bibitem[\protect\citeauthoryear{{Rice}, {Mayo}  \& {Armitage}}{{Rice}
  et~al.}{2010}]{rice10}
{Rice} W.~K.~M.,  {Mayo} J.~H.,   {Armitage} P.~J.,  2010, \mn@doi [\mnras]
  {10.1111/j.1365-2966.2009.15992.x}, \href
  {https://ui.adsabs.harvard.edu/abs/2010MNRAS.402.1740R} {402, 1740}

\bibitem[\protect\citeauthoryear{{Rice}, {Lopez}, {Forgan}  \& {Biller}}{{Rice}
  et~al.}{2015}]{rice15}
{Rice} K.,  {Lopez} E.,  {Forgan} D.,   {Biller} B.,  2015, \mn@doi [\mnras]
  {10.1093/mnras/stv1997}, \href
  {https://ui.adsabs.harvard.edu/abs/2015MNRAS.454.1940R} {454, 1940}

\bibitem[\protect\citeauthoryear{{Rodr{\'\i}guez}, {Loinard}, {D'Alessio},
  {Wilner}  \& {Ho}}{{Rodr{\'\i}guez} et~al.}{2005}]{rodriguez05}
{Rodr{\'\i}guez} L.~F.,  {Loinard} L.,  {D'Alessio} P.,  {Wilner} D.~J.,   {Ho}
  P. T.~P.,  2005, \mn@doi [\apjl] {10.1086/429223}, \href
  {https://ui.adsabs.harvard.edu/abs/2005ApJ...621L.133R} {621, L133}

\bibitem[\protect\citeauthoryear{{Santos}, {Israelian}  \& {Mayor}}{{Santos}
  et~al.}{2004}]{santos04}
{Santos} N.~C.,  {Israelian} G.,   {Mayor} M.,  2004, \mn@doi [\aap]
  {10.1051/0004-6361:20034469}, \href
  {https://ui.adsabs.harvard.edu/abs/2004A&A...415.1153S} {415, 1153}

\bibitem[\protect\citeauthoryear{{Schlaufman}}{{Schlaufman}}{2018}]{schlaufman18}
{Schlaufman} K.~C.,  2018, \mn@doi [\apj] {10.3847/1538-4357/aa961c}, \href
  {https://ui.adsabs.harvard.edu/abs/2018ApJ...853...37S} {853, 37}

\bibitem[\protect\citeauthoryear{{Stamatellos} \& {Whitworth}}{{Stamatellos} \&
  {Whitworth}}{2009}]{stamatellos09}
{Stamatellos} D.,  {Whitworth} A.~P.,  2009, \mn@doi [\mnras]
  {10.1111/j.1365-2966.2008.14069.x}, \href
  {https://ui.adsabs.harvard.edu/abs/2009MNRAS.392..413S} {392, 413}

\bibitem[\protect\citeauthoryear{{Stamatellos}, {Whitworth}, {Bisbas}  \&
  {Goodwin}}{{Stamatellos} et~al.}{2007}]{stamatellosetal07}
{Stamatellos} D.,  {Whitworth} A.~P.,  {Bisbas} T.,   {Goodwin} S.,  2007,
  \mn@doi [\aap] {10.1051/0004-6361:20077373}, \href
  {https://ui.adsabs.harvard.edu/abs/2007A&A...475...37S} {475, 37}

\bibitem[\protect\citeauthoryear{{Tobin}, {Hartmann}, {Chiang}, {Wilner},
  {Looney}, {Loinard}, {Calvet}  \& {D'Alessio}}{{Tobin}
  et~al.}{2012}]{tobin12}
{Tobin} J.~J.,  {Hartmann} L.,  {Chiang} H.-F.,  {Wilner} D.~J.,  {Looney}
  L.~W.,  {Loinard} L.,  {Calvet} N.,   {D'Alessio} P.,  2012, \mn@doi [\nat]
  {10.1038/nature11610}, \href
  {https://ui.adsabs.harvard.edu/abs/2012Natur.492...83T} {492, 83}

\bibitem[\protect\citeauthoryear{{Tobin} et~al.,}{{Tobin}
  et~al.}{2015}]{tobin15}
{Tobin} J.~J.,  et~al., 2015, \mn@doi [\apj] {10.1088/0004-637X/805/2/125},
  \href {https://ui.adsabs.harvard.edu/abs/2015ApJ...805..125T} {805, 125}

\bibitem[\protect\citeauthoryear{{Tokovinin}, {Thomas}, {Sterzik}  \&
  {Udry}}{{Tokovinin} et~al.}{2006}]{tokovinin06}
{Tokovinin} A.,  {Thomas} S.,  {Sterzik} M.,   {Udry} S.,  2006, \mn@doi [\aap]
  {10.1051/0004-6361:20054427}, \href
  {https://ui.adsabs.harvard.edu/abs/2006A&A...450..681T} {450, 681}

\bibitem[\protect\citeauthoryear{{Toomre}}{{Toomre}}{1964}]{toomre64}
{Toomre} A.,  1964, \mn@doi [\apj] {10.1086/147861}, \href
  {https://ui.adsabs.harvard.edu/abs/1964ApJ...139.1217T} {139, 1217}

\bibitem[\protect\citeauthoryear{{Veronesi}, {Paneque-Carre{\~n}o}, {Lodato},
  {Testi}, {P{\'e}rez}, {Bertin}  \& {Hall}}{{Veronesi}
  et~al.}{2021}]{veronesi2021}
{Veronesi} B.,  {Paneque-Carre{\~n}o} T.,  {Lodato} G.,  {Testi} L.,
  {P{\'e}rez} L.~M.,  {Bertin} G.,   {Hall} C.,  2021, \mn@doi [\apjl]
  {10.3847/2041-8213/abfe6a}, \href
  {https://ui.adsabs.harvard.edu/abs/2021ApJ...914L..27V} {914, L27}

\bibitem[\protect\citeauthoryear{{Vigan} et~al.,}{{Vigan}
  et~al.}{2017}]{Vigan17}
{Vigan} A.,  et~al., 2017, \mn@doi [\aap] {10.1051/0004-6361/201630133}, \href
  {https://ui.adsabs.harvard.edu/abs/2017A&A...603A...3V} {603, A3}

\bibitem[\protect\citeauthoryear{{Vigan} et~al.,}{{Vigan}
  et~al.}{2021}]{vigan21}
{Vigan} A.,  et~al., 2021, \mn@doi [\aap] {10.1051/0004-6361/202038107}, \href
  {https://ui.adsabs.harvard.edu/abs/2021A&A...651A..72V} {651, A72}

\bibitem[\protect\citeauthoryear{{Vorobyov} \& {Basu}}{{Vorobyov} \&
  {Basu}}{2010}]{vorobyov10}
{Vorobyov} E.~I.,  {Basu} S.,  2010, \mn@doi [\apjl]
  {10.1088/2041-8205/714/1/L133}, \href
  {https://ui.adsabs.harvard.edu/abs/2010ApJ...714L.133V} {714, L133}

\bibitem[\protect\citeauthoryear{{Wang}, {Xie}, {Barclay}  \& {Fischer}}{{Wang}
  et~al.}{2014}]{wang14}
{Wang} J.,  {Xie} J.-W.,  {Barclay} T.,   {Fischer} D.~A.,  2014, \mn@doi
  [\apj] {10.1088/0004-637X/783/1/4}, \href
  {https://ui.adsabs.harvard.edu/abs/2014ApJ...783....4W} {783, 4}

\bibitem[\protect\citeauthoryear{{Watkins}, {Bhattal}, {Boffin}, {Francis}  \&
  {Whitworth}}{{Watkins} et~al.}{1998a}]{watkinsetal98a}
{Watkins} S.~J.,  {Bhattal} A.~S.,  {Boffin} H.~M.~J.,  {Francis} N.,
  {Whitworth} A.~P.,  1998a, \mn@doi [\mnras]
  {10.1046/j.1365-8711.1998.01987.x}, \href
  {https://ui.adsabs.harvard.edu/abs/1998MNRAS.300.1205W} {300, 1205}

\bibitem[\protect\citeauthoryear{{Watkins}, {Bhattal}, {Boffin}, {Francis}  \&
  {Whitworth}}{{Watkins} et~al.}{1998b}]{watkinsetal98b}
{Watkins} S.~J.,  {Bhattal} A.~S.,  {Boffin} H.~M.~J.,  {Francis} N.,
  {Whitworth} A.~P.,  1998b, \mn@doi [\mnras]
  {10.1046/j.1365-8711.1998.01988.x}, \href
  {https://ui.adsabs.harvard.edu/abs/1998MNRAS.300.1214W} {300, 1214}

\bibitem[\protect\citeauthoryear{{Zucker} \& {Mazeh}}{{Zucker} \&
  {Mazeh}}{2002}]{zuckermazeh02}
{Zucker} S.,  {Mazeh} T.,  2002, \mn@doi [\apjl] {10.1086/340373}, \href
  {https://ui.adsabs.harvard.edu/abs/2002ApJ...568L.113Z} {568, L113}

\makeatother
\end{thebibliography}



\appendix


\bsp	
\label{lastpage}
\end{document}